\documentstyle[aps,eqsecnum,amssymb,graphicx,prd]{revtex}
\newcommand{\leftpartial}{\stackrel{\leftarrow}{\partial}}
\newcommand{\rightpartial}{\stackrel{\rightarrow}{\partial}}
\newcommand{\bothpartial}{\stackrel{\leftrightarrow}{\partial}}

\newcommand{\slbothpartial}{\not\!\!{\stackrel{\leftrightarrow}{\partial}}}
\newcommand{\Jcurrent}[5]{{J}^{#1 #2}_{\rm #5}\!\left( #3,#4 \right)} 
\newcommand{\Jcurrentdn}[4]{{J}_{#1 #2}\!\left( #3,#4 \right)} 
\newcommand{\Afield}[4]{{A}_{#1 #2}\!\left( #3,#4 \right)}
\newcommand{\Afieldup}[4]{{A}^{#1 #2}\!\left( #3,#4 \right)}
\newcommand{\Afieldname}[5]{{A}_{#1 #2}^{#5}\!\left( #3,#4 \right)}
\newcommand{\Afieldupname}[5]{{A}^{#1 #2}_{#5}\!\left( #3,#4 \right)}
\newcommand{\Dprop}[6]{{D}_{#1 #2 #3 #4}\!\left( #5,#6 \right)}
\newcommand{\Gprop}[2]{{G}\!\left( #1,#2 \right)}
\newcommand{\Sprop}[6]{{S}_{#1 #2 #3 #4}\!\left( #5,#6 \right)}

\newcommand{\Dmisc}[7]{{D}^{#7}_{#1 #2 #3 #4}\!\left( #5,#6 \right)}
\newcommand{\Dmiscup}[7]{{D}^{#7 #1 #2 #3 #4}\!\left( #5,#6 \right)}
\newcommand{\Gmisc}[3]{{G}^{#3}\!\left( #1,#2 \right)}
\newcommand{\Smisc}[7]{{S}^{#7}_{#1 #2 #3 #4}\!\left( #5,#6 \right)}

\newcommand{\Dplus}[6]{\Dmisc{#1}{#2}{#3}{#4}{#5}{#6}{+}}
\newcommand{\Gplus}[2]{\Gmisc{#1}{#2}{+}}
\newcommand{\Splus}[6]{\Smisc{#1}{#2}{#3}{#4}{#5}{#6}{+}}
\newcommand{\Dmin}[6]{\Dmisc{#1}{#2}{#3}{#4}{#5}{#6}{-}}
\newcommand{\Gmin}[2]{\Gmisc{#1}{#2}{-}}
\newcommand{\Smin}[6]{\Smisc{#1}{#2}{#3}{#4}{#5}{#6}{-}}
\newcommand{\Dgreat}[6]{\Dmisc{#1}{#2}{#3}{#4}{#5}{#6}{>}}
\newcommand{\Ggreat}[2]{\Gmisc{#1}{#2}{>}}
\newcommand{\Sgreat}[6]{\Smisc{#1}{#2}{#3}{#4}{#5}{#6}{>}}
\newcommand{\Dless}[6]{\Dmisc{#1}{#2}{#3}{#4}{#5}{#6}{<}}
\newcommand{\Gless}[2]{\Gmisc{#1}{#2}{<}}
\newcommand{\Sless}[6]{\Smisc{#1}{#2}{#3}{#4}{#5}{#6}{<}}
\newcommand{\Dgl}[4]{\Dmisc{}{}{#1}{#2}{#3}{#4}{\gtrless}}
\newcommand{\Dglup}[4]{\Dmiscup{}{}{#1}{#2}{#3}{#4}{\gtrless}}
\newcommand{\Sgl}[4]{\Smisc{}{}{#1}{#2}{#3}{#4}{\gtrless}}
\newcommand{\Ggl}[2]{\Gmisc{#1}{#2}{\gtrless}}

\newcommand{\Slg}[4]{\Smisc{}{}{#1}{#2}{#3}{#4}{\lessgtr}}
\newcommand{\Glg}[2]{\Gmisc{#1}{#2}{\lessgtr}}
\newcommand{\Dpm}[4]{\Dmisc{}{}{#1}{#2}{#3}{#4}{\pm}}
\newcommand{\Spm}[4]{\Smisc{}{}{#1}{#2}{#3}{#4}{\pm}}
\newcommand{\Gpm}[2]{\Gmisc{#1}{#2}{\pm}}
\newcommand{\Dnon}[6]{\Dmisc{#1}{#2}{#3}{#4}{#5}{#6}{0}}
\newcommand{\Gnon}[2]{\Gmisc{#1}{#2}{0}}
\newcommand{\Snon}[6]{\Smisc{#1}{#2}{#3}{#4}{#5}{#6}{0}}
\newcommand{\Dcaus}[6]{\Dmisc{#1}{#2}{#3}{#4}{#5}{#6}{\rm c}}
\newcommand{\Gcaus}[2]{\Gmisc{#1}{#2}{\rm c}}
\newcommand{\Scaus}[6]{\Smisc{#1}{#2}{#3}{#4}{#5}{#6}{\rm c}}

\newcommand{\Gacaus}[2]{\Gmisc{#1}{#2}{\rm a}}

\newcommand{\dn}[2]{{d}^{#1} #2 \:}
\newcommand{\dnpi}[2]{ \frac{ {d}^{#1} #2 }{ {(2\pi)}^{#1} } \:}
\newcommand{\dnpiinvar}[2]{ \frac{ {d}^{#1} #2 }{2 \left|{#2}_{0}\right|
                                    {(2\pi)}^{#1} } \:}
\newcommand{\intc}{\int_{\cal C}}
\newcommand{\deltaftn}[2]{ \delta^{#1} \! \left( #2 \right) }
\newcommand{\twopideltaftn}[2] {(2\pi)^{#1}\deltaftn{#1}{#2}}
\newcommand{\Smatrix}[1]{ {\left| S_{\rm #1} \right| }^2 }
\newcommand{\ReactRate}[3]{{\cal W}_{\rm #1}\left( #2,#3 \right)}
\newcommand{\EPdist}{ \frac{ d n_{\gamma} \!\left( x , q \right) }
	                  { \dn{3}{x} \dn{3}{q} d q^2} }
\newcommand{\EEdist}{ \frac{ d n_{e^{-}} \!\left( y , p \right) }
	                  { \dn{3}{y} \dn{3}{p} d p^2} }
\newcommand{\density}[3]{{#1} \left( #2,#3 \right)}
\newcommand{\denslabel}[4]{{#1}_{#4} \left( #2,#3 \right)}
\newcommand{\denslabelstar}[4]{{#1}^{*}_{#4} \left( #2,#3 \right)}
\newcommand{\ggtrless}[2]{\density{g^{\gtrless}}{#1}{#2}}
\newcommand{\ggtr}[2]{\density{g^{>}}{#1}{#2}}
\newcommand{\gless}[2]{\density{g^{<}}{#1}{#2}}
\newcommand{\dgtrless}[4]{\density{d^{\gtrless}_{#1 #2}}{#3}{#4}}
\newcommand{\dgtrlessup}[4]{\density{d^{\gtrless #1 #2}}{#3}{#4}}

\newcommand{\dless}[4]{\density{d^{<}_{#1 #2}}{#3}{#4}}
\newcommand{\sgtrless}[4]{\density{s^{\gtrless}_{#1 #2}}{#3}{#4}}
\newcommand{\sgtr}[4]{\density{s^{>}_{#1 #2}}{#3}{#4}}
\newcommand{\sless}[4]{\density{s^{<}_{#1 #2}}{#3}{#4}}
\newcommand{\pol}[2]{\epsilon^{#1} \!\left( #2 \right) }
\newcommand{\polstar}[2]{\epsilon^{* #1} \!\left( #2 \right) }
\newcommand{\poldn}[2]{\epsilon_{#1} \!\left( #2 \right) }
\newcommand{\polstardn}[2]{\epsilon^{*}_{#1} \!\left( #2 \right) }
\newcommand{\ME}[3]{\left< #1 \right| #2 \left| #3 \right>}

\newcommand{\ket}[1]{\left| #1 \right>}
\newcommand{\twoargftn}[3]{{\cal #1} \!\left( #2 , #3 \right) }

\newcommand{\dirslash}[1]{\not \! {#1}}
\begin{document}
\draft
\title{\begin{flushright}{\rm MSUCL-1090}\\[9mm]\end{flushright}
	Partons in Phase Space}
\author{
David A. Brown\thanks{email:dbrown@nscl.msu.edu}
and
Pawe{\l} Danielewicz\thanks{email:danielewicz@nscl.msu.edu}}
\address{National Superconducting Cyclotron Laboratory and\\
Department of Physics and Astronomy \\
Michigan State University \\ 
East Lansing, MI 48824}
\date{\today}

\maketitle                 

\begin{abstract}
Within QED, we examine several issues related to constructing a  
parton--model--based QCD transport theory.  We rewrite the QED analog of the
parton model, the Weizs\"acker--Williams Approximation, entirely in terms of 
phase--space quantities and we study the phase--space photon and electron
densities created by a classical point charge.  We find that the
densities take a distinctive ``source--propagator'' form.  This form does not
arise in a conventional derivation of the semiclassical transport equations 
because of the
overuse of the gradient approximation.  We do not apply the gradient
approximation and so derive the phase--space analog of the Generalized
Fluctuation--Dissipation Theorem.  Together, this theorem and the expression
for the phase--space
particle self--energies give a set of coupled phase--space evolution equations.
We illustrate how these evolution equations can be used perturbatively or to
derive semiclassical transport equations.  Our work relies on 
phase--space
propagators and sources, so we describe them in detail when calculating
the photon and electron phase--space densities.  We use these tools to discuss
the shape of a nucleon's parton cloud.
\end{abstract}

\pacs{PACS numbers: 24.10.Cn, 25.75.-q, 12.38.Mh}
%
%
\section{Introduction}
Primary hadronic collisions in a typical nuclear reaction at RHIC
will occur at $\sqrt{s}\sim 200A$ GeV.  Such a collision is so violent that
the partons, i.e. the quarks and gluons comprising the hadrons, will become
deconfined.  With hadronic densities exceeding 
the inverse volume of a typical hadron, the partons will remain
deconfined and are expected to form a quark-gluon plasma (QGP)
\cite{eskola93,harris00,nikitin86}.  Since transport
theory descriptions of nuclear collisions have proven successful at
lower energies, it is natural to attempt to describe 
the time evolution of the QGP using a transport model derived from QCD.
A transport model would describe the time evolution of the parton 
phase--space\footnote{By phase--space, we mean in 
space, time, momentum and energy (or invariant mass)
simultaneously.} 
densities throughout the collision.
The procedures for deriving semiclassical transport equations  
using time--ordered nonequilibrium methods are well developed 
\cite{pawel84,bezz72,klevin97,mrow90,mrow92}.
In fact, there have been several attempts at constructing a QCD transport model 
based on these procedures \cite{geiger96,henning,blaizot94}, but each of them
have their problems.  Chief among these problems is that one either treats the
soft long--range phenomena (in the case of \cite{blaizot94}) 
or one treats the hard short--distance phenomena (in the case of
\cite{geiger96}), but never both in the same framework.  
Normally when one discusses transport,
one assumes a separation between the interaction and the kinetic length scales.
If one relaxes this assumption then one may be able to treat {\em both} hard
and soft modes on equal footing.  We have not done this for QCD, but we have 
made several steps toward doing the analogous thing in QED.  Our techniques
also allow for a simple connection with the parton model.
This paper consists of three parts, each one using QED 
to describe different aspects of the problem of constructing a partonic 
transport theory.  In the end, we use our accumulated insight to discuss the
shape of the parton cloud of a nucleon.

Before outlining the paper, we must say a few words about our formalism.
In the first two sections, we use Feynman's formulation of perturbation 
theory.  In Feynman perturbation theory, one specifies the initial and 
final states of a 
reaction and calculates the probability of going from the initial to final
state.  Thus, it is the appropriate tool for calculating observables 
for simple processes (such as exclusive cross sections).  For this reason, we 
use Feynman perturbation theory to illustrate how the 
phase--space sources and 
propagators work and to calculate the reaction probability for some simple
processes.  We show that, in Feynman perturbation theory, the particle 
phase--space densities have a ``source--propagator'' form.  Namely, the 
densities are a convolution of the probability\footnote{Strictly speaking,
neither the phase--space sources nor propagators are probabilities as they can
be negative.  As with any other Wigner transformed quantities, they must be
smoothed over small phase--space volumes to render them positive definite.} 
to create a particle (the source) with the probability to propagate from the
creation point to the observation point (the propagator). 
We are not the first to consider writing transition probabilities in
phase--space:  Remler \cite{remler90} discusses simulating many--particle 
systems in phase--space.  Remler's work is not immediately
applicable to partons because it only applies to particles with large mass.
For more complicated processes, i.e. when we only know the initial 
conditions, we must resort to time--ordered nonequilibrium methods.  In 
the last two sections we use time--ordered methods to derive the
phase--space evolution equations, to derive the Generalized 
Fluctuation--Dissipation Theorem and to discuss the parton distributions of a 
nucleon.  The Generalized Fluctuation--Dissipation Theorem shows that, in
time--ordered field theory, the particle densities also have a
``source--propagator'' form.  It should come as no surprise that we
find similar forms for the particle phase--space densities since both
formalisms are are equivalent
descriptions of elementary processes.  For simple tree--type processes 
in the energy--momentum representation, the 
moduli of the Feynman and retarded (or advanced) single particle propagators 
are the same, so one can rewrite the 
reaction probability in terms of either \cite{Lehmann}.  
In fact, both sets of Feynman rules are special cases of the contour Feynman 
rules in Appendix \ref{sec:lagrangian}.     

In the parton model, a cross section is a folding of the Parton
Distribution Function (PDF) with the cross section for the partonic subprocess. 
The QED analog of the parton model is the Weizs\"acker--Williams Approximation 
\cite{ape77,quigg83} since a cross section in the 
Weizs\"acker--Williams Approximation  is a folding the effective photon
distribution with the cross section for the photon 
absorption subprocess\cite{WWA,jackson75,bert88}.   
In Section \ref{sec:pdist}, we write the 
Weizs\"acker--Williams Approximation in phase--space in several steps.  
First, we write the
reaction rate density for our ``partonic subprocess,''  namely the reaction
rate for absorbing a free photon.  By writing this rate in phase--space, we
also illustrate how we do our momentum--space to phase--space
conversions.  Next, we write the reaction probability for photon
exchange in phase--space.  Comparing the full reaction probability with the
reaction rate density for absorbing a photon, we identify the effective
phase--space photon distribution.  This photon distribution is the effective
photon number density in phase--space and it has the form of a phase--space 
source folded with a phase--space propagator.  We calculate the photon number
density surrounding a classical point charge and explain how the photon's 
phase--space propagator and phase--space source work.  Finally, we comment
on the implications of this section for the parton model.  We will find that we
understand how partons propagate and have an idea how to make the gluon 
distribution gauge invariant, but since our photon source is point--like we 
do not learn anything about the parton sources.

To find the Parton Distribution Functions, one can solve the 
parton evolution equations or equivalently one can sum up a class of parton 
ladder diagrams.  The simplest parton ladder has one rung corresponding to a
single partonic splitting.  
In Section \ref{sec:edist}, we study the QED analog of this process:
a virtual photon splitting into a virtual electron and on--shell positron. 
We start our analysis by generalizing the
phase--space Weizs\"acker--Williams Approximation to include electrons and
writing down the effective electron distribution.  This effective electron 
distribution takes the ``source--propagator'' form.  While this  ``partonic''
splitting leads to a complicated form of the electron source, the shape of the
source is mostly determined by the underlying photon (the ``parent parton'')
distribution.
We calculate the electron distribution explicitly for a classical point charge 
and discuss how the electron propagates from the source to the observation 
point.
We have another reason for studying electrons:  we can discuss a simple case 
where the ``source--propagator'' picture breaks down.  This is the case  of 
two virtual photons colliding to produce an electron positron pair.  
The ``source--propagator'' picture breaks down because the photon 
fields interfere on the length scale of the electron--positron creation region.
Nevertheless, discussing the process in phase--space gives us insight into the
reaction dynamics.  Finally, we comment on the implications of the results 
from this section for QCD parton densities.

In Feynman perturbation theory, both the photon and electron 
phase--space densities have a ``source--propagator'' form.  
This form does not usually arise when one 
uses time--ordered nonequilibrium methods because one usually derives 
transport theory only after making the gradient approximation.  
The gradient approximation
amounts to ignoring small--scale structure of the particle phase--space
densities, resulting in much simpler collision integrals \cite{pawel84,mrow92}. 
In Section \ref{sec:nthgen}, we follow essentially the standard semiclassical 
transport equation derivation, but never 
make the gradient approximation.  Thus, we arrive at the Generalized
Fluctuation--Dissipation Theorem which codifies the ``source--propagator''
picture of the particle densities.  Crucial inputs to the theorem are the 
phase--space sources; we
will discuss how to calculate them.  With the sources and the Generalized
Fluctuation--Dissipation Theorem, we derive a set of phase--space
evolution equations.   These evolution equations describe the
evolution of the system in phase--space from the distant past to the present,
including all ``partonic'' splittings, recombinations and scatterings. 
Furthermore, we can expand these evolution equations to get the
lowest order contributions to the particle densities or we can differentiate 
the evolution equations to get transport equations.  

As a practical application of this study, in Section \ref{sec:QCD}
we examine the coordinate
space structure of the parton cloud of a nucleon.  In principle, one should
Wigner transform the quark or gluon wavefunctions of a nucleon.
Since we do not know the quark or gluon
wavefunctions of a nucleon, such a specification is not possible and we must
result to model--building.  One might envision constructing a model 
phase--space parton density of a nucleon
by multiplying the momentum space density (the Parton Distribution Function) 
and the coordinate space density of the partons \cite{geigerPCM}.  This 
approximation neglects correlations between the momentum and position in the 
parton density which are present in the phase--space density 
\cite{genWiginfo,carruth83}.  One might insert these correlations using
uncertainty principle based arguments \cite{almueller89,geigerPCM}.  This has
intuitive appeal, but such a prescription is ad--hoc at best. 
We can approach this problem in a more systematic manner using 
some physical insight from the momentum--space renormalization--group 
improved parton model.  In this model, the parton densities  are 
calculated by evolving the parton densities in virtuality ($Q^2$) and in
longitudinal momentum fraction ($x$).  This evolution is equivalent to 
evaluating a certain class of ladder diagrams and these diagrams  
can be re-cast in the form of the phase--space Generalized 
Fluctuation--Dissipation 
Theorem.  Thus, we can discuss the parton phase--space densities of 
an hadron in the large--$Q^2$ limit or in the small--$x$ limit. 
We find that neither large--$Q^2$ partons nor small--$x$ partons extend
beyond the nucleon bag in the transverse direction.
However, we find that the large--$Q^2$ partons extend out an 
additional\footnote{The nucleon has 4--momentum $P_\mu=(P_0,P_L,\vec{0}_T)$.} 
$\hbar c/x P_L$  
from the bag surface in the longitudinal direction.  This is 
in line with what others have estimated \cite{almueller89,geigerPCM}.  
Furthermore, we estimate that the small--$x$ partons extend at least an
additional $\hbar c\sqrt{-q^2}$ from the bag 
so the small-$x$ parton cloud is substantially larger
than the large--$Q^2$ cloud.  

Throughout this paper we use natural units ($\hbar=c=1$) when
convenient, but we insert factors of $\hbar c$ whenever directly comparing a
length to an inverse momentum.  
The signature of the metric tensor is $(+,-,-,-)$.

%
%

\section{The Phase--Space Photon Density}
\label{sec:pdist}

One calculates a parton--model cross--section by folding a Parton Distribution
Function (PDF) with the cross section for the partonic subprocess.  One
follows a similar procedure for calculating the cross--section in the 
Weizs\"acker--Williams Approximation \cite{WWA,jackson75,bert88}: one folds
the effective photon distribution with the cross section for absorbing the
photon (shown in Fig. \ref{fig:dis}(b)) to obtain the full cross
section (shown in Fig. \ref{fig:dis}(a)).
We recast the Weizs\"acker--Williams Approximation in phase--space and, in 
the process, define the phase--space effective photon distribution.  
This phase--space photon density has the form of a 
phase--space source convoluted with a phase--space propagator.

Let us outline this section.  First we will compute the photon/current B reaction
rate in phase--space.  This is simply the probability for the probe particle, B, 
to interact with a free photon.  This calculation  is simple so we use it to 
illustrate how we rewrite everything in phase--space.   Second, we will
calculate the reaction probability for one--photon exchange.  
In the Weizs\"acker--Williams Approximation, the reaction rate is supposed 
to be the effective photon distribution folded with the photon/current B
reaction rate, so we can identify the 
the phase--space effective photon distribution.  The effective photon 
distribution is a gauge--independent effective number density of photons.
Next, we calculate 
the phase--space photon density surrounding a classical point charge.  This
calculation will highlight how the photon source and the propagator function in
phase--space.  Finally, we conclude this 
section with a brief discussion of the implications for a phase--space version 
of the parton model.

In all of our calculations, we find the reaction probability.  
Mapping our results 
to cross sections is trivial and is outlined in Appendix \ref{sec:crosssection}.
Since we are finding reaction probabilities, we work in Feynman perturbation 
theory.  

\subsection{Photon/Current B Reaction Rate}
\label{sec:pabsrate}

We start this subsection by finding the photon/current B reaction rate, 
$\ReactRate{\gamma B\rightarrow B'}{x}{q}$.   
This reaction rate plays the role of the ``partonic'' subprocess cross section
in a parton model calculation.
Our derivation demonstrates how to rewrite
the reaction probability completely in terms of phase--space quantities.  
The high point in this calculation occurs in
equation (\ref{eqn:itsaWT}) when we identify the Wigner transforms of B's 
current and of the photon field.  This type of identification lets us rewrite
the reaction probabilities in phase--space.   

To find 
$\ReactRate{\gamma B\rightarrow B'}{x}{q}$ we write the S--matrix for the 
process in \ref{fig:dis}(b):
\begin{eqnarray*}
S_{\gamma B\rightarrow B'}
&= & \int\dn{4}{x}\ME{0}{A^{\mu}(x)}{\vec{q},\lambda}
       \ME{B'}{j_{\mu}(x)}{B}\\
&= & \int\dn{4}{x}\dnpi{4}{k}e^{-i k\cdot x}\ME{0}{A^{\mu}(x)}{\vec{q},\lambda}
       \ME{B'}{j_{\mu}(k)}{B}. 
\end{eqnarray*}
Here $\ME{0}{A^{\mu} (x)}{\vec{q},\lambda}=\sqrt{\frac{4\pi}{2|q_0|V}}
\pol{\mu}{\lambda}e^{iq\cdot x}$ is the free photon wave function 
(with $q^2=0$) and $j_{\mu}$ is the current operator for the probe particle B.  
We leave both the initial and final states of $B$ unspecified so 
the final state may be a single particle or 
several particles (as in fig. \ref{fig:dis}(b)). 

We now square the S--matrix and average over photon polarizations:
\begin{eqnarray*}
\Smatrix{\gamma B\rightarrow B'}
& = & \int\dn{4}{x}\dn{4}{x'}
      \dnpi{4}{k}\dnpi{4}{k'}e^{-i(k\cdot x-k'\cdot x')}\\
&   & \times\displaystyle\frac{1}{2}\sum_{\lambda=\pm} 
      \ME{0}{A^{\mu}(x)}{\vec{q},\lambda} 
      \ME{\vec{q},\lambda}{A^{*\nu}(x')}{0} 
      \ME{B'}{j_{\mu}(k)}{B}\ME{B}{j^{\dagger}_{\nu}(k')}{B'}.
\end{eqnarray*}
On writing the coordinates and momenta in terms of the relative and average
quantities (i.e. $\tilde{k} = k-k'$ and $K=\frac{1}{2}(k+k')$),
and taking advantage of the momentum conserving delta
functions in the current matrix elements, $\Smatrix{\gamma B
\rightarrow B'}$  becomes      
\begin{equation}
\begin{array}{ccl}
\displaystyle
\Smatrix{\gamma B\rightarrow B'}=
& & \displaystyle\int\dn{4}{X}\dn{4}{\tilde{x}}
    \dnpi{4}{K}\dnpi{4}{\tilde{k}}
    e^{-i(K\cdot\tilde{x}+\tilde{k}\cdot X)}
    \frac{1}{2}\sum_{\lambda=\pm}
    \ME{0}{A^{\mu}(X+\tilde{x}/2)}{\vec{q},\lambda} \\
& & \times \ME{\vec{q},\lambda}{A^{*\nu}(X-\tilde{x}/2)}{0}
    \ME{B'}{j_{\mu}(K+\tilde{k}/2)}{B}
    \ME{B}{j^{\dagger}_{\nu}(K-\tilde{k}/2)}{B'}.
\end{array}
\label{eqn:itsaWT}
\end{equation}
There are two Wigner transforms in this equation:  the Wigner transform of the
photon field (the $\tilde{x}$ integral) and the Wigner transform of B's current
(the $\tilde{k}$ integral). 

Now we rewrite the S-matrix in terms of the phase--space quantities and define
the reaction rate density: 
\begin{equation}
\begin{array}{ccl}
\displaystyle
\Smatrix{\gamma B\rightarrow B'}
& = & \displaystyle\int\dn{4}{x}\dnpi{4}{k}\frac{\pi}{V|q_0|}\sum_{\lambda=\pm}
    \poldn{\mu}{\lambda}\polstardn{\nu}{\lambda}
    \twopideltaftn{4}{q-k}\Jcurrent{\mu}{\nu}{x}{k}{B}\\
& \equiv & \displaystyle\int\dn{4}{x}\ReactRate{\gamma B\rightarrow B'}{x}{q}.
\end{array}
\label{eqn:Ssquareabs}
\end{equation}
We also have defined the Wigner transform  of the current:
\begin{equation}
   \Jcurrent{\mu}{\nu}{x}{q}{B}\equiv\int\dnpi{4}{\tilde{q}}
      e^{-i\tilde{q}\cdot x}\ME{B'}{j^{\mu}(q+\tilde{q}/2)}{B}
      \ME{B}{j^{\dagger \nu}(q-\tilde{q}/2)}{B'}.
\label{eqn:current}
\end{equation}
Since B's Wigner current is proportional to the reaction rate, it is natural to
give them the same physical interpretation: as a ``probability''  
density,\footnote{Because the Wigner current is the Wigner transform of a quantum
object, it may not be positive definite \cite{genWiginfo,carruth83} so it can 
not be strictly interpreted as a probability.} for
absorbing a free photon with momentum $q$ at space--time point $x$.  Now, it
may not be clear where the spatial structure of the reaction rate comes from,
especially since the incident photon is completely delocalized in space (it is
in a momentum eigenstate).  To give the reaction rate spatial structure, we
must localize either
the initial or final states of $B$ with a wavepacket.  

\subsection{Photon Exchange}

In this section, we write the reaction rate for one--photon exchange 
(see Fig. \ref{fig:dis}(a)) in phase--space.  We do it two different ways:
in terms of the  Wigner transforms of the currents A and B and the photon 
propagator and in terms of the Wigner transform of the photon vector potential. 
The first form of the reaction rate has a clear physical interpretation in terms
of photon emission, propagation, and absorption.  However, it is the 
second form which can be brought into the form of a ``partonic'' 
cross--section.
 
The S--matrix for Fig. \ref{fig:dis}(a) is 
\begin{equation} 
   S_{\rm AB\rightarrow A'B'}=\int d^4x d^4y \ME{A'}{j^{A \mu}(x)}{A}
      \Dcaus{\mu}{\nu}{}{}{x}{y}\ME{B'}{j^{B \nu}(y)}{B}.
\label{eqn:ABgoes2AB}
\end{equation}
Taking the absolute square of this S--matrix and rewriting it in terms of 
Wigner transformed currents and propagators, we find 
\begin{equation}
\Smatrix{AB\rightarrow A'B'}=\int\dn{4}{y}\dn{4}{x}\dnpi{4}{q}
        \Jcurrent{\mu}{\nu}{y}{q}{A} \Dcaus{\mu}{\nu}{\mu'}{\nu'}{y-x}{q}
       	\Jcurrent{\mu'}{\nu'}{x}{q}{B}.
\label{eqn:photexch}
\end{equation}
Here, the Wigner transform of the photon propagator is
\[\begin{array}{ccl}
   \Dcaus{\mu}{\nu}{\mu'}{\nu'}{x}{q}
   &=& \displaystyle\int\dn{4}{\tilde{x}}e^{i \tilde{x}\cdot q}
      D^{c}_{\mu\nu}(x+\tilde{x}/2)
      D^{c*}_{\mu'\nu'}(x-\tilde{x}/2) \\
   &=& (4\pi)^2g_{\mu\nu}g_{\mu'\nu'}\Gcaus{x}{q}
\end{array}\]
and $\Gcaus{x}{q}$ is the Wigner transform of the scalar propagator. 
We derive $G^c(x,q)$ in Appendix \ref{append:prop}.  We discuss 
the Wigner transforms of the propagator and current when we study the photon
and electron distributions of a point charge.  

Equation (\ref{eqn:photexch}) has an obvious physical meaning:  
1)~current A makes a photon with momentum~$q$ at space--time point~$y$, 
2)~the photon propagates from~$y$ to~$x$ with momentum~$q$ and 
3)~current B absorbs the photon at space--time point~$x$.  
The spatial structure of the integrand of (\ref{eqn:photexch}) comes from
localizing either $A$ or $B$.

Now we take a detour and calculate the Wigner transform of the vector
potential of the current $A$.  In terms of the current density and propagator,
the vector potential is\footnote{Jackson actually
uses the retarded propagator to define the vector potential 
because he discusses classical fields.} \cite{jackson75}:
\begin{equation} 
   A^\mu(x)=\int\dn{4}{y}D^c_{\mu\nu}(x-y)J_A^\nu (y).
\label{eqn:vectpot}
\end{equation}
The Wigner transform of this is:
\begin{equation}
\begin{array}{ccl}
\Afield{\mu}{\nu}{x}{q} 
   &=&\displaystyle\int\dn{4}{\tilde{x}} e^{i\tilde{x}\cdot q} 
      A_\mu(x+\tilde{x}/2)A_\nu^{*}(x-\tilde{x}/2)\\ 
   &=&\displaystyle\int\dn{4}{y}\Jcurrent{\mu'}{\nu'}{y}{q}{A}
      \Dcaus{\mu'}{\nu'}{\mu}{\nu}{x-y}{q}.
\label{eqn:Afield}
\end{array}
\end{equation}
The Wigner transform of the vector potential has a ``source--propagator'' form.
Current $A$ (the photon source) creates the photon 
with momentum $q$ at
space--time point $y$ and the propagator takes the photon from $y$~to~$x$.  
Let us put this in equation (\ref{eqn:photexch}),
\begin{equation}
	|S_{AB\rightarrow A'B'}|^2=\int d^4 x \frac{d^4 q}{(2\pi)^4}
	A_{\mu\nu} (x,q) J^{\mu\nu}_B(x,q).
\label{eqn:halfway}
\end{equation}
Stated this way, the spatial structure of the integrand of this equation
comes from either localizing $B$ or from the spatial structure in the Wigner
transform of the photon vector potential.

Equation (\ref{eqn:halfway}) is close to the form of a cross section
in the parton model because current $B$ is proportional to the
photon/current B reaction rate (as discussed in the previous subsection) and 
the vector potential is proportional to the phase--space effective photon 
density (as discussed in the next subsection).

\subsection{The Weizs\"acker--Williams Approximation}

The effective photon distribution we derive  here is the phase--space
analog of Weizs\"acker--Williams' effective photon distribution.  Thus, it  
has the interpretation as the number of photons with virtuality $q^2$ in a 
unit cell of phase--space.  One could use it to calculate the reaction 
probability in Fig. \ref{fig:dis}(a) by folding it with the photon/target
reaction 
rate ${\cal W}_{\gamma B\rightarrow B'}(x,q)$.  In parton model terms, we take
the Parton Distribution Function (the effective photon distribution) and fold
it with the partonic subprocess (the photon/target reaction rate) to get the
reaction probability for the entire process.

We will derive the Weizs\"acker--Williams Approximation in several stages.  
First we 
decompose B's Wigner current into photon polarization vectors, allowing us 
to rewrite $\ReactRate{\gamma B\rightarrow B'}{x}{q}$ in terms of 
$\Jcurrent{\mu}{\nu}{x}{q}{B}$.  Knowing this, we identify 
the effective photon distribution.  In the final subsection we will discuss the
gauge independence of our effective photon distribution.

\subsubsection{Current Decomposition} 
 
If the photon probing $\Jcurrent{\mu}{\nu}{x}{q}{B}$ is sufficiently 
delocalized in space (i.e. $\partial_{\sigma} A_{\mu\nu}(x,q)
\ll q_{\sigma}A_{\mu\nu}(x,q)$), the momentum--space cutting 
rules tell us that we can expand $\Jcurrent{\mu}{\nu}{x}{q}{B}$ in terms of the
photon polarization vectors \cite{budnev75}:
\begin{equation}
\begin{array}{ccl}
   \displaystyle\Jcurrent{\mu}{\nu}{x}{q}{B} 
   & = & \displaystyle\sum_{\lambda=\pm}\pol{\mu}{\lambda}
      \polstar{\nu}{\lambda}J_{trans}(x,q)\\
   & + & \displaystyle\pol{\mu}{0}\polstar{\nu}{0}J_{scalar}(x,q)\\
   & + & \displaystyle\frac{q^{\mu}q^{\nu}}{q^2}J_{long}(x,q).
\end{array}
\label{eqn:hadrontensor}
\end{equation}
Here, $\epsilon_\mu(0)$ is the scalar (i.e. time--like) polarization vector:
$\epsilon_\mu(0)=p_{B\mu}-q_\mu q\cdot p_B/q^2$, where $p_B$ is the momentum 
of B.  The transverse polarization vectors,
$\epsilon_\mu(\pm)$, span the hyperplane perpendicular to $\epsilon_\mu(0)$
and $q_\mu$.  Now, if $A_{\mu\nu}(x,q)$ is not delocalized, then
Eq.~(\ref{eqn:hadrontensor}) should be modified to include
gradients\footnote{These gradients come from Wigner transforming terms
proportional to the relative photon momentum.} in $x$.  However, if we were to
include those gradients here, we could not map $\Jcurrent{\mu}{\nu}{x}{q}{B}$ 
to ${\cal W}_{\gamma B\rightarrow B'}$. 

Since $\pol{\mu}{\lambda}\polstardn{\mu}{\lambda'}=\delta_{\lambda\lambda'}$,
it is simple to find the separate currents in (\ref{eqn:hadrontensor}) 
in terms of $\Jcurrent{\mu}{\nu}{x}{q}{B}$:
\[
   J_{scalar}(x,q)=\poldn{\mu}{0}\polstardn{\nu}{0}
   \Jcurrent{\mu}{\nu}{x}{q}{B}
\]
and 
\[
   J_{trans}(x,q)=\frac{1}{2}\sum_{\lambda=\pm}\poldn{\mu}{\lambda}
      \polstardn{\nu}{\lambda}\Jcurrent{\mu}{\nu}{x}{q}{B}
\]
The longitudinal piece, $J_{long}(x,q)$, vanishes due to current conservation. 

\subsubsection{The Effective Photon Distribution}

If we insert (\ref{eqn:hadrontensor}) 
into equation (\ref{eqn:photexch}), the 
reaction probability is a sum of two terms:
\begin{equation}
\begin{array}{ccl}
   \Smatrix{AB\rightarrow A'B'} 
   & = & \displaystyle\int\dn{4}{x}\dnpi{4}{q}\Afield{\mu}{\nu}{x}{q}
      \sum_{\lambda=\pm}\pol{\mu}{\lambda}\polstar{\nu}{\lambda}
      J_{trans}(x,q)\\
   & + & \displaystyle\int\dn{4}{x}\dnpi{4}{q}\Afield{\mu}{\nu}{x}{q}
      \pol{\mu}{0}\polstar{\nu}{0}J_{scalar}(x,q).
\end{array}
\label{eqn:factorized}
\end{equation} 
The two terms in (\ref{eqn:factorized}) describe transverse 
and scalar photon exchange between currents A and B, respectively.  

Noting that if 
$J_{trans}(x,q)$ has  a weak $q^2$ dependence,\footnote{The reader 
should note here that the reaction rates are for photons
with any $q^2$, while in Section \ref{sec:pabsrate} the reaction rate was for 
on--shell photons only.} then
$J_{trans}(x,q)\propto\ReactRate{\gamma B\rightarrow B'}{x}{q}$.  
In other words, $J_{trans}(x,q)$ is proportional to the reaction rate for the
``partonic'' subprocess.
Therefore, the transverse term of (\ref{eqn:factorized}) can be written as
\begin{equation}
   \Smatrix{AB\rightarrow A'B'}=\frac{1}{4\pi}\int \dn{4}{x}
     \frac{Vd^3 q}{(2\pi)^3} \frac{d q^2}{2\pi} 
      \EPdist\ReactRate{\gamma B\rightarrow B'}{x}{q},
\label{eqn:wwapprox}
\end{equation}
provided we identify the transverse effective photon distribution as 
\begin{equation}
   \EPdist=\sum_{\lambda=\pm}\pol{\mu}{\lambda}\polstar{\nu}
   {\lambda}\Afield{\mu}{\nu}{x}{q}.
\label{eqn:wwapprox2}
\end{equation}
We can make a similar identification  with the scalar term.
This effective photon distribution is the spin summed photon Wigner function.
In other words it is the phase--space number density of 
photons at time $x_0$ per unit $q^2$.   This effective photon distribution 
is the QED analog of the
phase--space Parton Distribution Function.   This generalizes the 
Weizs{\"a}cker--Williams method to phase--space.

While classical derivations of the Weizs{\"a}cker--Williams method begin with
finding the photon power spectra from the Poynting flux
\cite{jackson75,WWA}, a quantum mechanical
derivation follows along the lines of what we do here
\cite{budnev75,bert88}.  Were we
to perform the spatial integrals in (\ref{eqn:factorized}), we would find that
the exponentials in the Wigner transforms conspire to make several
delta functions.  The resulting delta function integrations are trivial and we
would quickly recover the momentum--space result.

Now, multiplying the photon phase--space density by the projection 
tensor $\sum_{\lambda=\pm}\pol{\mu}{\lambda}\polstar{\nu}{\lambda}$ in
(\ref{eqn:wwapprox2}) does {\em not} 
render the photon distribution gauge invariant, unlike in momentum space 
\cite{budnev75}, because the photon distribution is not completely 
delocalized.  When $\Afield{\mu}{\nu}{x}{q}$ undergoes a gauge 
transformation, terms proportional to $q_{\mu}$ are removed but terms 
proportional to $\partial/\partial x^{\mu}$ are not. 
A gauge invariant virtual photon distribution is 
introduced below.  This gauge invariant
distribution reduces to (\ref{eqn:wwapprox2}) when $A_{\mu\nu}(x,q)$ is
sufficiently delocalized.

\subsubsection{Gauge Issues}
\label{sec:gauge}

Parton densities are supposed to be gauge invariant but our 
effective photon distribution is gauge dependent.
In this section, we discuss how $\Afield{\mu}{\nu}{x}{q}$ transforms 
under a change of gauge, determine the gauge invariant part of 
$\Afield{\mu}{\nu}{x}{q}$, and state how the gauge invariant part of 
$\Afield{\mu}{\nu}{x}{q}$ is related to the effective photon phase--space 
distribution.  

In the energy--momentum representation,
gauge transforming the photon field adds an arbitrary function 
in the direction of the photon momentum to the photon vector potential:
$A_{\mu}(q)\longrightarrow A_{\mu}(q) + q_{\mu} f(q)$.  
Because components of $A_{\mu}(q)$ in the direction of
$q_{\mu}$ are gauge dependent, we can write $A_{\mu}(q)$ as a sum of 
the gauge independent and dependent parts:
\[A_{\mu}(q)=A_{\mu}^{\|}(q)+A_{\mu}^{\bot}(q)\]
where
$A_{\mu}^{\|}(q)=\frac{q_{\mu}q_{\nu}}{q^2}A^{\nu}(q)$ is the gauge
dependent part of $A_{\mu}(q)$ and $A_{\mu}^{\bot}(q)=A_{\mu}(q)-
A_{\mu}^{\|}(q)$ is the gauge independent part.  Wigner transforming
the photon field gives us a term that is gauge independent and
terms which are gauge dependent: 
\begin{equation}
\begin{array}{ccl}
   \Afield{\mu}{\nu}{x}{q} 
   & = & \displaystyle\int\dnpi{4}{\tilde{q}} e^{-ix\cdot\tilde{q}}
      \left[A_{\mu}^{\|}(q+\tilde{q}/2)+A_{\mu}^{\bot}(q+\tilde{q}/2)\right]
      \left[A_{\nu}^{\|}(q-\tilde{q}/2)+A_{\nu}^{\bot}(q-\tilde{q}/2)\right]^{*}\\
   & \equiv & \displaystyle \Afieldname{\mu}{\nu}{x}{q}{\bot\bot}
      +\Afieldname{\mu}{\nu}{x}{q}{\bot\|}+\Afieldname{\mu}{\nu}{x}{q}{\|\bot}
      +\Afieldname{\mu}{\nu}{x}{q}{\|\|}.
\end{array}
\label{eqn:proj1}
\end{equation}
The only gauge independent piece of $\Afield{\mu}{\nu}{x}{q}$ is
$\Afieldname{\mu}{\nu}{x}{q}{\bot\bot}$.  We do the
integrals in (\ref{eqn:proj1}) and 
identify the tensor that projects off the gauge dependent
part of $\Afield{\sigma}{\rho}{x}{q}$:
\begin{equation}
\begin{array}{ccl}
   \Afieldname{\mu}{\nu}{x}{q}{\bot\bot} 
   & = & \displaystyle(g_{\mu\sigma}-h^{+}_{\mu\sigma})
      (g_{\nu\rho}-h^{-}_{\nu\rho})\Afieldup{\sigma}{\rho}{x}{q} \\
   & \equiv & \displaystyle{\cal P}_{\mu\nu\sigma\rho}
      \Afieldup{\sigma}{\rho}{x}{q}
\end{array}
\label{eqn:proj2}
\end{equation}
where
\begin{equation}
   h^{\pm}_{\mu\nu}=\frac{(q\pm i\partial /2)_{\mu}(q\pm i\partial /2)_{\nu}}
   {(q\pm i\partial /2)^2}.
\label{eqn:htensor}
\end{equation}
This projector must be understood as a series in $q_\mu$ and $\partial_\mu$, so
can only really be used when 
$\partial_\sigma A_{\mu\nu}(x,q) < q_\sigma A_{\mu\nu}(x,q)$.  
Now, the statement of current conservation for a general 
$\Jcurrentdn{\mu}{\nu}{x}{q}$ is
\begin{equation}
   (q\pm i\partial /2)^{\mu}\Jcurrentdn{\mu}{\nu}{x}{q}=
   (q\pm i\partial /2)^{\nu}\Jcurrentdn{\mu}{\nu}{x}{q}=0.
\end{equation}
So, as expected, current conservation ensures that only the gauge independent 
part of $\Afield{\mu}{\nu}{x}{q}$ appears in the reaction probability.     

With $\Afieldname{\mu}{\nu}{x}{q}{\bot\bot}$ in hand, we can postulate the 
gauge invariant photon distribution:
\begin{equation}
   \EPdist=\sum_{\lambda=\pm}\pol{\mu}{\lambda}\polstar{\nu}
   {\lambda}\Afieldname{\mu}{\nu}{x}{q}{\bot\bot}.
\label{eqn:wwapprox3}
\end{equation}
This reduces to (\ref{eqn:wwapprox2}) if the photon field varies slowly in 
space (i.e. we neglect the gradients 
$\partial_{\sigma}\Afield{\mu}{\nu}{x}{q}\ll q_{\sigma}\Afield{\mu}{\nu}{x}{q}$),
as we now show.  Neglecting the derivatives in 
(\ref{eqn:htensor}), the projection tensor in (\ref{eqn:proj2}) reduces to
\begin{equation}
\begin{array}{rcl}
   {\cal P}_{\mu\nu\sigma\rho}&\approx& 
   \left(g_{\mu\sigma}-\frac{q_{\mu}q_{\sigma}}{q^2}\right)
   \left(g_{\nu\rho}-\frac{q_{\nu}q_{\rho}}{q^2}\right)\\
   &=&\left(\sum_{\lambda=\pm,0}\poldn{\mu}{\lambda}
   \polstardn{\sigma}{\lambda}\right)^{*}
   \left(\sum_{\lambda'=\pm,0}\poldn{\nu}{\lambda'}
   \polstardn{\rho}{\lambda'}\right).
\end{array}
\label{eqn:approxproj}
\end{equation}
Since the~polarization~vectors form a~complete~basis in~Minkowski~space, i.e.
$\sum_{\lambda=\pm,0}\poldn{\mu}{\lambda}\polstardn{\nu}{\lambda}+
\frac{q_{\mu}q_{\nu}}{q^2}=g_{\mu\nu}$.
Putting (\ref{eqn:approxproj}) in equation
(\ref{eqn:wwapprox3}), we arrive back at (\ref{eqn:wwapprox2}).

The tactic of projecting out the gauge dependent parts of
the photon distribution works mainly because of the simple form of the $U(1)$
gauge transformation.  Nevertheless, a variant of this technique probably could
be applied to gluons.

\subsection{Photon Phase--Space Density of Classical Point Charge}
\label{sec:classEPD}

We now calculate the density for the simple 
case of a classical point charge radiating photons.  If we localize the
source's wavepacket and view it on a length scale larger than its localization
scale, we can treat its density as a delta function.   
Thus, the shape of the photon distribution is determined the photon 
propagation and we can use this calculation to illustrate how
partons propagate in phase--space.  We explain that, despite our use of 
Feynman perturbation theory, the photon propagates via the Wigner
transform of the retarded (time--ordered) propagator.  As such, 
the photon propagates a distance of roughly $R_\|=1/|q_L|$ in the
direction parallel to the photon 3--momentum and $R_\perp=1/\sqrt{|q^2|}$ in 
the direction perpendicular to the photon 3--momentum.
We demonstrate this behavior by plotting the coordinate--space distribution
of photons with $q^2\ll q_0^2$ (making the photons collinear with the source) 
and with $q^2 \sim q_0^2$.  In Appendix~\ref{append:static} we examine the
additional case of a static point charge (i.e. $\vec{v}=0$).  This case is not
relevant for partons as a parton source must be taken in the limit
$|\vec{v}|\rightarrow c$.

\subsubsection{Classical Current}

For our source
current, we assume the source particle's wavepacket is localized on the
length scales that our photons can resolve so we can replace its 
current with the current of a point particle  
(we discuss when this replacement is valid in Appendix \ref{append:current}).
The source particle follows a classical trajectory $x_{\mu}=x_{0}v_{\mu}$ with  
four--velocity $v_{\mu}=(1,v_L,\vec{0}_T)$, $v_L \approx c$ 
and $\gamma=1/(1-v_L^2)\gg 1$.  Ignoring the recoil caused by 
photon emission, the current of the point charge 
is \cite{jackson75} 
\[j_{\mu}(x)=e v_{\mu} \: \deltaftn{3}{\vec{x}-x_0\vec{v}}.\]
The Wigner transform of this is the classical Wigner current:
\begin{equation}
\begin{array}{rl}
   \Jcurrent{\mu}{\nu}{x}{q}{classical} 
   & = \displaystyle \int \dn{4}{\tilde{x}} e^{iq\cdot\tilde{x}}
      j_\mu(x+\tilde{x}/2) j_\nu^\dagger(x-\tilde{x}/2)\\
   & = 2\pi \alpha_{em} \: v_{\mu} v_{\nu}\: \delta (q\cdot v)
      \deltaftn{3}{\vec{x}-x_0\vec{v}}.
\end{array}
\label{eqn:classcurrent}
\end{equation}
Here $e^2=\alpha_{em}$ is the QED coupling constant.

The current has several easy to interpret
features.  The first delta function sets $q\cdot v=0$.  This ensures that the 
emitted photons are space--like and that current is conserved.  It also 
insures that, when $q^2\rightarrow 0$, the
photons become collinear with the emitting particle ($q_0 = v_L q_L \approx q_L$
making $q^2\approx \vec{q}_T^2\approx 0$).  This delta function arises because
we neglect the recoil of the point charge as it emits a
photon.  The second delta function insures
that the source is point--like and follows its classical trajectory.  

This source has one other feature of note:  it allows for emission of both
positive and negative energy photons. In the following work, we consider 
creating only positive energy photons so we insert a factor
of $2\theta(q_0)$ in (\ref{eqn:classcurrent}).  This amounts to constraining 
the source's initial energy to be greater its final energy.

\subsubsection{Retarded Propagator}
\label{sec:retprop}

For our phase--space propagator, we take the Wigner transform of the retarded
propagator instead of the Wigner transform of the Feynman propagator.  
This replacement is legal since the reaction probability may be expressed
in terms of either propagator, provided one uses the same asymptotic states
in both cases.  Both propagators will give different particle densities at
intermediate stages but the final state particle densities converge as 
time goes to $\pm \infty$. 
We use the retarded propagator here because it leads to a more transparent
interpretation fo the particle densities.  
We discuss the Feynman propagator in our discussion of the electron 
distribution of the point charge.

The Wigner transform of the retarded propagator gives the weight for 
a particle with four--momentum $q_\mu$ to propagate across the space--time 
separation $\Delta x_\mu = x_\mu-y_\mu$.
In the Lorentz gauge, the retarded photon propagator,
$D^+_{\mu\nu\mu'\nu'}(\Delta x, q)$, is proprtional to the retarded
scalar propagator $G^+(\Delta x,q)$:
\begin{equation}
	D^+_{\mu\nu\mu'\nu'}(\Delta x, q)=g_{\mu\nu}g_{\mu'\nu'}G^+(\Delta x,q) 
\end{equation} 
The retarded scalar propagator is
\[\Gplus{\Delta x}{q}=\frac{1}{\pi}\theta (\Delta x_0)\theta (\Delta x^2)\theta 
      (\lambda^2) \frac{\sin{(2\sqrt{\lambda^2})}}{\sqrt{\lambda^2}} \]
and it is derived in Appendix \ref{append:prop}.  In this expression, the 
Lorentz invariant $\lambda^2$ is $\lambda^2=(\Delta x\cdot q)^2-q^2 \Delta x^2$.  

Let us now estimate how far the retarded propagator can send a particle with 
the momentum $q_\mu=(q_0,q_L,\vec{0})$.  
First, the retarded propagator has two theta functions, one that enforces 
causality and one that forces propagation inside the light cone.  
The rest of the interesting features of the propagator are tied up in the 
dependence on $\lambda^2$.  Since 
$\Gplus{\Delta x}{q}\propto \theta (\lambda^2) 
\sin{(\sqrt{\lambda^2})}/{\sqrt{\lambda^2}}$, 
the particle can not propagate farther than the inequalities
$0\le \sqrt{\lambda^2} \lesssim 1$ allow.  To see what these constraints mean, 
we investigate the $q^2>0$, $q^2<0$, and $q^2=0$ cases separately.  

To study the $q^2>0$ case, we position ourselves in the frame where 
$q'_\mu=(q'_0,\vec{0})$.
In this frame, the $\lambda^2$ constraint translates into a restriction on the 
spatial distance a particle can propagate:
\[
   0\le {q'_0}^2 \Delta\vec{x}'^2\lesssim 1.
\]
Combined with the light--cone constraint, $\Delta\vec{x}'$ is constrained to 
\begin{mathletters}
\begin{equation}
   |\Delta\vec{x}'| \lesssim \left\{ 
   \begin{array}{cl}
      \Delta x'_0 & \mbox{for small} \; \Delta x'_0 < 1/|q'_0| \\
      1/|q'_0| & \mbox{for large} \; \Delta x'_0 > 1/|q'_0|.
   \end{array}
   \right.
\end{equation} 
To find a cutoff for $\Delta x'_0$, we realize that, for a given $q'_\mu$, the
propagator gives the ``probability'' distribution for propagating across the
space--time displacement $\Delta x'_\mu$.  Thus, we can integrate 
$\Gplus{\Delta x'}{q'}$ over all space and over time up to some cutoff time
$\tau$, giving us the total ``probability'' for propagating to time $\tau$.  
We find that the propagation probability becomes unimportant for $\tau\gtrsim
1/|q'_0|$ and this sets a cutoff in $\Delta x'_0$: 
\begin{equation}
	\Delta x'_0 \lesssim 1/|q'_0|. 
\end{equation}
\label{eqn:constraintsa}
\end{mathletters}
Together, these three constraints define the space--time region where the 
particle can propagate.
When we move back to the frame with 
$q_\mu=(q_0,q_L,\vec{0}_T)$, the region contracts in the temporal and
longitudinal directions.  From Eq.~(\ref{eqn:constraintsa}), the 
limits of the propagation region are 
\begin{mathletters}
\begin{eqnarray}
	|\Delta\vec{x}_T|&\lesssim& R_\perp=\frac{1}{\sqrt{|q^2|}}\\
	|\Delta x_L|&\lesssim& R_\|=\displaystyle\frac{1}{|q_L|}\\
	|\Delta x_0|&\lesssim& R_0=\displaystyle\frac{1}{|q_0|}
\end{eqnarray}
\end{mathletters}

We study the $q^2<0$ case in a similar manner.  In the frame with
$q'_\mu=(0,q'_L,\vec{0}_T)$, the $\lambda^2$  constraint implies 
\[
	0\le {q'}_L^2 (\Delta {x'}_0^2-\Delta {x'}_T^2)\lesssim 1.
\]
Combining this with the light--cone constraint immediately gives us a limit on
$\Delta x'_L$:
\begin{mathletters}
\begin{equation}
	|\Delta x'_L|\lesssim \frac{1}{|q'_L|}.
\end{equation}
As with the $q^2>0$ case, we can integrate the propagator to find the total
``probability'' for propagating to the time $\tau$.  In this case, the
propagation probability is important only for  $\tau\lesssim 1/|q'_L|$, giving
us a limit in $\Delta x'_0$ of:
\begin{equation}
	|\Delta x'_0|\lesssim \frac{1}{|q'_L|}.
\end{equation} 
The limit on $|\Delta \vec{x}'_T|$ then follows directly from the light--cone
constraint:
\begin{equation}
	|\Delta \vec{x}'_T|\lesssim \frac{1}{|q'_L|}.
\end{equation} 
\end{mathletters}
Again, these constraints define a the space--time region where the particle can
propagate.  
Boosting back to the frame with $q_\mu=(q_0,q_L,\vec{0}_T)$, again the
longitudinal and temporal spread gets Lorentz contracted:
\begin{mathletters}
\begin{eqnarray}
	|\Delta \vec{x}_T | &\lesssim& R_\perp =\displaystyle\frac{1}
	{\sqrt{|q^2|}}\\
	|\Delta x_L | &\lesssim& R_\| =\displaystyle\frac{1}{|q_0|}\\
	|\Delta x_0 | &\lesssim& R_0 =\displaystyle\frac{1}{|q_L|}.
\end{eqnarray}
\label{eqn:retlimits}
\end{mathletters}

Now we study the $q^2=0$ case.  With $q^2=0$, $\lambda^2$ becomes
\begin{equation}
	\lambda^2=|\Delta x\cdot q|
	=|q_0||\Delta\vec{x}\cdot\hat{q}-\Delta x_0|\lesssim 1.
\label{eqn:followsclasspath}
\end{equation}
On other words, high energy particles tend to follow their classical path 
while low energy particles can deviate from their classical path.  
Expression (\ref{eqn:followsclasspath})
then gives a measure of the deviation from the classical path.

\subsubsection{Phase--Space Effective Photon Distribution}

Now we put these elements together into the photon density.  We concentrate 
our efforts on $A_{\mu\nu}(x,q)$ because all of the spatial dependence of the
photon distribution is tied up in the Wigner transform of the vector potential.
Inserting the classical
current and the retarded photon propagator in the 
Lorentz gauge into (\ref{eqn:Afield}), we find
\begin{equation}
   \Afield{\mu}{\nu}{x}{q} = 4\pi \alpha_{em} \: v_{\mu} v_{\nu}
      \theta (q_0) \delta (q\cdot v)\int\dn{4}{y}\Gplus{x-y}{q} \:
      \deltaftn{3}{\vec{y}-y_0\vec{v}}.
\label{eqn:retintoA}
\end{equation}
The delta function integrals 
in equation~(\ref{eqn:retintoA}) are trivial, 
however the remaining proper time integral can not be done analytically.
We find
\begin{equation}
   \Afield{\mu}{\nu}{x}{q}=\frac{(8\pi)^2 \alpha_{em} \gamma 
      \theta(q_0)\delta(q\cdot v)}
      {\sqrt{-q^2}}v_{\mu}v_{\nu}\twoargftn{A}{2|x\cdot q|}
      {2\sqrt{-q^2\gamma^2\left((x\cdot v)^2-x^2 v^2\right)-(x\cdot q)^2}},
\label{eqn:Amunu}
\end{equation}  
where the dimensionless function $\twoargftn{A}{a}{b}$ is given by
\begin{equation}
   \twoargftn{A}{a}{b}=\int_{a}^{\infty}d\tau \frac{\sin{\tau}}
      {\sqrt{b^2+\tau^2}}.
\label{eqn:dimlessA}
\end{equation}  

There are two interesting cases that are easy to explore: that of 
photons nearly collinear to the source particle 
(i.e. $q_0\approx q_L\gg |\vec{q}_T|$), and that of photons with a large 
transverse momentum (i.e. $|\vec{q}_T|\sim q_0,q_L$).  Since there is a 
$1/\sqrt{-q^2}$ singularity in the photon density and nearly on--shell
photons (i.e. $q^2\rightarrow 0$) are collinear, there will be many more
collinear photons than any other kind.  

Two plots, representative
of collinear photons and high $\vec{q}_T$ photons, are shown in
Fig. \ref{fig:photondist}.  The left is a plot of the
dimensionless function $\twoargftn{A}{a}{b}$ for collinear photons with
$q_{\mu}=(m_{e},m_e/v_L,\vec{0}_T)$.  On the right is a plot of photons with
transverse momentum comparable to their transverse momentum and energy, 
$q_{\mu}=(m_{e},m_e/v_L,0.56 {\rm MeV/c},0)$.    
The characteristic energy scale of QED is $m_e$, so we choose this scale for 
the momenta to plot.  In both plots, we chose $v_L=0.9c$ to illustrate the 
Lorentz contraction of the distribution.
The oscillations exhibited by both photon distributions are expected for
a Wigner transformed density \cite{genWiginfo,carruth83}.  To obtain an 
equivalent classical distribution, one should smear this distribution over a 
unit volume of phase--space.  

Both cuts through the photon distribution show Lorentz
contraction.  For the collinear photons, this contraction occurs in
the longitudinal direction.  We can account for the contraction with the 
behavior of the retarded propagator.  We expect that the  
width will be $\sim R_\| = \hbar c/|q_0|$ parallel to  $\vec{q}$
and $\sim R_\perp = \hbar c/\sqrt{|q^2|}$ perpendicular to $\vec{q}$.  
For the collinear photons, $\vec{q}$ is in the longitudinal direction and
$q_0=\gamma\sqrt{|q^2|}$ so $\sim R_L= \hbar c/\gamma\sqrt{|q^2|}$.  
In other words, the collinear photon distribution is a
``Lorentz contracted onion'' centered on the moving point source.  
The inner layers of this ``onion'' correspond to higher $|q^2|$ photons.  
However, we must emphasize that the contraction is
{\em not} due to the movement of the source, but rather due to kinematics of 
the photon's creation and the propagation of the photon. 
To illustrate this point, one only needs to look at the high transverse
momentum photons: their distribution is tilted.  In the case plotted on the
right in Fig. \ref{fig:photondist}, the photon momentum points 
$45^{\circ}$ to the longitudinal direction, coinciding with the tilt
of the distribution .
Furthermore, the width of the distribution is $\sim R_\| = \hbar c/|q_0|$ 
along this tilted axis and $\sim R_\perp = \hbar c/\sqrt{|q^2|}$ 
perpendicular to this tilted axis.

\subsubsection{Comment on the Gauge Dependence of the Effective Photon 
   Distribution of a Point Charge}

Now, $A_{\mu\nu}(x,q)$ is a gauge dependent
object and the physically interesting object, the effective photon distribution,
is gauge invariant.  One might ask whether the interesting features of
$A_{\mu\nu}(x,q)$ disappear under a gauge transform.  To see whether this
happens, one must insert $A_{\mu\nu}(x,q)$ into equation (\ref{eqn:wwapprox3});
the only things in  (\ref{eqn:wwapprox3}) that could significantly 
alter shape of the distribution (\ref{eqn:Amunu}) are the gradients. 
Now because the photon source is extremely localized (it is a delta function),
the shape of the photon distribution comes solely from the propagator.  Since
the propagator varies significantly on length scale comparable to $1/q_\mu$,
derivatives of $A_{\mu\nu}(x,q)$ are always comparable in size to $q_\mu$ and
any expansion of the gauge projector in Eq.~(\ref{eqn:proj2}) will not converge.
So, we must conclude that our photon distribution can not be made gauge
invariant.
Now, had we {\em not} used a point source for our photons, 
the integration over the source could 
smooth the photon distribution so that it varies slower in space.
In that case, our distribution could be rendered gauge invariant.

\subsection{What the Photons Tell Us about the Partons}

Parton model cross sections can be written in phase--space as a folding of the
phase--space parton distribution function with the reaction rate for the
partonic sub--process.
The phase--space parton distribution functions are the spatial
number density of partons with a certain momentum.  The parton distribution
functions have a ``source--propagator'' form and can be defined in a gauge
invariant manner.  If the phase--space parton source produces only positive
energy partons or if we use time--ordered field theory then the partons
propagate from their source using the Wigner transform of the retarded
propagator.  This retarded propagator propagates off--shell partons up to
roughly $\sim R_\| = \hbar c /{\rm min}(|q_0|,|\vec{q}|)$
parallel to the parton three--momentum and 
$\sim R_\perp = \hbar c/\sqrt{|q^2|}$ perpendicular 
to the parton 3--momentum.    Both of these estimates are valid only in frames
with $q_0, \vec{q} \neq 0$.  When either $q_0=0$ or $\vec{q}=0$, propagation is 
cut off at $\sim R_{\|,\perp} = \hbar c/\sqrt{|q^2|}$.  On--shell (i.e.
$q^2=0$) partons tend to follow their classical trajectory, with deviations
from that trajectory of order $\sim 1/|q_0|$.

Despite what we have learned, we know next to nothing about 
parton sources in phase--space.  We use a point source while
a nucleon has spatial structure on the length scales of
interest.   Furthermore, partons radiate other partons and this
alters the source.  We gain more insight into the phase--space sources 
in the next few sections. 

%
%
\section{Phase--Space Electron Density}
\label{sec:edist}

In the parton model, the Parton Distribution Functions can be found by summing
a class of ladder diagrams and the simplest of these has only one rung,
corresponding to a single partonic splitting.
One can see the QED analog of the first rung of such a ladder in 
Fig.~\ref{fig:splitting}(a).
Probing the electron distribution occurs in three steps:
1)~a virtual photon splits into and electron--positron pair with the positron
on--shell, 2)~the virtual electron propagates from the splitting point toward 
the probe particle and 3)~the electron interacts with the probe.  
We can give this rate a parton--model like form by 
associating steps 1 and 2 with the effective electron distribution
and step 3 with the electron/probe reaction rate (see Fig.
\ref{fig:splitting}(b)).  

Let us outline this section.  In Subsection~\ref{sec:3a}, we demonstrate 
that the reaction rate for the virtual electron exchange process in 
Fig.~\ref{fig:splitting}(b) factorizes in phase--space, giving the reaction 
rate a parton model like form.  In the process, the electron 
phase--space density acquires the ``source--propagator'' form.
In Subsection~\ref{sec:3b}, we calculate the electron distribution of a point
charge.  The electron source shape is determined mostly by the shape of the
parent photon distribution.  The electron can equivalently propagate with the
retarded or Feynman phase--space propagators.  
We assume the
electrons are massless throughout this section because we only know the form
of the massive propagators in the high mass limit ($m_e\gg
p_0,|\vec{p}|$).  This limit is irrelevant for QCD as $\Lambda_{QCD} \sim m_q
\ll p_0,|\vec{p}|$.  For completeness, we calculate the electron 
distribution in the high--mass limit in Appendix \ref{append:massive}.
In Subsection~\ref{sec:probeedist},  we discuss an apparent failure of the
``source--propagator'' picture: lepton pair production in the strong
field produced by two point charges.  Because the photon fields of the two point
charges interfere, it is not possible to clearly isolate the
source or probe and we can not
factorize the square S--matrix into an electron distribution and electron/probe
interaction.  Nevertheless, we can still discuss the process in phase--space,
even though we cannot write down the electron distribution.
Finally, in Subsection~\ref{sec:3d} we discuss the implications of this 
section.  The
discussions of splitting, of the massless propagators, and of a failure of
factorization are all relevant for partons.  

\subsection{Factorization and the Effective Electron Phase--Space Distribution}
\label{sec:3a}

First we show that the process in Fig.~\ref{fig:splitting} can be factorized 
in phase--space, giving a parton model--like form.
The S--matrix for the process in Fig.~\ref{fig:splitting}(a) is:
\begin{equation}
	S_{\rm \gamma B\rightarrow \bar{e} B'} = 
		\int \dn{4}{x}\dn{4}{y}
		A_{\mu}(x)\psi_{\bar{e}}(x,s) e\gamma^{\mu} S^c(x-y)
		{\cal V}_{\rm B e\rightarrow B'} (y).
\label{eqn:eSmatrix}
\end{equation}
The spatial structure of the electron source comes from localizing the
photon vector potential, $A_\mu(x)$.  The spatial structure of the ``partonic''
subprocess comes from ${\cal V}_{\rm B e\rightarrow B'} (y)$, the 
electron/probe interaction in Fig.~\ref{fig:splitting}(b).
In Eq.~(\ref{eqn:eSmatrix}),
 $\psi_{\bar{e}}(x,s)=\int\dnpi{4}{k}v(k,s) e^{i k\cdot x}
\frac{f^{*}(k)}{\sqrt{2 k_0 V}}$ is the final positron
wavepacket and $S^{c}(x-y)=\int\dnpi{4}{p} e^{-i p
\cdot (x-y)}\frac{\not{p}+m_e}{p^2+m_e^2+i\epsilon}$ is the electron
Feynman propagator.  

We square $S_{\rm \gamma B\rightarrow \bar{e} B'}$ and write it in
terms of phase--space quantities:
\begin{equation}
\begin{array}{ccl}
       \displaystyle\Smatrix{\gamma B\rightarrow \bar{e} B'}&=&\alpha_{em}
               \displaystyle\int\dn{4}{x}\dn{4}{y}
               \dnpi{4}{p}\dnpi{4}{q}\dnpi{4}{k}\Afield{\mu}{\nu}{x}{q}
               f(x,k)\twopideltaftn{4}{k+p-q}\\
       & & \displaystyle\times {\rm Tr} \left\{ \frac{1}{2} 
               (\dirslash{k}-m)\gamma^{\mu}
               S^{c}(y-x,p){\cal V}_{\rm B e\rightarrow B'}(y,p)
               \gamma^{\nu}\right\}
\end{array}
\label{eqn:wignerSelec}
\end{equation}
Here, $S^{c}(y-x,p)$ is the Wigner transform of the electron Feynman propagator
and it can be written in terms of the Wigner transform of the scalar Feynman
propagator, $\Gcaus{x}{q}$:
\[\begin{array}{rl}
   \Scaus{\alpha}{\alpha'}{\beta}{\beta'}{x}{p}
   &= \displaystyle \int \dnpi{4}{\tilde{p}} e^{-i\tilde{p}\cdot x}
      S^c_{\alpha\beta}(p+\tilde{p}/2)\bar{S}^c_{\alpha'\beta'}(p-\tilde{p}/2)\\
   &= (\dirslash{p}+i\dirslash{\partial}+m_e)_{\alpha\beta}
      (\dirslash{p}-i\dirslash{\partial}+m_e)_{\alpha'\beta'}\Gcaus{x}{p}.
\end{array}\]
Also in Eq.~(\ref{eqn:wignerSelec}), ${\cal V}_{\rm B e\rightarrow B'}(y,p)$ is
the Wigner transform of the electron/probe interaction and $f(x,k)$ is the
phase--space density of final state positrons.

Since the positron rung in Fig.~\ref{fig:splitting} is cut, we can put the final 
positron in a momentum eigenstate\footnote{This makes the positron momentum 
weight--function $f^{*}(k)\propto \delta^4 (k-k_f)$, with $k_f^2=m_e^2$, and 
the positron phase--space density $f(x,k)=(2V|k_{f0}|)^{-1}
\twopideltaftn{4}{k-k_f}$.} and 
sum over the positron final states and spin.  
Furthermore, we can separate off the spinor structure of the electron 
propagator and shift
the derivatives to act on ${\cal V}_{\rm B e\rightarrow B'}$.   
In the end we find
\begin{equation}
\begin{array}{ccl}
	\Smatrix{\gamma B\rightarrow \bar{e} B'}&=&\displaystyle\alpha_{em}
		\int\dn{4}{x}\dn{4}{y}\dnpi{4}{p}\dnpi{4}{q}\dnpiinvar{3}{k_f}\\
	& & \displaystyle\times\Afield{\mu}{\nu}{x}{q}
		\Gcaus{y-x}{p}\twopideltaftn{4}{k_f+p-q}\\
	& & \displaystyle\times {\rm Tr} \left\{ \frac{1}{2} 
		(\dirslash{k}_f-m)\gamma^{\mu}
		(\dirslash{p}+i\dirslash{\partial}/2+m_e)
		{\cal V}_{\rm B e\rightarrow B'}(y,p)
		(\dirslash{p}-i\dirslash{\partial}/2+m_e)\gamma^{\nu}\right\}.
\end{array}
\label{eqn:elecSmatrix}
\end{equation}

Since ${\cal V}_{\rm B e\rightarrow B'}(y,p)$ is separated from the electron
propagator, the reaction probability factorizes.  
We could explicitly calculate the rate for the ``partonic'' subprocess
$eB\rightarrow B'$, but we are only interested in the shape of the distribution 
as a function of electron momentum.  We can guess the form of the electron
density just by looking at equation (\ref{eqn:elecSmatrix}),
without performing the explicit rate density calculation. 
The electron density is
\begin{equation}\begin{array}{ccl}
   \displaystyle\EEdist &\propto &\displaystyle\alpha_{em} \int \dn{4}{x} 
   	\Gcaus{y-x}{p} 
      \int \dnpiinvar{3}{k} \Afield{\mu}{\nu}{x}{k+p}\\
      &\equiv &\displaystyle\int \dn{4}{x} \Gcaus{y-x}{p} \Sigma(x,p).
\end{array}\label{eqn:elecdist}\end{equation}
Here, $\Gcaus{x}{p}$ is the Wigner transform of the scalar propagator.
Equation (\ref{eqn:elecdist}) has the ``source--propagator''
form: the integral of $\Afield{\mu}{\nu}{x}{k+p}$ over the 
positron momentum, $\Sigma$, plays the role of the ``partonic'' source.  
Because the emitted positron is in a momentum eigenstate, the spatial 
structure of the source comes solely
from the parent photon's phase--space distribution.

At this stage, we see several important features of the source.
First, we note the $d^3k/|k_0|$ in the positron momentum integral.
This factor weights positron emission toward small $k_0$.  In a typical parton
ladder, ordering momenta to maximize the contribution from this particular 
singularity
leads to the so--called BFKL evolution equations\cite{genQCD,laenen94}.
Second, we note that the entire spatial
dependence of the electron source comes from the parent photon distribution. 
These two points are especially important for Section~\ref{sec:QCD} so 
they are elaborated on in the next subsection.

\subsection{The Effective Electron Distribution of a Classical Point Charge}
\label{sec:3b}

Our main interest is with how the ``parton ladder'' (in our case, the source 
is only one rung of the ladder) shapes the electron distribution.  First, we
discuss the electron's source and how both the parent photon distribution and 
the cut positron rung effect it.  Second, we discuss the interplay of 
the electron creation and propagation.
Because the electron has positive energy, we
can use either the retarded or Feynman phase--space propagator.  
We choose to use the retarded propagator but, for completeness,
we describe the Feynman propagator.  

\subsubsection{The Electron Source}
\label{sec:esource}

For our electron source, we choose the photon distribution of
Eq.~(\ref{eqn:Amunu}).  This is not a parton--like parent distribution as the
photon source is point--like.  Nevertheless, we can use it to illustrate the
general features that one expects from the electron source in 
Eq.~(\ref{eqn:elecdist}).  In particular, we discuss the $1/|k_0|$ singularity
from the positron rung and we detail both the shape of the source and how this
shape depends on the photon distribution.  

Up to irrelevant constants, the electron source is 
\begin{equation}
   \Sigma(x,p) \propto \alpha_{em}\int \dn{4}{k} \theta(-k_0)
      \theta(p_0+k_0)\deltaftn{}{k^2-m_e^2} \deltaftn{}{q\cdot v} 
      \frac{\twoargftn{A}{a}{b}}{\sqrt{-(k+p)^2}}
\label{eqn:edist_w1}
\end{equation} 
where $a=2|x\cdot (k+p)|$ and $b=2\sqrt{-(k+p)^2\gamma^2
((x\cdot v)^2-x^2v^2)-(x\cdot (k+p))^2}$.  The longitudinal and temporal
positron momentum integrals can be done with the delta functions, leaving
the transverse momentum integrals:
\begin{equation}
\begin{array}{rl}
   \Sigma(x,p) \propto & \displaystyle\alpha_{em}\theta(p\cdot v) 
      \int_{|\vec{k}_T|<k_{T {\rm max}}}\frac{\dn{2}{k_T}}
      {\sqrt{k_{T {\rm max}}^2-k_T^2}} \\
      & \times\displaystyle\left\{\frac{\theta(p_0+k_{0+})
      	 \twoargftn{A}{a_{+}}{b_{+}}}{\sqrt{-(k_{+}+p)^2}} 
      +\frac{\theta(p_0+k_{0-})\twoargftn{A}{a_{-}}{b_{-}}}
      	{\sqrt{-(k_{-}+p)^2}} \right\}.
\end{array}
\label{eqn:edist_w3}
\end{equation}
Here, $k_{T {\rm max}}^2=\gamma^2 (p\cdot v)^2-m_e^2$. 
The two roots of the positron momentum are given by
\begin{equation}
\begin{array}{rcl}
   k_{0 \pm} & = & -\gamma(\gamma p\cdot v\mp v_L\sqrt{k_{T{\rm max}}^2-
	\vec{k}_T^2})\\
   k_{L \pm} & = & -\gamma(\gamma v_L p\cdot v\mp \sqrt{k_{T{\rm max}}^2-
	\vec{k}_T^2}).
\end{array}
\end{equation}

Now, in a parton ladder we expect to find a factor of $d^3k/|k_0|$ for each
cut rung.  Here is no exception, one can see that $d^4k \theta (-k_0)
\delta(k^2-m^2_e)$ gives us this factor.  However, because we neglect the 
recoil of the source, we have an additional $\delta(q\cdot v)$ and the 
factor becomes $d^2k/\sqrt{k_{T{\rm max}}^2-k_T^2}$.  
Because $|\vec{k}_T|<k_{T{\rm max}}=\gamma 
p\cdot v\ll|k_{0\pm}|,|k_{L\pm}|$, this singularity forces the positrons to be
anti--collinear with the point charge.  Furthermore, because $q^2\approx0$
(because of the $1/\sqrt{-q^2}$ singularity) the electrons are collinear with 
the point charge.

As in a parton ladder, the shape of $\Sigma$ comes froms the parent's
distribution.  In the case at hand, we can actually estimate the 
$\left<q_\mu\right>$ that gives the dominant contribution to $\Sigma$.  
Because $k_L$ and $k_0$ are fixed by the delta functions and
$\vec{k}_T$ is bounded by $k_{T {\rm max}}$, we can estimate the average
positron recoil momentum.  
The average $\vec{k}_T$ is given by 
$\left< |\vec{k}_T| \right> \approx \frac{\sqrt{3}}{2}k_{T {\rm max}}$.
In general, for
$v_L\approx 1$, the average $\left< k_{\mp\mu} \right>$ is given by
\[
\left< k_{\mp\mu} \right>=\gamma^2 (p\cdot v) \left(-1\pm \frac{1}{2}, 
-1\pm \frac{1}{2}, \cos(\theta_T)/\gamma,\sin(\theta_T)/\gamma\right)
\]   
For the purposes of illustration, we choose to emit the positron in the 
direction $\hat{k}_T\cdot\hat{x}_T =\cos(\theta_T)=\frac{1}{\sqrt{2}}$.  
By momentum conservation, the dominant photon momentum is
$\left<q_{\pm\mu}\right>=p_\mu+\left<k_{\pm\mu}\right>$.

On the left in Fig.~\ref{fig:source}, we plot the electron source for 
$p_{\mu}=(2.0, 2.05, \vec{0}_T)$~MeV/c electrons from a point 
charge moving to the right with $v_L=0.9c$.  We choose this $p_\mu$ because 
it is both collinear with the point charge and because it is space--like 
($p^2<0$).  Our source can emit both $p^2>0$ and $p^2\le 0$ electrons, however 
the typical parton in a parton ladder
is either space--like or on--shell.  On the right in Fig.~\ref{fig:source}, we
also plot the photon distribution corresponding to the dominant 
$\left<q_\mu\right>$.  Note
that both the source and the photon
distribution have approximately the same width in both the longitudinal and
transverse directions.  The tilt in the photon distribution gets averaged away
in the $\vec{k}_T$ integrals in equation (\ref{eqn:edist_w3}).

\subsubsection{Electron Density Using the Retarded Propagator}

Now we put elements of the electron distribution together.
In Eq.~(\ref{eqn:edist_w3}), we need the Wigner transform of the
Feynman propagator.  However, since the electrons have positive energy we can 
replace the Feynman propagator with the retarded propagator.  We discuss the
retarded propagator in Section \ref{sec:retprop} and we 
describe the phase--space Feynman propagator in the next
subsection.

We are interested in electrons that have momenta that are both
space--like and collinear with the source (for comparison with partons), so we
plot the coordinate space distribution of electrons with 
$p_\mu=(2.0, 2.05, \vec{0}_T)$~MeV/c in Fig.~\ref{fig:edist4}.
The point source
is moving to the right with velocity $0.9c$.  Both the source and the
underlying photon distribution for these electrons is shown in
Fig.~\ref{fig:source}.  To perform the four--dimensional spatial integral in 
Eq.~(\ref{eqn:edist_w3}), we use a Monte--Carlo integration scheme 
\cite{press89}.  This integration scheme, being probabilistic by nature,
returns both the integral at a point and the error on the integral at
that point.  The nonzero data
points never had a relative error greater than 20\%, but due to this error, the
location of the zero contours is uncertain by $\sim 30$~fm. 

Comparing the electron distribution with the source, we see that
the electron distribution is elliptical with longitudinal and transverse 
widths comparable to what one
expects by adding the source width in Fig.~\ref{fig:source}a. 
to our estimates for the propagation length
in Eqs.~(\ref{eqn:retlimits}).
Unlike the electron source distribution, the electron distribution
is not symmetric about $x_L=0$. 
This is caused by the positron recoil because, were there no positron recoil, 
we would have a delta function to insure $p_0=p_L v_L$ (as we found for the
photons).  Because of the positron recoil, the delta function is
widened and the additional spread in energy causes the electron to
preferentially propagate forward.

\subsubsection{The Feynman Propagator}
\label{subsec:feynprop}

Even though we choose situations where we can avoid using the phase--space 
Feynman propagator, we should describe how it works.   While the 
Feynman propagator propagates a
particle with a given momentum (say $p_\mu=(p_0,p_L,\vec{0}_T)$) across a
space--time displacement $\Delta x_\mu=(\Delta x_0, \Delta x_L, \Delta
\vec{x}_T)$, it does so in a manner very different from the retarded
propagator.  The Feynman propagator is
\[\begin{array}{rl}
   \Gcaus{\Delta x}{p}= & \displaystyle \frac{1}{4\pi} \left[ 
   	\,{\rm sgn}(\Delta x^2) + 
      {\rm sgn}(p^2) + 2\:{\rm sgn}(\Delta x\cdot p)\right] \\
    & \times \displaystyle \left\{ \theta (\lambda^2)
      \frac{\sin{(2\sqrt{\lambda^2})}}{\sqrt{\lambda^2}} -
      \theta (-\lambda^2)\frac{\exp{(-2\sqrt{-\lambda^2})}}
		{\sqrt{-\lambda^2}} \right\},
\end{array}\]
The combination of the sign functions in the square brackets can be 
rewritten in a more transparent form:
\[ 
	[\ldots]=\left\{ 
		\begin{array}{ll}
			 4& \mbox{ if } \Delta x\cdot p,\: p^2, \:\Delta x^2 >0\\
			-4& \mbox{ if } \Delta x\cdot p,\: p^2, \:\Delta x^2 <0\\
			2\:{\rm sgn}(\Delta x\cdot p)& \mbox{ if } p^2,\: x^2 \mbox{ have 
			opposite sign}
		\end{array}
	\right.
\]
Thus, particles with time--like momentum tend to travel forward
in time and inside the light--cone and particles with space--like momentum tend
to travel backwards in time outside the light cone.  
Also as one might expect, anti--particles with time--like momentum tend 
to travel backwards in time inside the light--cone
and anti--particles with space--like momentum tend 
to travel forwards in time outside the light--cone.

The rest of the interesting features of the Feynman propagator are tied up
in the dependence on the Lorentz invariant $\lambda^2=(\Delta x\cdot p)^2 -
\Delta x^2 p^2$.   
As with the retarded propagator in Subsection~\ref{sec:classEPD}, we will 
study the $p^2>0$, $p^2<0$, and $p^2=0$ cases separately.  

To study the $p^2>0$ case, we boost to the frame where $p'_\mu=(p'_0,\vec{0})$.
In this frame, $\lambda^2={p'}^2_0 |\Delta \vec{x}'|^2 \ge 0$, so only the sine
term contributes.  The sine term is greatest for $\sqrt{\lambda^2}\lesssim 1$
so we have the following limit on the spatial propagation distance:
\begin{mathletters}
\begin{equation}
	|\Delta \vec{x}'| \lesssim \frac{1}{|p'_0|}.
\end{equation}
As with the retarded propagator, we can compute the total ``probability'' to
propagate to certain time.  This calculation gives us the following limit on
the temporal propagation distance:
\begin{equation}
	|\Delta x_0| \lesssim \frac{1}{|p'_0|}.
\end{equation}
\end{mathletters}
Boosting the space--time region defined by these constraints back to the frame
with $p_\mu=(p_0,p_L,\vec{0}_T)$, we find the following constraints:
\begin{mathletters}
\begin{eqnarray}
	|\Delta\vec{x}_T|&\lesssim& R_\perp=\frac{1}{\sqrt{|p^2|}}\\
	|\Delta x_L|&\lesssim& R_\|=\displaystyle\frac{1}{|p_L|}\\
	|\Delta x_0|&\lesssim& R_0=\displaystyle\frac{1}{|p_0|}
\end{eqnarray}
\end{mathletters}
These limits are exactly the same as the ones we found for the retarded
propagator in Subsection~\ref{sec:classEPD}.

To study the $p<0$ case, we boost to the $p'_\mu=(0,p'_L,\vec{0}_T)$ frame.
In this frame, $\lambda^2={p'}_L^2(\Delta {x'}_0^2-\Delta {x'}_T^2)$.    
Inside the light--cone, the exponential term disappears and we get a
constraint on $\lambda^2$:
\[
	0\geq \lambda^2={p'}_L^2(\Delta {x'}_0^2-\Delta {x'}_T^2)\lesssim 1.
\]
We can integrate to find the total ``probability'' to
propagate to a certain time, giving us a limit on $\Delta x'_0$:
\begin{mathletters}
\begin{equation}
	|\Delta x'_0|\lesssim\frac{1}{|p'_L|}.
\end{equation}
Using the $\lambda^2$ and light--cone constraints, we find similar limits on 
$\Delta \vec{x}'_T$ and $\Delta x'_L$:
\begin{eqnarray}
	|\Delta x'_L|&\lesssim&\displaystyle\frac{1}{|p'_L|} \\
	|\Delta \vec{x}'_T|&\lesssim&\displaystyle\frac{1}{|p'_L|}.
\end{eqnarray}
\end{mathletters}
Boosting back to the $p_\mu=(p_0,p_L,\vec{0}_T)$ frame, we find
\begin{mathletters}
\begin{eqnarray}
	|\Delta\vec{x}_T|&\lesssim& R_\perp=\frac{1}{\sqrt{|p^2|}}\\
	|\Delta x_L|&\lesssim& R_\|=\displaystyle\frac{1}{|p_0|}\\
	|\Delta x_0|&\lesssim& R_0=\displaystyle\frac{1}{|p_L|},
	\label{eqn:feynlimits}
\end{eqnarray}
\end{mathletters}
which is what we found for the retarded propagator.
Now, outside of the light--cone the situation is more complicated and we must
integrate the propagator in the various directions to find limits.
We find:
\begin{mathletters}
\begin{eqnarray}
	|\Delta x'_0|&\lesssim&\displaystyle\frac{1}{|p'_L|} \\
	|\Delta x'_L|&\lesssim&\displaystyle\frac{1}{|p'_L|} \\
	|\Delta \vec{x}'_T|&\lesssim&\displaystyle\frac{1}{|p'_L|}.
\end{eqnarray}
\end{mathletters}
When we boost back to the frame with
$p_\mu=(p_0,p_l,\vec{0}_T)$, we find the result in Eq.~(\ref{eqn:feynlimits}).

Finally, we investigate the $p^2=0$ case.  With $p^2=0$, $\lambda^2$ becomes
\begin{equation}
	0\le\lambda^2=|\Delta x\cdot p|
	=|p_0||\Delta\vec{x}\cdot\hat{p}-\Delta x_0|\lesssim 1
\label{eqn:alsofollowsclasspath}
\end{equation}
because the exponential term does not contribute on the light cone.
On other words the Feynman propagator functions exactly like the retarded
propagator: high energy particles tend to follow their classical path 
while low energy particles can deviate from their classical path.  
Expression (\ref{eqn:alsofollowsclasspath})
then gives a measure of the deviation from the classical path.

We find that, despite the different boundary conditions on the
two propagators, both the Feynman and retarded propagators send particles the
same distances.  
This is probably no surprise since a calculation done using Feynman's
formulation of perturbation theory must give the same results as the same
calculation done using time--ordered perturbation theory.

\subsection{Failure of Factorization and the ``Source--Propagator'' Picture}
\label{sec:probeedist}

In this section, we investigate electron--positron pair production
in the strong field of two point charges.   One
might visualize this interaction as a virtual photon from one point
charge probing the virtual electron distribution of another point
charge.    Thus, 
the electron distribution would appear factorized from the virtual 
electron--virtual photon collision process.
However, we will show that this picture is incorrect because
the photon fields interfere with one another on length scales
comparable to the size of pair production region.  Of course, this also means
that our ``source--propagator'' picture fails here.  Nevertheless, we 
can still formulate the problem in phase space and discuss the 
interplay of the interaction length and particle production length scales.

\subsubsection{Interference of Photon Fields}

We can write down the S--matrix corresponding to the process in Figs. 
\ref{fig:2phot1} and \ref{fig:2phot2} using the same procedures used in the 
previous sections.  To lowest order in the coupling strength, we obtain:
\begin{equation}
\begin{array}{ccl}
   S_{12\rightarrow 1'2'e\bar{e}} &=& \displaystyle 
   \alpha_{em}\int\dn{4}{x_1}\dn{4}{x_2}
   \dnpi{4}{k_1}\dnpi{4}{k_2}\dnpi{4}{p}\\
   &  & \displaystyle\times \frac{f^{*}(k_1,k_2)}{\sqrt{2k_{1,0}V}
   \sqrt{2k_{2,0}V}}e^{ik_1\cdot x_1+ik_2\cdot x_2+ip\cdot (x_1-x_2)}\\
   & & \displaystyle\times\Lambda_{\mu \nu}(k_1,s_1,k_2,s_2,p)
   \left\{ A_{1}^{\mu}(x_1) A_{2}^{\nu}(x_2) + 
   A_{2}^{\mu}(x_1) A_{1}^{\nu}(x_2)\right\}.
\end{array}
\label{eqn:2photSmatrix}
\end{equation}
Here $x_1$ and $x_2$ are the interaction points of the photons and 
should not be confused with the classical source particles $1$~and~$2$.
We have already separated the $\gamma\gamma e\bar{e}$ effective vertex
\[
   \Lambda_{\mu \nu}(k_1,s_1,k_2,s_2,p)
   = \bar{u}(k_1,s_1)\gamma_{\mu}iS^{c}(p)\gamma_{\nu}v(k_2,s_2).
\] 
In $\Lambda_{\mu \nu}(k_1,s_1,k_2,s_2,p)$,  $S^{c}(p)$ is the momentum--space 
Feynman electron propagator.  The final state electron--positron wavepacket is 
$f^{*}(k_1,k_2)$ and we will assume the final $e\bar{e}$ pair to be free 
and use the free wavepacket from Appendix \ref{append:current}.  The 
reader should note that we can already see the photons interfering 
in Eq.~(\ref{eqn:2photSmatrix}).

As usual, we can rewrite Eq.~(\ref{eqn:2photSmatrix}) in terms of 
Wigner transformed quantities.  However, due to the photon fields interfering, 
the structure of the cross terms are complicated.  The  
$\Smatrix{12\rightarrow 1'2'e\bar{e}}$ is:
\begin{equation}
\begin{array}{ccl}
\Smatrix{12\rightarrow 1'2'e\bar{e}}&=&\alpha_{em}^2
    \displaystyle\int\dn{4}{R}\dn{4}{r}
    \dnpi{4}{k_1}\dnpi{4}{k_2}\dnpi{4}{q_1}\dnpi{4}{q_2}\dnpi{4}{p}\\
& & \displaystyle\times f(R-r/2,k_1,R+r/2,k_2)\\
& & \displaystyle\times \Lambda_{\mu\mu'\nu\nu'}(k_1,k_2,p,r)
    \twopideltaftn{4}{q_1+q_2-k_1-k_2}\\
& & \displaystyle\times\left\{ \twopideltaftn{4}{q_1-k_1+p} 
    \Afieldupname{\mu}{\mu'}{R-r/2}{q_1}{1}
    \Afieldupname{\nu}{\nu'}{R+r/2}{q_2}{2}\right.\\
& & \displaystyle+\left.  \twopideltaftn{4}{q_1-k_2-p}
    \Afieldupname{\nu}{\nu'}{R+r/2}{q_1}{1}
    \Afieldupname{\mu}{\mu'}{R-r/2}{q_2}{2} \right.\\
& & \displaystyle+\left. \int\dn{4}{\tilde{r}} \exp{[i\tilde{r}\cdot(-p+
    \frac{k_1-k_2}{2})-ir\cdot(q_1-q_2)]}\right.\\
& & \displaystyle\times\left.\Afieldupname{\nu}{\mu'}{R-\tilde{r}/4}{q_1}{1}
    \Afieldupname{\mu}{\nu'}{R+\tilde{r}/4}{q_2}{2}\right.\\
& & \displaystyle+\left.\int\dn{4}{\tilde{r}} \exp{[i\tilde{r}\cdot(-p+
    \frac{k_1-k_2}{2})+ir\cdot(q_1-q_2)]}\right.\\
& & \displaystyle\times\left.\Afieldupname{\mu}{\nu'}{R+\tilde{r}/4}{q_1}{1}
    \Afieldupname{\nu}{\mu'}{R-\tilde{r}/4}{q_2}{2}\right\}
\end{array}
\label{eqn:misceqn}
\end{equation}
This equation could look simpler if, in the interference terms, 
we Wigner transformed $A_1$ together with $A_2$.  However then we would have 
a virtual electron being emitted by some interference field and then 
reabsorbed by another interference field and the resulting equations would 
be impossible to interpret using our photon distributions.
In Eq.~(\ref{eqn:misceqn}), we neglect $\tilde{k}$ relative to $k$ in the 
effective vertex and in the factors of $(2k_{0}V)$ because the final 
state wave packets are sharply peaked in momentum.  In 
Eq.~(\ref{eqn:misceqn}), $R$
is the center of the interaction points $x_1$ and $x_2$ and $r$ is the 
space--time separation of these points. The final state Wigner 
density is 
\[\begin{array}{rl}
	\displaystyle f(x_1,k_1,x_2,k_2)=&	
	\displaystyle\frac{1}{(2k_{1,0}V)}\frac{1}{(2k_{2,0}V)}
	\int\dnpi{4}{\tilde{k}_1}\dnpi{4}{\tilde{k}_2}
	e^{-i\tilde{k}_1 x_1-i\tilde{k}_2 x_2} \\
	&\displaystyle\times f^{*}(k_1+\tilde{k}_1/2,k_2+\tilde{k}_2/2)
	f(k_1-\tilde{k}_1/2,k_2-\tilde{k}_2/2) 
\end{array}\]
and the Wigner transform of the effective vertex is
\[
	\Lambda_{\mu\mu'\nu\nu'}(k_1,k_2,p,r)=\int\dnpi{4}{\tilde{p}} 
	e^{i\tilde{p}\cdot r} \Lambda_{\mu\nu}(k_1,k_2,p+\tilde{p}/2)
	\Lambda^{*}_{\mu'\nu'}(k_1,k_2,p-\tilde{p}/2).
\]  
We can write the effective vertex in terms of the scalar Feynman 
propagator, 
\begin{equation}
\begin{array}{ccl}
   \Lambda_{\mu\mu'\nu\nu'}(k_1,k_2,p,r) & = & \bar{u}(k_1,s_1)\gamma_{\mu}
   (\dirslash{p}+\frac{i}{2}\dirslash{\partial}+m_e)\gamma_{\nu}v(k_2,s_2)\\
   & & \times \bar{v}(k_2,s_2)\gamma_{\nu'}(\dirslash{p}-
	\frac{i}{2}\dirslash{\partial}+m_e)
	\gamma_{\mu'}u(k_1,s_1)\Gcaus{r}{p}\\
	& \equiv & \lambda_{\mu\mu'\nu\nu'}(k_1,k_2,p,r) \Gcaus{r}{p}.
\end{array}
\end{equation}

We simplify the reaction probability by summing over the final state
electron and positron spins.  We simplify things even further by working in
the ultrarelativistic limit, namely when $v_1^2\approx v_2^2\approx 0$. 
Under these approximations,  we find
\begin{equation}
\label{eqn:noname}
\begin{array}{ccl}
\Smatrix{12\rightarrow 1'2'e\bar{e}}&=& \alpha_{em}^2 
    \displaystyle\int\dn{4}{R}\dn{4}{r}
    \dnpi{4}{k_1}\dnpi{4}{k_2}\dnpi{4}{q_1}\dnpi{4}{q_2}\dnpi{4}{p}\\
& & \displaystyle\times f(R-r/2,k_1,R+r/2,k_2)\\
& & \displaystyle\times \sum_{\rm spins}\lambda_{\mu\mu'\nu\nu'}(k_1,k_2,p,r)
    \Gcaus{r}{p}\twopideltaftn{4}{q_1+q_2-k_1-k_2}\\
& & \displaystyle\times\left\{ \twopideltaftn{4}{q_1-k_1+p} 
    \Afieldupname{\mu}{\mu'}{R-r/2}{q_1}{1}
    \Afieldupname{\nu}{\nu'}{R+r/2}{q_2}{2}\right.\\
& & \displaystyle+\left.  \twopideltaftn{4}{q_1-k_2-p}
    \Afieldupname{\nu}{\nu'}{R+r/2}{q_1}{1}
    \Afieldupname{\mu}{\mu'}{R-r/2}{q_2}{2} \right.\\
& & \displaystyle+\left. 2 \int\dn{4}{\tilde{r}} \cos{[\tilde{r}\cdot(-p+
    \frac{k_1-k_2}{2})-r\cdot(q_1-q_2)]}\right.\\
& & \displaystyle\times\left.\Afieldupname{\nu}{\mu'}{R-\tilde{r}/4}{q_1}{1}
    \Afieldupname{\mu}{\nu'}{R+\tilde{r}/4}{q_2}{2}\right\}.
\end{array}
\end{equation}
Given the relatively simple form of this equation, one would think that we
could identify the exchanged electron's phase--space density.  In fact, 
if we use free particle distributions for the 
final state electron and positron and sum over final states, we can 
identify  the virtual electron distribution 
(equation (\ref{eqn:elecdist})) in the direct terms. 
However,  we can not make the same identification in the
interference term and factorization is not possible here.   
We might find factorization again if we had several point charges as one can
envision a situation with many photon sources screening the photons (a plasma 
for instance).  The photon field might then 
be an incoherent superposition of photon fields.  In the absence of photon
interference, we might be able to define an effective electron distribution.

\subsubsection{The $e\bar{e}$ Production Region vs. the Interaction Region}

With equation (\ref{eqn:noname}), we can discuss 
the various length scales of the
problem.  First the $e\bar{e}$ production region is set by the shape and 
size of the photon distributions.
Second, the two photon interaction region's size depends on the 
mass and virtuality of the exchanged electron.  

First, take the virtual photon distribution of the classical point charge
from Section~\ref{sec:classEPD}.   Now, the lowest energy and momentum that 
each of the interacting photons can have is\footnote{The 
distribution of photons with $q=(m_e,m_e/v_L,\vec{0}_T)$ is shown in 
Fig.~\ref{fig:photondist}.} $q=(m_e,m_e/v_L,\vec{0}_T)$. 
Because the high energy or far off--shell photons are closer to the point 
charge then their lower energy and nearly on--shell cousins, photons 
with the minimum $q_\mu$ have the largest distributions.   So, the 
geometrical overlap of the high energy 
portions of the virtual photon distribution sets the size of 
the $e\bar{e}$ production region.  In Fig.~\ref{fig:overlap} we illustrate
this:  the two ellipses represent the edge of the photon distribution and the
shaded region is the region where the $e\bar{e}$ pairs can be created.

Now, the size of the two photon interaction itself is determined by how 
far the exchanged electron can travel between the vertices in Fig. 
\ref{fig:2phot2}.  For this, we look at the phase--space electron propagator. 
Assuming massive electrons,\footnote{The $m_e=0$ case is uninteresting because
the two photon interaction always extends over the entire $e\bar{e}$ production
region.} we use Remler's causal propagator.  Here the phase--space 
``probability''  for propagating between two space--time points drops 
like $e^{-2m_{e}\tau}$ for space--like electrons and like 
$\sin{2\tau (\sqrt{p^2}\pm m_{e})}/(\sqrt{p^2}\pm m_{e})$ for 
time--like electrons. The proper time along the electron 4--momentum is $\tau$.
In the direction transverse to the electron four--momentum, the ``probability'' 
is zero.  Thus,  the interaction region has a
characteristic length scale of $\approx 1/m_{e}$.  This is 
comparable to the width of the photon distributions, so there is no scale 
separation.  Typically one requires the interaction length scale to be much
smaller than the characteristic length scale of the particle density
in order to justify the gradient expansions and allow for a transport 
description.  Because our approach does not rely on the gradient expansions,
a transport description may still be possible.


\subsection{What the Electrons Tell Us About the Partons}
\label{sec:3d}

In this section, we learned several things about the massless parton 
phase--space densities.  First, owing to the fact that the simplest parton
ladder contains one rung representing a single partonic splitting, we
learned how both the parent parton and cut rung affect the parton
distribution.  The shape of the parent parton distribution determines
the spatial structure of the parton source.  The rung of the parton ladder 
segment gets cut, putting that parton on shell.  The integral over final 
states of this parton is weighted toward giving it a low $k_0$. 
Second, partons propagate to the same distances with the Feynman propagator
that we found for the retarded propagator, despite the difference in the
boundary conditions of the two propagators.
Finally, we learned that the ``source--propagator'' picture of parton densities 
fails when the source particle and probe particle interact, even  
through quantum interference. 
Nevertheless, we can still discuss the process in phase--space with the
phase--space sources and propagators, even if the densities have no clear
meaning.

%
%
\section{QED Transport Theory}
\label{sec:nthgen}

The ``source--propagator'' picture of particle densities seems both common and
physically intuitive.  In this section, we see how this picture arises 
in time--ordered nonequilibrium theory by  
deriving the Generalized Fluctuation--Dissipation Theorem in phase--space.  
The derivation of this theorem mirrors the steps often used to derive the 
semiclassical transport equations.  Namely, we derive the
Kadanoff--Baym equations and formally solve them to get the Generalized
Fluctuation--Dissipation Theorem.  Unlike other derivations of 
the transport equations, we do not perform the gradient 
expansion.  Instead,
we  Wigner transform the Fluctuation-Dissipation Theorem.  
If we then insert the Wigner transformed self--energies into the
Fluctuation--Dissipation Theorem, we get a set of phase--space evolution
equations for the particle densities.  These evolution equations describe the
complete evolution of the system in phase--space from some time in the distant
past to the present, including all parton splittings, recombinations and
scatterings.  In principle, the equations are nonperturbative, but we
can expand them perturbatively.  We demonstrate this by  recalculating the
photon and electron distributions of Sections \ref{sec:pdist} and 
\ref{sec:edist}.  Following this, we derive transport equations from the
phase--space evolution equations.  Finally, we describe how the results of this
section can be applied to parton transport.

For those familiar with the common steps in deriving semiclassical transport 
equations from the Kadanoff--Baym equations, we suggest skipping past 
Section \ref{sec:standardstuff} to Section \ref{sec:newstuff}.

\subsection{Green's Functions}
\label{sec:greens}

Our derivations begin with the contour Green's function which we
define as
\begin{mathletters}
\begin{equation}
	i\Gprop{x}{y}=\left< \tilde{T} \hat{\phi}(x) \hat{\phi}^{*}(y) \right>
\label{eqn:contgreens1}
\end{equation}
for scalar particles,
\begin{equation}
	i\Dprop{\mu}{\nu}{}{}{x}{y}=\left< \tilde{T} \hat{A}_{\mu}(x) 
	\hat{A}_{\nu}(y) \right>-\left< \hat{A}_{\mu}(x) \right>
	\left< \hat{A}_{\nu}(x) \right>
\label{eqn:contgreens2}
\end{equation}
for photons and
\begin{equation}
	i\Sprop{\alpha}{\beta}{}{}{x}{y}=\left< \tilde{T} \hat{\psi}_{\alpha}(x) 
		\hat{\bar{\psi}}_{\beta}(y) \right>
\label{eqn:contgreens3}
\end{equation}
\end{mathletters}
for fermions.  The $\left<\ldots\right>={\rm Tr}(\rho\ldots)/{\rm Tr}(\rho)$
is a trace over the system's density matrix, specified at 
time $t_0 \rightarrow -\infty$.  The field 
operators are taken in the Heisenberg picture. $\tilde{T}$ denotes 
ordering along the contour shown in Fig. \ref{fig:contour}.  
This ordering can be written as
\begin{equation}
	\tilde{T} A(x) B(y) \equiv \theta(x_0,y_0) A(x) B(y) \pm
	\theta(y_0,x_0) B(y) A(x). 
\label{eqn:ordering}
\end{equation}
The upper sign refers to bosons and the lower sign to fermions.
The contour theta function is defined as
\[\theta(x_0,y_0) = \left\{
	\begin{array}{ll}
	1 & \mbox{if $x_0$ is later on the contour than $y_0$}\\
	0 & \mbox{otherwise}
	\end{array}
\right. \]

In addition to the contour Green's functions (\ref{eqn:contgreens1}--
\ref{eqn:contgreens3}), we define the $>$ and $<$ Green's functions:
\begin{mathletters}
\begin{eqnarray}
	i\Ggreat{x}{y} & = & \left< \hat{\phi}(x)\hat{\phi}^{*}(y) \right>
\label{eqn:gldensity}\\
	i\Dgreat{\mu}{\nu}{}{}{x}{y} & = & \left< \hat{A}_{\mu}(x) \hat{A}_{\nu}(y)
		\right> - \left< \hat{A}_{\mu}(x) \right> \left< \hat{A}_{\nu}(y) 
		\right>\\
	i\Sgreat{\alpha}{\beta}{}{}{x}{y} & = & \left< \hat{\psi}_{\alpha}(x)
		\hat{\bar{\psi}}_{\beta}(y) \right>\\
	i\Gless{x}{y} & = & \left< \hat{\phi}^{*}(y)\hat{\phi}(x) \right>\\
	i\Dless{\mu}{\nu}{}{}{x}{y} & = & \left< \hat{A}_{\nu}(y)
		\hat{A}_{\mu}(x) \right> - \left< \hat{A}_{\mu}(x) \right> 
		\left< \hat{A}_{\nu}(y) \right>\\
	i\Sless{\alpha}{\beta}{}{}{x}{y} & = & - \left< \hat{\bar{\psi}}_{\beta}(y)
		\hat{\psi}_{\alpha}(x) \right>
\end{eqnarray}
\end{mathletters}
These Green's functions are hermitian and contain the complete single--particle
information of the system.
For example, setting $x=y$ gives us the single particle density matrix.
Furthermore, Wigner transforming in the relative coordinate, we find the
off--mass shell generalization of the 
Wigner function for the particles:  
\[
	f(x,p)=i\Gless{x}{p}=\int\dn{4}{(x-y)}e^{i(x-y)\cdot
	p}i\Gless{x}{y}=\int\dn{4}{(x-y)}e^{i(x-y)\cdot
	p}\left< \hat{\phi}^{*}(y)\hat{\phi}(x)\right>.
\]
We identify $f(x,p)$ with the number density of particle 
(or antiparticles) per unit volume in phase--space per unit invariant mass 
at time $x_0$:
\[
   f(x,p)=\frac{dn(x,p)}{d^3x \:d^3p \:dp^2}
\]
The off--shell Wigner function is related to the 
conventional Wigner function, $f_0(x,\vec{p})$, through the invariant mass
integration:
\[
   f_0 (x,\vec{p})=\frac{dn(x_0,\vec{x},\vec{p})}{d^3x \:d^3p}=
   \int^\infty_{-\infty} dp^2 f(x,p).
\]

In terms of the $\gtrless$ Green's functions, the contour Green's function can 
be written as
\begin{equation}
\Gprop{x}{y}=\theta(x_0,y_0)\Ggreat{x}{y}+\theta(y_0,x_0)\Gless{x}{y}
\label{eqn:greenrelation1}
\end{equation}
for both fermions and bosons.
Furthermore, because of the equal time commutation relations, $\Ggreat{x}{y}
= \Gless{y}{x}$ and $\Gprop{x}{y}=\Gprop{y}{x}$, for both fermions and 
bosons.

We define several auxiliary Green's functions in terms of the $>$ and $<$ 
Green's functions:  the retarded and advanced Green's 
functions and the Feynman and anti--Feynman propagators.
We write only the equations for the scalar particles.  
For the retarded and advanced propagators, we have
\begin{equation}
\Gmisc{x}{y}{\pm}=\pm\theta(\pm(x_0-y_0))(\Ggreat{x}{y}-\Gless{x}{y}).
\label{eqn:greenrelation2}\\
\end{equation}
For the Feynman and anti--Feynman propagators, we have: 
\begin{mathletters}
\begin{eqnarray}
\Gcaus{x}{y}&=&\theta(x_0-y_0)\Ggreat{x}{y}+\theta(y_0-x_0)\Gless{x}{y},
\label{eqn:greenrelation3}\\
\Gacaus{x}{y}&=&\theta(y_0-x_0)\Ggreat{x}{y}+\theta(x_0-y_0)\Gless{x}{y}.
\label{eqn:greenrelation4}
\end{eqnarray}
\end{mathletters}
One can also obtain these Feynman and anti--Feynman propagators by restricting
the arguments of the contour propagators to be on one side of the contour in 
Fig. \ref{fig:contour}.  


\subsection{Conventional Transport Theory}
\label{sec:standardstuff}

In this subsection, we follow the standard derivation of the transport
equations up to the point where we find the Generalized
Fluctuation--Dissipation Theorem.  The procedure is as follows: 1) find the
Dyson--Schwinger Equations for the contour Green's functions, 2) apply the free
field equations of motion to get the Kadanoff--Baym equations and 3) solve the
Kadanoff--Baym equations to get the Generalized Fluctuation--Dissipation 
Theorem.

\subsubsection{Dyson--Schwinger Equations}

The Dyson--Schwinger equations encapsulate all of the nonperturbative effects 
in the field theory that are possible with only two--point functions. 
We can write the Dyson--Schwinger equations for the photons, electrons
and scalars using the contour Green's functions:
   \begin{mathletters}
\label{eqn:DSEall} 
\begin{eqnarray}
\Gprop{1}{1'} & = & \Gnon{1}{1'} + 
      \intc d2\: d3\: \Gnon{1}{2} Q(2,3) \Gprop{3}{1'} 
      \label{eqn:DSE1}\\
\Dprop{\mu}{\nu}{}{}{1}{1'} & = & \Dnon{\mu}{\nu}{}{}{1}{1'} + 
      \intc d2\: d3\: \Dnon{\mu}{\mu'}{}{}{1}{2} \Pi^{\mu'\nu'}(2,3) 
      \Dprop{\nu'}{\nu}{}{}{3}{1'} \label{eqn:DSE2} \\
\Sprop{\alpha}{\beta}{}{}{1}{1'} & = & \Snon{\alpha}{\beta}{}{}{1}{1'} + 
\intc d2\: d3\: \Snon{\alpha}{\alpha'}{}{}{1}{2} \Sigma^{\alpha'\beta'}(2,3) 
\Sprop{\beta'}{\beta}{}{}{3}{1'} \label{eqn:DSE3}
\end{eqnarray}
\end{mathletters}
In these equations, we 
represent the coordinates by their index, i.e. $x_1\rightarrow 1$.
We present the corresponding diagrams in Figs. \ref{fig:DSE}(a-c).
In equations (\ref{eqn:DSE1}-\ref{eqn:DSE3}), the non--interacting 
contour Green's functions have a $0$ superscript.

The self--energies describe all of the branchings and recombinations possible
for the photons, electrons and scalars.  The self--energies are:
\begin{mathletters}
\label{eqn:selfen}
\begin{eqnarray}
Q(1,1') & = & i (eZ\bothpartial^\mu) \intc d2\: d3\:
   \Gprop{1}{3} \Gamma_{\gamma \phi\phi}^{\nu}(2,3,1')
   \Dprop{\mu}{\nu}{}{}{1}{3} \nonumber \\
& &+i (2i \alpha_{em} Z^2 g^{\mu\nu}) \intc d2\: d3\: d4\: \Gprop{1}{2}
   \Gamma_{\gamma\gamma \phi\phi}^{\mu'\nu'}(2,3,4,1')
   \Dprop{}{}{\mu}{\mu'}{1}{3}\Dprop{}{}{\nu}{\nu'}{1}{4}
   \label{eqn:selfen1}\\
& &+Q_{\rm MF} (1)\deltaftn{4}{1,1'} \nonumber \\
\Pi_{\mu\nu}(1,1') & = & -i (-i e (\gamma_{\mu})_{\alpha\beta})
   \intc d2\: d3\: \Sprop{}{}{\alpha}{\alpha'}{1}{2}
   \Gamma_{\gamma e, \nu}^{\alpha' \beta'}(2,3,1')
   \Sprop{}{}{\beta'}{\beta}{3}{1} \nonumber\\
& & +i (eZ \bothpartial_\mu) \intc d2\: d3\: \Gprop{1}{2}
   \Gamma_{\gamma \phi\phi, \nu}(2,3,1')\Gprop{3}{1} 
   \label{eqn:selfen2}\\
& & +i (2 i \alpha_{em} Z^2 g_{\mu\mu'}) \intc d2\: d3\: d4\:
   \Gprop{1}{2}\Gprop{3}{1}
   \Gamma_{\gamma\gamma \phi\phi,\nu}^{\nu'}(2,3,4,1')
   \Dprop{\mu'}{\nu'}{}{}{1}{4} \nonumber \\
& &+\Pi_{\rm MF}(1) g_{\mu\nu} \deltaftn{4}{1,1'} \nonumber \\
\Sigma_{\alpha \beta}(1,1') & = & i (-i e 
   (\gamma^{\mu})_{\alpha \alpha'}) \intc d2\: d3\: 
   \Sprop{\alpha'}{\beta'}{}{}{1}{2}
   \Gamma_{\gamma e}^{\beta' \beta,\nu}(2,3,1')
   \Dprop{\mu}{\nu}{}{}{1}{3}
   \label{eqn:selfen3}\\
& & + \Sigma_{\rm MF}(1)
   \delta_{\alpha \beta}\deltaftn{4}{1,1'} \nonumber
\end{eqnarray}
\end{mathletters}
In Figs. \ref{fig:selfenergy}(a-c), we show all the diagrams corresponding 
to the non--mean--field terms in equations 
(\ref{eqn:selfen1}-\ref{eqn:selfen3}).  We define the contour delta function 
$\deltaftn{4}{x,y}$ by
\[\deltaftn{4}{x,y}=\left\{ 
\begin{array}{ll}
\deltaftn{4}{x-y} & \mbox{for $x_0$, $y_0$ on the upper branch}\\
0 & \mbox{for $x_0$, $y_0$ on different branches}\\
-\deltaftn{4}{x-y} & \mbox{for $x_0$, $y_0$ on the lower branch}\\
\end{array}
\right. \]

\subsubsection{Kadanoff--Baym Equations}

The free--field contour Green's functions satisfy the equations of
motion:
\begin{mathletters}
\label{eqn:greeneqmot}
\begin{eqnarray}
	(\partial_{x}^2+M^2)\Gnon{x}{y}&=&\deltaftn{4}{x,y}\\
	\partial_{x}^2\Dnon{}{}{\mu}{\nu}{x}{y}&=&4\pi g_{\mu\nu}\deltaftn{4}{x,y}\\
	(i\dirslash{\partial}_{x}-m_e)\Snon{}{}{\alpha}{\beta}{x}{y}&=&
		\delta_{\alpha\beta}\deltaftn{4}{x,y}
\end{eqnarray}
\end{mathletters}
Combining these with the Dyson--Schwinger equations, we have
\begin{mathletters}
\label{eqn:fullgreeneqmot}
\begin{eqnarray}
(\partial_{1}^2+M^2)\Gprop{1}{1'} & = & \deltaftn{4}{1,1'} + 
      \intc d2\: Q(1,2) \Gprop{2}{1'} \\
\partial_{1}^2\Dprop{\mu}{\nu}{}{}{1}{1'} & = &
      4\pi g_{\mu\nu}\deltaftn{4}{1,1'}+
      4\pi\intc d2\: \Pi_{\mu}^{\;\;\nu'}(1,2) 
      \Dprop{\nu'}{\nu}{}{}{2}{1'} \\
(i\dirslash{\partial}_{1}-m_e)\Sprop{\alpha}{\beta}{}{}{1}{1'} & = &
      \delta_{\alpha\beta}\deltaftn{4}{1,1'}+ 
      \intc d2\: \Sigma_{\alpha\beta'}(1,2) \Sprop{\beta'}{\beta}{}{}{2}{1'}. 
\end{eqnarray}
\end{mathletters}
There is a conjugate set of equations (\ref{eqn:greeneqmot}),
(\ref{eqn:fullgreeneqmot}) with the differential operators acting on $1'$.

Restricting $t_1$ and $t_{1'}$ to lie on different sides of the
time contour in Fig. \ref{fig:contour}, we arrive at the
Kadanoff--Baym equations.  
\begin{mathletters}
\label{eqn:KBEall}
\begin{equation}
\begin{array}{ccl}
(\partial_{1}^2+M^2)\Ggl{1}{1'}&=&\displaystyle\int \dn{3}{x_2} Q_{\rm MF}
   (\vec{x}_1,\vec{x}_2,t_1) \Ggl{\vec{x}_2,t_1}{1'}\\
& &+\displaystyle\int^{t_1}_{t_0} d2\: \left(Q^{>}(1,2)-Q^{<}(1,2)\right) 
   \Ggl{2}{1'} \\
& &+\displaystyle\int^{t_1'}_{t_0} d2\: Q^{\gtrless}(1,2)\left(
   \Ggreat{2}{1'}-\Gless{2}{1'}\right) 
\end{array}
\end{equation}
\begin{equation}
\begin{array}{ccl}
\frac{1}{4\pi}\partial_{1}^2\Dgl{\mu}{\nu}{1}{1'}&=&\displaystyle
    \int \dn{3}{x_2}
    \Pi_{\rm MF}(\vec{x}_1,\vec{x}_2,t_1)
    \Dgl{\mu}{\nu}{\vec{x}_2,t_1}{1'}\\
& & +\displaystyle\int^{t_1}_{t_0} d2\: 
    \left(\Pi^{>\:\:\:\nu'}_{\:\:\:\mu}(1,2)
    -\Pi^{<\:\:\:\nu'}_{\:\:\:\mu}(1,2)\right) \Dgl{\nu'}{\nu}{2}{1'} \\
& & +\displaystyle\int^{t_1'}_{t_0} d2\: 
    \Pi^{\gtrless\:\:\:\nu'}_{\:\:\:\mu}(1,2)
    \left(\Dgreat{}{}{\nu'}{\nu}{2}{1'}-\Dless{}{}{\nu'}{\nu}{2}{1'}\right) 
\end{array}
\end{equation}
\begin{equation}
\begin{array}{ccl}
(i\dirslash{\partial}_{1}-m_e)\Sgl{\alpha}{\beta}{1}{1'}&=&
    \displaystyle\int \dn{3}{x_2} \Sigma_{\rm MF}(\vec{x}_1,\vec{x}_2,t_1)
    \Sgl{\alpha}{\beta}{\vec{x}_2,t_1}{1'}\\
& & +\displaystyle\int^{t_1}_{t_0} d2\: \left(\Sigma^{>}_{\alpha\beta'}(1,2)
    -\Sigma^{<}_{\alpha\beta'}(1,2)\right) \Sgl{\beta'}{\beta}{2}{1'} \\
& & +\displaystyle\int^{t_1'}_{t_0} d2\: \Sigma^{\gtrless}_{\alpha\beta'}(1,2)
    \left(\Sgreat{}{}{\beta'}{\beta}{2}{1'}-
    \Sless{}{}{\beta'}{\beta}{2}{1'}\right) 
\end{array}
\end{equation}
\end{mathletters}
Here the $>$ and $<$ self--energies have the same meaning relative to
the contour self--energy as the $>$ and $<$ Green's functions do
relative to the contour Green's functions.  Again, there is a set
of conjugate equations with the differential operators acting on $1'$.

\subsubsection{Generalized Fluctuation--Dissipation Theorem}

Now we define the retarded and advanced self--energies for
scalars:
\begin{equation} Q^{\pm}(1,2)=Q_{\rm MF}\deltaftn{}{t_1,t_2}
\pm\theta\left( \pm(t_1-t_2)\right) \left( Q^{>}(1,2)-Q^{<}(1,2) \right)
\label{eqn:pmselfen}\end{equation}
The photon polarization tensor and electron self--energy are defined
in a similar manner.

Using these, we simplify the Kadanoff--Baym equations:
\begin{mathletters}
\label{eqn:KBEall2}
\begin{eqnarray}
(\partial_{1}^2+M^2)\Ggl{1}{1'} &=&
   \displaystyle\int^{\infty}_{t_0} d2\: Q^{+}(1,2) \Ggl{2}{1'}\\
   &&+\displaystyle\int^{\infty}_{t_0} d2\: Q^{\gtrless}(1,2)
   \Gmin{2}{1'}\\
\frac{1}{4\pi}\partial_{1}^2\Dgl{\mu}{\nu}{1}{1'}&=&
   \displaystyle\int^{\infty}_{t_0} d2\: \Pi^{+\:\:\:\nu'}_{\:\:\:\mu}(1,2)
   \Dgl{\nu'}{\nu}{2}{1'} \\
   &&+\displaystyle\int^{\infty}_{t_0} d2\: 
   \Pi^{\gtrless\:\:\:\nu'}_{\:\:\:\mu}(1,2)
   \Dmin{}{}{\nu'}{\nu}{2}{1'}\\ 
(i\dirslash{\partial}_{1}-m_e)\Sgl{\alpha}{\beta}{1}{1'}&=&
    \displaystyle\int^{\infty}_{t_0} d2\: \Sigma^{+}_{\alpha\beta'}(1,2)
    \Sgl{\beta'}{\beta}{2}{1'} \\
    &&+\displaystyle\int^{\infty}_{t_0} d2\: \Sigma^{\gtrless}_{\alpha\beta'}(1,2)
    \Smin{}{}{\beta'}{\beta}{2}{1'}
\end{eqnarray}
\end{mathletters}
If we subtract the $>$ equations from the $<$ equations and
multiply the resulting equations by $\pm\theta(\pm(t_1-t_{1'}))$,
we get a second set of differential equations:
\begin{mathletters}
\label{eqn:KBEall3}
\begin{eqnarray}
(\partial_{1}^2+M^2)\Gpm{1}{1'} &=& \deltaftn{4}{1-1'}
   +\int^{\infty}_{t_0} d2\: Q^{\pm}(1,2)\Gpm{2}{1'}\\
\frac{1}{4\pi}\partial_{1}^2\Dpm{\mu}{\nu}{1}{1'} &=& \deltaftn{4}{1-1'}
   +\int^{\infty}_{t_0} d2\: \Pi^{\pm\:\:\:\nu'}_{\:\:\:\mu}(1,2)
   \Dpm{\nu'}{\nu}{2}{1'}\\ 
(i\dirslash{\partial}_{1}-m_e)\Spm{\alpha}{\beta}{1}{1'}&=&
    \deltaftn{4}{1-1'}
    +\int^{\infty}_{t_0} d2\: \Sigma^{\pm}_{\alpha\beta'}(1,2)
    \Spm{\beta'}{\beta}{2}{1'}
\end{eqnarray}
\end{mathletters}

Solving the initial value problem posed by equations
(\ref{eqn:KBEall2}) using equations (\ref{eqn:KBEall3}),  
we find:
\begin{mathletters}
\label{eqn:FDT}
\begin{eqnarray}
\Ggl{1}{1'} &=& \int^{\infty}_{t_0} d2\: \int^{\infty}_{t_0} d3\:
   \Gplus{1}{2}Q^{\gtrless}(2,3)\Gmin{3}{1'}\\
   & & + \int \dn{3}{x_2}\dn{3}{x_3} \Gplus{1}{\vec{x}_2,t_0}
   \Ggl{\vec{x}_2,t_0}{\vec{x}_3,t_0}\Gmin{\vec{x}_3,t_0}{1'}
   \nonumber\\
\Dgl{\mu}{\nu}{1}{1'} &=& \int^{\infty}_{t_0} d2\:
   \int^{\infty}_{t_0} d3\:\Dplus{}{}{\mu}{\mu'}{1}{2}
   \Pi^{\gtrless\:\:\:\mu'\nu'}(2,3)\Dmin{}{}{\nu'}{\nu}{3}{1'}\\ 
   & & + \int \dn{3}{x_2}\dn{3}{x_3}
   \Dplus{}{}{\mu}{\mu'}{1}{\vec{x}_2,t_0}
   \Dglup{\mu'}{\nu'}{\vec{x}_2,t_0}{\vec{x}_3,t_0}
   \Dmin{}{}{\nu'}{\nu}{\vec{x}_3,t_0}{1'}
   \nonumber\\
\Sgl{\alpha}{\beta}{1}{1'}&=&\int^{\infty}_{t_0} d2\:
   \int^{\infty}_{t_0} d3\: \Splus{}{}{\alpha}{\alpha'}{1}{2}
   \Sigma^{\gtrless}_{\alpha'\beta'}(2,3)
   \Smin{}{}{\beta'}{\beta}{3}{1'}\\
   & & + \int \dn{3}{x_2}\dn{3}{x_3}
   \Splus{}{}{\alpha}{\alpha'}{1}{\vec{x}_2,t_0}
   \Sgl{\alpha'}{\beta'}{\vec{x}_2,t_0}{\vec{x}_3,t_0}
   \Smin{}{}{\beta'}{\beta}{\vec{x}_3,t_0}{1'}
   \nonumber
\end{eqnarray}
\end{mathletters}
These equations are the Generalized Fluctuation--Dissipation Theorem.
They describe the evolution of a density fluctuation (given by the $>$
and $<$ Green's functions) from $t_0$ to $t_1$.  

\subsection{Phase--Space Generalized Fluctuation--Dissipation Theorem}
\label{sec:newstuff}

We  now translate the fluctuation
dissipation equations (\ref{eqn:FDT}) into phase--space.  We only
do so for the scalar equation because the photon and electron equations
are similar.  First we  extend the
integration region to cover all time:
\[
\begin{array}{ccl}

\Ggl{x_1}{x_{1'}} &=& \displaystyle\int \dn{4}{x_2}\: \dn{4}{x_3}\:
   \Gplus{x_1}{x_2}Q^{\gtrless}(x_2,x_3)\Gmin{x_3}{x_{1'}}\\
& & + \displaystyle\lim_{t_0\rightarrow -\infty} \int \dn{4}{x_2}\dn{4}{x_3}
   \deltaftn{}{t_0-x_{2 0}}\deltaftn{}{t_0-x_{3 0}}\\
& & \times\Gplus{x_1}{x_2}\Ggl{x_2}{x_3}\Gmin{x_3}{x_{1'}}.
\end{array}
\]

Next, we Wigner transform in the relative variable $x_1-x_{1'}$:
\begin{equation}
\begin{array}{ccl}
   \Ggl{x}{p}&=&{\displaystyle\int \dn{4}{x'}\dnpi{4}{p'}
      \tilde{G}^{+}(x,p;x',p')Q^{\gtrless}(x',p')}\\
   & &+ {\displaystyle\lim_{x'_0\rightarrow-\infty}\int\dn{3}{x'}\dnpi{4}{p'} 
      \tilde{G}^{+}(x,p;x',p')\Ggl{x'}{\vec{p}'}}
\end{array}
\label{eqn:inteqn}
\end{equation}
We recognize the Wigner transforms of the self--energy 
and initial particle density:
\begin{equation}
   Q^{\gtrless}(x,p)=\int\dn{4}{\tilde{x}} e^{ip\cdot\tilde{x}}
      Q^{\gtrless}(x+\tilde{x}/2,x-\tilde{x}/2)
\end{equation}
and
\begin{equation}\begin{array}{ccl}
   \lefteqn{\displaystyle\deltaftn{}{t_0-x_0}\Ggl{x}{\vec{p}}=}\\
   & & \displaystyle\int\dn{4}{\tilde{x}}
      e^{ip\cdot\tilde{x}}\deltaftn{}{t_0-(x_0+\tilde{x}_0/2)}
      \deltaftn{}{t_0-(x_0-\tilde{x}_0/2)}\Ggl{x+\tilde{x}/2}{x-\tilde{x}/2}.
\end{array}\end{equation}
The delta functions render the initial density independent of $p_0$.  
We have also defined the retarded propagator in phase--space:
\begin{equation}
   \tilde{G}^{+}(x,p;y,q)=\int\dn{4}{x'}\dn{4}{y'}e^{i(p\cdot x'-q\cdot y')}
   \Gplus{x+x'/2}{y+y'/2}\Gmin{x-x'/2}{y-y'/2}
\end{equation}
At this point, one usually applies the gradient approximation to equation 
(\ref{eqn:inteqn}), eliminating the $\dn{4}{x'}$ integral.  We do 
not do this.  

Next, we assume the translational invariance of the advanced and retarded
propagators.  This is reasonable at lowest order in the coupling
since the free field advanced and retarded propagators are
translationally invariant.  Making this approximation, the retarded
propagator in phase--space becomes
\begin{equation}\begin{array}{ccl}
   \displaystyle
   \tilde{G}^{+}(x,p;y,q)&=&{\displaystyle\twopideltaftn{4}{p-q}\int
   \dn{4}{z}e^{ip\cdot z}G^{+}(x-y+z/2)\left(G^{+}(x-y-z/2)\right)^{*}}\\
   &\equiv & \twopideltaftn{4}{p-q}\Gplus{x-y}{p}.
\end{array}\end{equation}
We will use $\Gplus{x-y}{p}$ in all subsequent calculations. 
In practice, we will only use the lowest order contribution to $\Gplus{x-y}{p}$.
This  means that we dress the $\gtrless$ propagators but not the $\pm$ 
propagators when we iterate equation (\ref{eqn:FDT}).  
Thus, our particles propagate as though they are in the vacuum.  In Appendix
\ref{append:prop} we calculate the lowest order contribution to 
$\Gplus{x-y}{p}$.  

Repeating this for the photons and electrons,
we arrive at the phase--space Generalized Fluctuation--Dissipation Theorem:
\begin{mathletters}
\label{eqn:evolve1}
\begin{eqnarray}
\Ggl{x}{p}&=&\int\dn{4}{y}\Gplus{x-y}{p} Q^{\gtrless}(y,p)\nonumber\\
	&&+ \lim_{y_0\rightarrow -\infty}\int\dn{3}{y}\Gplus{x-y}{p}
	\Ggl{y}{\vec{p}}\\
\Dgl{\mu}{\nu}{x}{p}&=&\int\dn{4}{y}\Dplus{\mu}{\nu}{\mu'}{\nu'}{x-y}{p}
	\Pi^{\gtrless \mu'\nu'}(y,p)\nonumber\\
	&&+\lim_{y_0\rightarrow -\infty}\int\dn{3}{y}
	\Dplus{\mu}{\nu}{\mu'}{\nu'}{x-y}{p}\Dglup{\mu'}{\nu'}{y}{\vec{p}}\\
\Sgl{\alpha}{\beta}{x}{p}&=&\int\dn{4}{y}
	\Splus{\alpha}{\beta}{\alpha'}{\beta'}{x-y}{p}
	\Sigma^{\gtrless}_{\alpha'\beta'}(y,p)\nonumber\\
	&&+\lim_{y_0\rightarrow -\infty}\int\dn{3}{y}
	\Splus{\alpha}{\beta}{\alpha'}{\beta'}{x-y}{p}
	\Sgl{\alpha'}{\beta'}{y}{\vec{p}}.
\end{eqnarray}
\label{eqn:GFDT}
\end{mathletters}
These equations describe the evolution of the particle phase--space
densities from $y_0\rightarrow -\infty$ to the time $x_0$, including
particle creation and absorption through the particle self--energies.
They clearly have the ``source--propagator'' form, but also contain 
information about the initial particle density.  
The derivation of these equations does not rely on the form of the 
self--energies and the general form is shown diagrammatically in Fig.
\ref{fig:GFDThm}.  Thus, these equation can be re--applied to QCD.  We 
exploit this fact when we discuss the shape of a nucleon's parton cloud.

\subsection{Phase--Space Evolution Equations}
\label{sec:evolve}

The first step toward getting the phase--space evolution equations
from the Generalized Fluctuation--Dissipation Theorem is to calculate 
the self--energies (i.e. the sources).  
To do this, we insert equations 
(\ref{eqn:pmselfen}) and (\ref{eqn:greenrelation1}) into the self--energy 
equations and keep only the lowest order approximation to the vertex 
functions.  Thus, we assume that the interaction 
time is much smaller than the other time scales in the problem.  So, 
we arrive at the creation and absorption rates:
\begin{mathletters}
\label{eqn:selfen4}
\begin{eqnarray}
Q^{\gtrless}(1,1')&=&i \alpha_{em} Z^2 \bothpartial_{1\mu} \Ggl{1}{1'}
   \bothpartial_{1'\nu} \Dglup{\mu}{\nu}{1}{1'}
   +Q_{\rm MF}^{\gtrless}(1) \deltaftn{4}{1-1'}\\
\Pi^{\gtrless}_{\mu\nu}(1,1')&=& i \alpha_{em} {\rm Tr}\left\{
   \gamma_\mu\Sgl{}{}{1}{1'}\gamma_\nu\Slg{}{}{1}{1'}
   \right\}
   +i \alpha_{em} Z^2 \bothpartial_{1\mu} \Ggl{1}{1'}
   \bothpartial_{1'\nu} \Glg{1}{1'}\\
   &+&\Pi_{\rm MF}^{\gtrless}(1)g_{\mu\nu}\deltaftn{4}{1-1'}\nonumber\\
\Sigma^{\gtrless}_{\alpha\beta}(1,1')&=&-i \alpha_{em}
	(\gamma_\mu)_{\alpha\alpha'}
   \Sgl{\alpha'}{\beta'}{1}{1'}(\gamma_\nu)_{\beta'\beta}
   \Dglup{\mu}{\nu}{1}{1'}
   +\Sigma_{\rm MF}^{\gtrless}(1)\delta_{\alpha\beta}\deltaftn{4}{1-1'}
\end{eqnarray}
\end{mathletters}
We have neglected the second scalar term in the polarization tensor
and the second photon term in the scalar self--energy
because they enter with a factor $\alpha_{em}^2$ which is higher order than 
the other terms we kept.  

The self--energies in (\ref{eqn:selfen4}) can be Wigner transformed.
Taking care to integrate the derivative scalar couplings by parts, we arrive at
\begin{mathletters}
\label{eqn:selfen5}
\begin{eqnarray}
Q^{\gtrless}(x,p)&=&i \alpha_{em} Z^2\int\dnpi{4}{q_1}\dnpi{4}{q_2}
   (q_1+q_2-i\bothpartial/2)_{\mu} \Ggl{x}{q_1}\\
   & & \times(q_1+q_2-i\bothpartial/2)_{\nu} \Dglup{\mu}{\nu}{x}{q_2}
   \twopideltaftn{4}{p-(q_1+q_2)}
   +Q_{\rm MF}^{\gtrless}(x) \nonumber\\
\Pi^{\gtrless}_{\mu\nu}(x,p)&=&i \alpha_{em} \int\dnpi{4}{q_1}\dnpi{4}{q_2}
   {\rm Tr}\left\{\gamma_\mu\Sgl{}{}{x}{q_1}\gamma_\nu\Sgl{}{}{x}{q_2}
   \right\}\twopideltaftn{4}{p-(q_1+q_2)}\nonumber\\
   & &+i \alpha_{em} Z^2\int\dnpi{4}{q_1}\dnpi{4}{q_2} 
   (q_1+q_2+i\bothpartial/2)_{\mu}\Ggl{x}{q_1}\\
   & & \times(q_1+q_2+i\bothpartial/2)_{\nu}\Ggl{x}{q_2}
   \twopideltaftn{4}{p-(q_1+q_2)}
   +\Pi_{\rm MF}^{\gtrless}(x)g_{\mu\nu}\nonumber\\
\Sigma^{\gtrless}_{\alpha\beta}(x,p)&=&-i \alpha_{em} 
	(\gamma_\mu)_{\alpha\alpha'}\int\dnpi{4}{q_1}\dnpi{4}{q_2}
   \Sgl{\alpha'}{\beta'}{x}{q_1}(\gamma_\nu)_{\beta'\beta}
   \Dglup{\mu}{\nu}{x}{q_2}\\
   & &\times\twopideltaftn{4}{p-(q_1+q_2)}
   +\Sigma_{\rm MF}^{\gtrless}(x)\delta_{\alpha\beta}\nonumber
\end{eqnarray}
\end{mathletters}

We can rewrite these equations directly in terms of 
the particle and antiparticle densities\footnote{A.k.a. the particle and
anti--particle Wigner functions.} to make their structure explicit.
In accordance with equations (\ref{eqn:gldensity}-f), we define the particle 
densities as follows:
\begin{mathletters}
\begin{equation}
i\Ggl{x}{p}=\ggtrless{x}{p}={\mbox{scalar densities}}
\end{equation}
\begin{equation}
i\Dgl{\mu}{\nu}{x}{p}=\dgtrless{\mu}{\nu}{x}{p}={\mbox{photon densities}}
\end{equation}
\begin{equation}
\pm i\Sgl{\alpha}{\beta}{x}{p}=\sgtrless{\alpha}{\beta}{x}{p}
={\mbox{lepton densities}}.
\end{equation}
\end{mathletters}
Here the positive energy part of the $>$ Green's functions correspond to the 
density for emission of $(|p_0|,\vec{p})$ quanta, while the negative energy 
part corresponds to the density for absorption of $(-|p_0|,\vec{p})$ quanta.
Similarly the positive energy part of the $<$ Green's functions correspond 
to absorbing $(|p_0|,\vec{p})$ quanta and the negative energy part to 
emission of $(-|p_0|,\vec{p})$ quanta.  Thus,  
$\theta(p_0)\gless{x}{p}$ is the scalar density and 
$\theta(-p_0)\ggtr{x}{p}$ is the antiscalar density.  We can make similar
identifications for the photon\footnote{The $d^{<}_{\mu\nu}(x,p)$ in this 
section is the Wigner transform of the vector potential, $A_{\mu\nu}(x,q)$, 
of Section \ref{sec:pdist}, as we demonstrate in Subsection 
\ref{sec:nonequilibedist}.} and lepton densities.  

Now, combining (\ref{eqn:evolve1}) and (\ref{eqn:selfen5}) and inserting 
the particle densities, we arrive at
\begin{mathletters}
\label{eqn:evolve2}
\begin{eqnarray}
\ggtrless{x}{p}&=&\int\dn{4}{y}\dnpi{4}{q_1}\dnpi{4}{q_2}\Gplus{x-y}{p} 
   \twopideltaftn{4}{p-(q_1+q_2)}\nonumber\\
   & & \times \alpha_{em} Z^2 (q_1+q_2-i\bothpartial/2)^{\mu}\ggtrless{y}{q_1} 
   (q_1+q_2-i\bothpartial/2)^{\nu}\dgtrless{\mu}{\nu}{y}{q_2}\\
   & & + \int\dn{4}{y}\Gplus{x-y}{p} iQ_{\rm MF}^{\gtrless}(y)\nonumber\\
   & & + \lim_{y_0\rightarrow -\infty}\int\dn{3}{y}\Gplus{x-y}{p}
   \ggtrless{y}{\vec{p}}\nonumber\\
\dgtrless{\mu}{\nu}{x}{p}&=&\int\dn{4}{y}\dnpi{4}{q_1}\dnpi{4}{q_2}
   \Dplus{\mu}{\nu}{\mu'}{\nu'}{x-y}{p}\twopideltaftn{4}{p-(q_1+q_2)}
   \nonumber\\
   & & \times \left\{ \alpha_{em} {\rm Tr}\left[\gamma^{\mu'} 
   \sgtrless{}{}{y}{q_1}\gamma^{\nu'}\sgtrless{}{}{y}{q_2}\right]\right.
   \nonumber\\
   & & + \left. \alpha_{em} Z^2 (q_1+q_2+i\bothpartial/2)^{\mu'} 
   \ggtrless{y}{q_1}
   (q_1+q_2+i\bothpartial/2)^{\nu'}\ggtrless{y}{q_2}\right\}\\
   & & + \int\dn{4}{y}\Dplus{\mu}{\nu}{\mu'}{\nu'}{x-y}{p}
   i\Pi_{\rm MF}^{\gtrless}(y)g^{\mu'\nu'}\nonumber\\
   & & + \lim_{y_0\rightarrow -\infty}\int\dn{3}{y}
   \Dplus{\mu}{\nu}{\mu'}{\nu'}{x-y}{p}\dgtrlessup{\mu'}{\nu'}{y}{\vec{p}}
   \nonumber\\
\sgtrless{\alpha}{\beta}{x}{p}&=&\int\dn{4}{y}\dnpi{4}{q_1}\dnpi{4}{q_2}
   \Splus{\alpha}{\beta}{\alpha'}{\beta'}{x-y}{p}
   \twopideltaftn{4}{p-(q_1+q_2)}\nonumber\\
   & &\times \alpha_{em} (\gamma^\mu)_{\alpha'\alpha''} 
   \sgtrless{\alpha''}{\beta''}{y}{q_1}(\gamma^\nu)_{\beta''\beta'} 
   \dgtrless{\mu}{\nu}{y}{q_2}\\
   & & + \int\dn{4}{y}\Splus{\alpha}{\beta}{\alpha'}{\beta'}{x-y}{p}
   \left(\pm i\Sigma_{\rm MF}^{\gtrless}(y)\right)\delta_{\alpha'\beta'}
   \nonumber\\
   & & + \lim_{y_0\rightarrow -\infty}\int\dn{3}{y}
   \Splus{\alpha}{\beta}{\alpha'}{\beta'}{x-y}{p}
   \sgtrless{\alpha'}{\beta'}{y}{\vec{p}}.\nonumber
\end{eqnarray}
\end{mathletters} 
These equations are the most important result of this section.  They
simultaneously describe all ``partonic'' splittings, recombinations and
scatterings from the distant past to the present.  Note that these splittings
and recombinations occur in all cells of coordinate--space.  This is very 
different from the conventional approach where particles interact only 
when they are within $\sqrt{\sigma_{\footnotesize TOT}}$ of each other
\cite{ko87,geigerPCM,klevin97}.  

Equations (\ref{eqn:evolve2}) are the
phase--space QED analog of Mahklin's evolution equations \cite{makhlin95}.  A
QCD version of the phase--space evolution equations should reduce to Makhlin's
equations when integrating out the coordinate dependence.  
Geiger \cite{geiger96} has derived a set of QCD transport equations based on
Makhlin's work.  While his derivation is very similar to our derivation of the
phase--space evolution equation,  he does use the gradient approximation to
simplify his collision integrals.  The QCD version of the transport equations
we derive in Section \ref{sec:transport} would reduce to his semiclassical
equations if one applies this approximation.

There are several ways to solve equation (\ref{eqn:evolve2}) 
but we propose only two methods in the following 
subsections.  The first method is a perturbative scheme which
we will use to derive the time--ordered version of the results of sections 
\ref{sec:pdist}--\ref{sec:edist}.  The
second method is to derive transport equations from equation 
(\ref{eqn:evolve2}).  

\subsection{Perturbative Solution to the Phase--Space Evolution Equations}
\label{sec:nonequilibedist}

We can perform a perturbative expansion on equations (\ref{eqn:evolve2}) and 
get the leading contributions to the particle densities.  We show this for
both the photons and electrons surrounding a classical (scalar) point charge. 

We begin by stating the initial conditions\footnote{Unlike Feynman 
perturbation theory, we can only specify the initial particle
densities here.} for the particle densities and
listing the other assumptions used here.
The initial electron and photon densities (at $y_0\rightarrow -\infty$) are
$\sgtrless{\alpha}{\beta}{-\infty,\vec{y}}{\vec{p}}=
\dgtrless{\mu}{\nu}{-\infty,\vec{y}}{\vec{p}}=0$.
We also take the initial scalar densities to be
$\ggtr{-\infty,\vec{y}}{\vec{p}}=0$ and $\gless{y_0,\vec{y}}{\vec{p}}=
{\cal N} \theta(p_0)\deltaftn{3}{\vec{p}-\vec{p_i}}\delta(p^2-M^2)
\deltaftn{3}{x_0\vec{p}/p_0-\vec{x}}$. 
To get this form for $g^{<}$, we localize the 
initial scalar wavepacket in momentum as discussed in
Appendix~\ref{append:classdens}.  In addition to assuming these densities, we
must also neglect the mean field and drop the gradients in 
the scalar--photon coupling.  

\subsubsection{Photons}

Since the scalar field only couples to the photons, the lowest order
contribution to the photon density comes from the photons directly coupling to
the initial scalar density.  The cut diagram for this process is in Fig.
\ref{fig:noneqEPD}.  For positive energy 
photons, the density is
\[\begin{array}{ccl}
\dless{\mu}{\nu}{x}{p}&=&\displaystyle\int\dn{4}{y}\dnpi{4}{q_1}\dnpi{4}{q_2}
   \Dplus{\mu}{\nu}{\mu'}{\nu'}{x-y}{p}\twopideltaftn{4}{p-(q_1+q_2)}\\
   & & \times \alpha_{em} Z^2 (q_1+q_2)^{\mu'} \gless{y}{q_1}
   (q_1+q_2)^{\nu'}\gless{y}{q_2}.
\end{array}\]
Now, $\Ggreat{x}{p} = \Gless{x}{-p}$ because 
$\gtrless$ propagators obey the relation $\Ggreat{x}{y}
= \Gless{y}{x}$.  Thus, we can 
switch one of the $\gless{y}{q}$ to $\ggtr{y}{-q}$, changing it from an 
initial state antiscalar to a final state scalar.  Doing so, we have 
\begin{equation}
\label{eqn:noneqEPD1}
\begin{array}{ccl}
\dless{\mu}{\nu}{x}{p}&=&\displaystyle\int\dn{4}{y}\dnpi{4}{q_1}\dnpi{4}{q_2}
   \Dplus{\mu}{\nu}{\mu'}{\nu'}{x-y}{p}\twopideltaftn{4}{p-(q_1-q_2)}\\
   & & \times \alpha_{em} Z^2 (q_1-q_2)^{\mu'} \gless{y}{q_1}
   (q_1-q_2)^{\nu'}\ggtr{y}{q_2}.
\end{array}
\end{equation}

Now we can bring $\alpha_{em} Z^2 (q_1-q_2)^{\mu'} \gless{y}{q_1}
(q_1-q_2)^{\nu'}\ggtr{y}{q_2}$ into the form of the Wigner transform of 
the scalar current.  To do this, we  take the final state scalar to be free and
sum over all possible final momentum.\footnote{We perform this calculation in 
detail in Appendix \ref{append:current}.}  
Doing so, Eq.~(\ref{eqn:noneqEPD1}) becomes
\[
	\dless{\mu}{\nu}{x}{p}=\int\dn{4}{y}\Dplus{\mu}{\nu}{\mu'}{\nu'}{x-y}{p}
	\Jcurrent{\mu'}{\nu'}{y}{p}{classical}.
\]
Thus, $\dless{\mu}{\nu}{x}{p}$ can be identified with the 
$\Afield{\mu}{\nu}{x}{p}$ in equation (\ref{eqn:Afield}).

\subsubsection{Electrons}

Since the electrons only couple to the photons, the lowest order contribution
to the electron density comes from a photon splitting into electron--positron
pairs.  The cut diagram for this 
is shown in Fig. \ref{fig:noneqEED}.  From 
equation (\ref{eqn:evolve2}c) we have:
\[\begin{array}{ccl}
\sless{\alpha}{\beta}{x}{p}&=&\displaystyle\int\dn{4}{y}\dnpi{4}{q_1}
   \dnpi{4}{q_2}\Splus{\alpha}{\beta}{\alpha'}{\beta'}{x-y}{p}
   \twopideltaftn{4}{p-(q_1+q_2)}\\
   & &\displaystyle\times \alpha_{em} (\gamma^\mu)_{\alpha'\alpha''} 
   \sless{\alpha''}{\beta''}{y}{q_1}(\gamma^\nu)_{\beta''\beta'} 
   \dless{\mu}{\nu}{y}{q_2}.\\
\end{array}\]
Using $\sless{\alpha}{\beta}{x}{q}=\sgtr{\beta}{\alpha}{x}{-q}$, we find
\begin{equation}\label{eqn:noneqEED1}\begin{array}{ccl}
\sless{\alpha}{\beta}{x}{p}&=&\displaystyle\int\dn{4}{y}\dnpi{4}{q_1}
   \dnpi{4}{q_2}\Splus{\alpha}{\beta}{\alpha'}{\beta'}{x-y}{p}
   \twopideltaftn{4}{p-(-q_1+q_2)}\\
   & &\displaystyle\times \alpha_{em} (\gamma^\mu)_{\alpha'\alpha''} 
   \sgtr{\beta''}{\alpha''}{y}{q_1}(\gamma^\nu)_{\beta''\beta'} 
   \dless{\mu}{\nu}{y}{q_2}.\\
\end{array}\end{equation}
Taking the initial photon density from (\ref{eqn:noneqEPD1})
and taking $\sgtr{\beta''}{\alpha''}{y}{q_2}$ to be a 
final state positron, we recover equation 
(\ref{eqn:elecdist}).  However here all of the 
propagators are retarded while the electron propagator in equation 
(\ref{eqn:elecdist}) is causal.  

\subsection{Transport Equations}
\label{sec:transport}

In this section, we find a set of transport equations from the integral 
equations in (\ref{eqn:evolve2}).  We write the two equations of 
motion for the phase--space retarded propagator.  Applying these equations to
the phase--space evolution equations, we derive two sets of coupled 
integro--differential equations.
The first set of equations are the transport equations and the second set 
are the ``constraint'' equations of Mr\'{o}wczy\'{n}ski and 
Heinz~\cite{mrow92,zhuang96}.  The 
``constraint'' equations describe the mass shift of the particles in medium.  

The equation of motion for the non--interacting 
retarded massless scalar propagator is
\[\partial^2G^{+}(x)=\deltaftn{4}{x}.\]
The conjugate equation is 
\[\partial^2(G^{+}(x))^{*}=\deltaftn{4}{x}.\]
Multiplying both sides of the first equation by $(G^{+}(y))^{*}$,  
both sides of the 
second equation $G^{+}(y)$ and Wigner transforming in the relative 
space--time coordinate, we find two equations:
\begin{mathletters}
\begin{eqnarray}
	(k+ i\partial/2)^2\Gplus{x}{k}&=&\int\dn{4}{x'}e^{ix'\cdot k}
		\left(G^{+}(x- x'/2)\right)^{*} \deltaftn{4}{x+ x'/2}\\
	(k- i\partial/2)^2\Gplus{x}{k}&=&\int\dn{4}{x'}e^{ix'\cdot k}
		\left(G^{+}(x+ x'/2)\right) \deltaftn{4}{x- x'/2}
\end{eqnarray}
\end{mathletters}
Inserting the retarded propagator in the energy--momentum representation 
(with $m_e=0$) and adding and subtracting the $+$ and $-$
equations, we find the equations of motion for the retarded propagator:
\begin{mathletters}
\begin{eqnarray}
\label{eqn:retprop1}
k\cdot\partial\Gplus{x}{k}&=&\frac{2}{\pi}\theta (x_0)\delta (x^2)
\sin{(2 x\cdot k)}\\
\label{eqn:retprop2}
(\partial^2/4-k^2)\Gplus{x}{k}&=&\frac{2}{\pi}\theta (x_0)\delta (x^2)
\cos{(2 x\cdot k)}.
\end{eqnarray}
\end{mathletters}
Taylor series expanding the sine or cosine and keeping only the lowest order 
is equivalent to performing the gradient expansion.

Now, we apply the $k\cdot\partial$ and 
$(\partial^2/4-k^2)$ operators to the particle densities in equation 
(\ref{eqn:evolve2}).  On the right hand side, these differential operators
act on the retarded propagators, so we can use their equations of motion to 
simplify the results.  For scalars we get
\begin{mathletters}
\label{eqn:transport1}
\begin{eqnarray}
\begin{array}{ccl}
\lefteqn{p\cdot\partial\ggtrless{x}{p}=\displaystyle
   \int\dn{4}{y}\dnpi{4}{q_1}\dnpi{4}{q_2}
   \frac{2}{\pi}\theta(x_0-y_0)\delta((x-y)^2)}\\
   & &\displaystyle\times\sin{(2(x-y)\cdot p)} 
   \twopideltaftn{4}{p-(q_1+q_2)}\\
   & & \displaystyle\times \alpha_{em} Z^2 
   (q_1+q_2-i\bothpartial/2)^{\mu}\ggtrless{y}{q_1} 
   (q_1+q_2-i\bothpartial/2)^{\nu}\dgtrless{\mu}{\nu}{y}{q_2}\\
   & & + \displaystyle\int\dn{4}{y} \frac{2}{\pi}\theta(x_0-y_0)\delta((x-y)^2)
   \sin{(2(x-y)\cdot p)} iQ_{\rm MF}^{\gtrless}(y)\\
   & & + \displaystyle\lim_{y_0\rightarrow -\infty}\int\dn{3}{y}
   \frac{2}{\pi}\theta(x_0-y_0)\delta((x-y)^2)
   \sin{(2(x-y)\cdot p)}\ggtrless{y}{\vec{p}}
\end{array}\\
\begin{array}{ccl}
\lefteqn{(\partial^2/4-k^2)\ggtrless{x}{p}=\displaystyle\int\dn{4}{y}
	\dnpi{4}{q_1}
   \dnpi{4}{q_2}\frac{2}{\pi}\theta(x_0-y_0)\delta((x-y)^2)}\\
   & & \times \displaystyle\cos{(2(x-y)\cdot p)} 
   \twopideltaftn{4}{p-(q_1+q_2)}\\
   & & \times \displaystyle\alpha_{em} Z^2 (q_1+q_2-i\bothpartial/2)^{\mu}
   \ggtrless{y}{q_1}
   (q_1+q_2-i\bothpartial/2)^{\nu}\dgtrless{\mu}{\nu}{y}{q_2}\\
   & & + \displaystyle\int\dn{4}{y} \frac{2}{\pi}\theta(x_0-y_0)\delta((x-y)^2)
   \cos{(2(x-y)\cdot p)} iQ_{\rm MF}^{\gtrless}(y)\\
   & & + \displaystyle\lim_{y_0\rightarrow -\infty}\int\dn{3}{y}
   \frac{2}{\pi}\theta(x_0-y_0)\delta((x-y)^2)
   \cos{(2(x-y)\cdot p)}\ggtrless{y}{\vec{p}}
\end{array}
\end{eqnarray}
\end{mathletters}
Now because of the delta functions, the boundary conditions at 
$y_0\rightarrow -\infty$ only contribute when
$|\vec{x}-\vec{y}|$ goes to $\infty$, implying that we need 
$\ggtrless{x}{p}$  as $\vec{x}\rightarrow \infty$.  The 
densities are zero here, so they drop out from these equations.

The transport equations for the photons and electrons are 
\begin{mathletters}
\label{eqn:transport2a}
\begin{eqnarray}
\lefteqn{p\cdot\partial\dgtrless{\mu}{\nu}{x}{p} = 
   \displaystyle\frac{2}{\pi}\int\dn{4}{y}
   \theta(x_0-y_0)\delta((x-y)^2)\sin{(2(x-y)\cdot p)}}\hspace*{.25in}\nonumber  \\
& & \times\left\{ \displaystyle\int\dnpi{4}{q_1}\dnpi{4}{q_2}
   \twopideltaftn{4}{p-(q_1+q_2)}\left\{ \alpha_{em} {\rm Tr}\left[\gamma^{\mu} 
   \sgtrless{}{}{y}{q_1}\gamma^{\nu}\sgtrless{}{}{y}{q_2}\right]\right.\right.\\
& & + \left.\left.\alpha_{em} Z^2 (q_1+q_2+i\bothpartial/2)^{\mu} 
   \ggtrless{y}{q_1}(q_1+q_2+i\bothpartial/2)^{\nu}\ggtrless{y}{q_2}
   \right\}\right.\nonumber\\
& & \left. +  g_{\mu\nu}i\Pi_{\rm MF}^{\gtrless}(y)\right\}\nonumber\\
\lefteqn{p\cdot\partial\sgtrless{\alpha}{\beta}{x}{p} = 
   \displaystyle\frac{2}{\pi}(\dirslash{p}+i\dirslash{\partial})_{\alpha\alpha'}
   (\dirslash{p}-i\dirslash{\partial})_{\beta\beta'}}\hspace*{.25in}\nonumber\\
& &\times\displaystyle\int\dn{4}{y}\theta(x_0-y_0)\delta((x-y)^2)
	\sin{(2(x-y)\cdot p)}\nonumber\\
& & \times\left\{ \displaystyle\int\dnpi{4}{q_1}\dnpi{4}{q_2}
   \twopideltaftn{4}{p-(q_1+q_2)}\right.\\
& & \times\displaystyle\left.\alpha_{em} (\gamma^\mu)_{\alpha'\alpha''} 
   \sgtrless{\alpha''}{\beta''}{y}{q_1}(\gamma^\nu)_{\beta''\beta'} 
   \dgtrless{\mu}{\nu}{y}{q_2}\right.\nonumber\\
& & + \left. \delta_{\alpha'\beta'}\left(\pm i\Sigma_{\rm MF}^{\gtrless}(y)
   \right)\right\}.\nonumber
\end{eqnarray}
\end{mathletters}
These equations almost have the form of the Boltzmann equation:  the left side
clearly is the Boltzmann transport operator and the right side is almost the
collision integrals.  
If we were to expand the (co)sines in the collision integrals and keep only the 
lowest term, we would recover the collision integrals.
Furthermore, if we were to do this same approximation to the QCD version of
(\ref{eqn:transport2a}) we would arrive at Geiger's semiclassical QCD transport
equations \cite{geiger96}.

We also state the constraint equations:
\begin{mathletters}
\label{eqn:transport2b}
\begin{eqnarray}
\lefteqn{(\partial^2/4-k^2)\dgtrless{\mu}{\nu}{x}{p} = 
   \displaystyle \frac{2}{\pi}\int\dn{4}{y}
   \theta(x_0-y_0)\delta((x-y)^2)\cos{(2(x-y)\cdot p)}}
   \hspace*{.25in}\nonumber\\
& & \times\left\{ \displaystyle\int\dnpi{4}{q_1}\dnpi{4}{q_2}
   \twopideltaftn{4}{p-(q_1+q_2)}\left\{ \alpha_{em} {\rm Tr}\left[\gamma^{\mu} 
   \sgtrless{}{}{y}{q_1}\gamma^{\nu}\sgtrless{}{}{y}{q_2}\right]\right.\right.\\
& & + \left.\left.\alpha_{em} Z^2 (q_1+q_2+i\bothpartial/2)^{\mu} 
   \ggtrless{y}{q_1}(q_1+q_2+i\bothpartial/2)^{\nu}\ggtrless{y}{q_2}
   \right\}\right.\nonumber\\
& & \left. +  g_{\mu\nu}i\Pi_{\rm MF}^{\gtrless}(y)\right\}\nonumber\\
\lefteqn{(\partial^2/4-k^2)\sgtrless{\alpha}{\beta}{x}{p} = 
   \displaystyle\frac{2}{\pi}(\dirslash{p}+i\dirslash{\partial})_{\alpha\alpha'}
   (\dirslash{p}-i\dirslash{\partial})_{\beta\beta'}}\hspace*{.25in}\nonumber\\
& & \times\displaystyle\int\dn{4}{y}\theta(x_0-y_0)\delta((x-y)^2)
	\cos{(2(x-y)\cdot p)}\nonumber\\
& & \times\left\{ \displaystyle\int\dnpi{4}{q_1}\dnpi{4}{q_2}
   \twopideltaftn{4}{p-(q_1+q_2)}\right.\\
& & \times\displaystyle\left.\alpha_{em} (\gamma^\mu)_{\alpha'\alpha''} 
   \sgtrless{\alpha''}{\beta''}{y}{q_1}(\gamma^\nu)_{\beta''\beta'} 
   \dgtrless{\mu}{\nu}{y}{q_2}\right.\nonumber\\
& & + \left. \delta_{\alpha'\beta'}\left(\pm i\Sigma_{\rm MF}^{\gtrless}(y)
   \right)\right\}\nonumber.
\end{eqnarray}
\end{mathletters}
The constraint equations give rise to the in--medium mass shift for the photons 
and electrons because, were we to derive the constraint equation 
for massive particles, we would find that $(\partial^2/4-k^2)\rightarrow
(\partial^2/4-k^2+m^2)$.  Thus, the RHS of the constraint equations can be
interpreted as an ``in--medium'' mass.  Note that, even in the
presence of this ``in--medium'' mass, particles scatter onto the light--cone. 
This is not a  surprise since the particles are massless.  Also note 
that the theta functions enforce the retarded time--ordering.  Finally,
we have not written the various constants in terms of there
renormalized values.  Dressing the particle densities by solving the evolution
equations (which are nonperturbative) should be equivalent to using
renormalized couplings.

\subsection{Implications for Parton Transport Theory}

The ``source--propagator'' picture must be valid for partons since the
phase--space Generalized Fluctuation--Dissipation Theorem does not depend on 
the form of the self--energies.  
So, if we find the QCD self--energies and define the parton distributions
appropriately, we could construct  phase--space parton evolution equations.  
Of course, for these
equations to have meaning, one must understand the role of $\Lambda_{QCD}$.  
$\Lambda_{QCD}$ is usually interpreted as a momentum cut--off in perturbation 
theory; as one approaches $\Lambda_{QCD}$, nonperturbative effects increase
and perturbation treatments break down.  This interpretation may not be
appropriate for several reasons.  First, the phase--space evolution
equations are non--perturbative objects, so there should be no cut--off in
momentum.  Second, it is not clear whether $\Lambda_{QCD}$ should viewed as a
cut--off in momentum or whether $1/\Lambda_{QCD}$ should be viewed as a
cut--off in coordinates.  In fact, it may be that $1/\Lambda_{QCD}$ is simply a
characteristic length scale for QCD bound states (i.e. hadrons) so that
higher--order correlations (i.e. four and six point functions) need to be built
into the Dyson--Schwinger equations.  Presumably this would require 
some understanding of  hadronization.  Nevertheless, in the absence of a 
phase--space evolution equation, we can still use the
Generalized Fluctuation--Dissipation Theorem as insight to build models.  This 
is what we do in the next section.

%
%

\section{Parton Cloud of a Nucleon}
\label{sec:QCD}

We cannot calculate the phase--space Parton Distribution Functions
without a set of QCD phase--space evolution
equations.  Nevertheless, there is significant work calculating the
Parton Distribution Functions in momentum--space and many of these results
can be translated into phase--space.  In  particular, we show that 
the Leading Logarithm Approximation works in phase--space.  Using the 
momentum ordering in the Leading Logarithm Approximation and 
a simple model of the nucleon we estimate the size of the sea parton
distribution as a function of parton momentum.

\subsection{Parton Model and Leading Logarithm Approximation}

The parton model rests on a two simple 
assumptions: $\alpha_s \ll 1$ (so perturbation theory is valid) and the 
parton lifetime is much larger than parton interaction time \cite{genQCD}.  
Both of these
conditions are necessary to factorize a cross section in a momentum--space 
calculation.  Typically the parton distribution functions are calculated using 
either DGLAP, BFKL, or GLR evolution, all of which are equivalent to applying 
a Leading Logarithm Approximation (LLA).  In the LLA, we assume the parton is
produced in a cascade represented by the ladder diagram in Fig. 
\ref{fig:noneqNthgen}.  The probability of emitting the $n^{th}$ parton with 
longitudinal momentum fraction $x_n$ and transverse momentum $q_{nT}^2$ from
this cascade is \cite{laenen94}
\begin{equation}
dP_n=\frac{N_c \alpha_s}{\pi} \frac{d x_n}{x_n} \frac{d q_{nT}^2}{q_{nT}^2}.
\label{eqn:emitprob}
\end{equation} 
Thus, by ordering the momentum properly as we go down the ladder, we can pick 
up the largest logarithmic contributions to the $n^{th}$ parton's density.

Most hadron colliders probe regions where the data are well described with
parton distribution functions calculated within the
Dokshitzer--Gribov--Lipatov--Altarelli--Parisi (DGLAP) evolution scheme.
DGLAP evolution is equivalent the Leading Logarithm Approximation in $1/q^2$
(LLA($Q^2$)).  New experiments at HERA are beginning to see evidence that
Balitsky--Fadin--Kuraev--Lipatov (BFKL) type evolution is necessary to describe
the parton distribution functions at small--$x$ \cite{H195}.   BFKL--type
physics is believed to be responsible for the rise in the number of partons as
$x\rightarrow 0$, however this rise can also be partially described by
DGLAP--type physics \cite{laenen94,nikitin86,H195}.  BFKL evolution is 
equivalent the Leading Logarithm Approximation in $1/x$ (LLA($x$)).  
Unlike DGLAP and BFKL evolution, Gribov--Levin--Ryskin (GLR) type evolution does 
not have a simple momentum ordering because one sums terms with
varying powers of $1/x$ and $1/q^2$ \cite{gribov83,laenen94}.  Because of the 
simplicity of the ladder 
structure and the momentum ordering needed to pick up the largest
contributions, we will discuss both DGLAP and BFKL
type partons in phase--space.

We can apply the parton model and LLA in phase--space if both are modified 
appropriately.  Assume that we are working in a regime where $\alpha_s \ll 1$, 
so we can apply phase--space perturbation theory, and assume that all 
elementary particles are massless.  Assume also that the probe is localized on
the length scale of the parton cloud.  This assumption is equivalent to saying 
the parton lifetime is large compared to the interaction time.

Now, if we find the same singularities in both phase--space and
momentum--space, then we know that the LLA will give the
dominant contribution to the particle densities in phase--space.
The Generalized Fluctuation 
Dissipation Theorem tells us that the parton density has the form 
\begin{equation}
g^\gtrless (x,p)=\int\dn{4}{y}G^{+}(x-y,p)\Sigma^\gtrless (y,p).
\label{eqn:neGeneric}
\end{equation}
The self--energy, $\Sigma^{\gtrless}$, is given by the parton ladder 
in Fig.~\ref{fig:noneqNthgen} and 
the $n^{th}$ segment of $\Sigma^{\gtrless}$ is shown in Fig. \ref{fig:rung}.
In momentum--space, the cut rung gives a $d^3k/|k_0|$ 
which leads to the $dx/x$ in equation (\ref{eqn:emitprob}).
To see how the factor of $d^3k/|k_0|$ arises in phase--space, one needs only
look at the  electron source in Section \ref{sec:esource}.  The electron 
source has exactly
the form of the segment in Fig. \ref{fig:rung} and in that calculation we
found exactly this factor of $d^3k/|k_0|$.
The fact that we find the same factor of $d^3k/|k_0|$ in both the
energy--momentum representation and in phase--space simply reflects the fact
that the cut parton density is $\theta(k_0)\delta(k^2)$ in both cases 
and we sum over final parton states.
The factor of $d q^2/q^2$
occurs in equation (\ref{eqn:emitprob}) because of the integration over the 
leg's propagator $1/q^2$. In phase--space, the $1/q^2$ pole is tied up in the 
Wigner transform of the retarded propagator, but it is still there:
\[
   \Gplus{x}{q}=\int\dnpi{4}{q'} e^{-i x \cdot q'}
   \frac{1}{(q+q'/2)^2+i\epsilon(q_0+q_0'/2)}
   \frac{1}{(q-q'/2)^2-i\epsilon(q_0-q_0'/2)}.
\] 
Thus, this segment of the parton ladder produces the same divergencies in 
phase--space and  momentum--space.  Whatever orderings are needed to produce 
the leading contributions in momentum space will produce the same leading 
contributions in phase--space.

Our self--energy has the same ladder structure as the electron source in
Section \ref{sec:edist}, so we know the spatial structure of the $n^{th}$
parton's source is given by the $n-1^{th}$ parton's distribution.  Iterating
back to the $0^{th}$ parton (a valence quark), we see that the shape of the
valence distribution sets the shape of the sea parton source.
We take the valence quark wavefunction to be uniformly spread throughout 
a bag with radius $R_{bag}$.  Since we are 
interested in high--energy collisions, we take the nucleon bag to be moving to 
the  right with 4--momentum $p_\mu=(P_0,P_L,\vec{0}_T)$ with 
$P_0\approx P_L \gg M_N$.  Thus, this nucleon has 4--velocity  
$v_\mu=(1,v_L,\vec{0}_T)$ and the bag is contracted in the 
longitudinal direction by a factor of $\gamma=1/\sqrt{1-v_L^2} \gg 1$.
We assume the partons lose memory of the original valence quark
momentum as one goes down the ladder.   
Thus, any momentum/coordinate correlations in the 
source function should be washed out by the spatial 
integrations in equation (\ref{eqn:neGeneric}).  One might expect that 
the sea partons forget the shape of the nucleon bag as well, but we show
that the partons cannot propagate far enough from the original source for this
to happen.

\subsection{Large--$Q^2$ (DGLAP) Partons}
 
In the large--$Q^2$ regime,  the parton density is low but 
$\alpha_s (Q^2) \ln (Q^2/\Lambda_{QCD}^2) \gtrsim 1$.  Here the largest 
contribution to the leading log ladder comes from large $Q^2$ 
logarithms.\footnote{$Q^2$ can be taken as the typical momentum 
scale of the process.  In the case of a DIS probe, this is the momentum 
transferred by the probe.}  To get the largest contributions 
from these logs, we order the momenta as we move down the ladder:
\[
   -q_{n}^2 \gg -q_{n-1}^2 \gg \ldots \gg -q_{1}^2 \gg 1/R^2_{bag}
   \approx \Lambda_{QCD}^2.
\]  
Here $q_{i}^2$ is the virtuality of the $i^{th}$ leg.  The kinematics at each 
leg--rung vertex ensure that the momentum fraction carried by each leg is 
also ordered:
\[
   1 \ge x_1 \ge \ldots \ge x_{n-1} \ge x_{n}.
\]
Whether a rung or leg is a quark or gluon is irrelevant, provided $k^2=0$ and 
the $q^2$ ordering holds.  Now, given that the proton has longitudinal 
momentum $P_L$ 
and the rungs and legs are massless, each generation of partons must have energy
$q_{n0}\approx x_n P_L$ and transverse momentum of $q_T^2\approx -q^2 \ll x^2
P_L^2$. 

Let us discuss the general features of the parton cloud.  
The retarded propagator  
lets the $n^{th}$ parton propagate out to 
$R_{n\perp}\sim \hbar c/\sqrt{-q_n^2}$ 
transverse to the parton momentum and to 
$R_{n\|}\sim \hbar c/q_{n0}= \hbar c/x_n P_L$ 
parallel to the the parton momentum.  Since $xP_L \gg p_T$, the
parton momentum is approximately parallel to the nucleon momentum.  Because 
$R_{n\perp}\ll R_{bag}$, the partons can never get far from the bag in the
transverse direction and 
transverse spread of the partons will dominated by the bag size: 
$\Delta R_{T} \sim R_{bag}$.  
On the other hand, the longitudinal spread of the partons is roughly given by 
$\Delta R_{L} \sim R_{bag}/\gamma +\hbar c/x P_L$, so can be dominated 
by the longitudinal 
propagation distance $R_\|$ if $x\ll M_N R_{bag}/\hbar c$.  In fact, for very 
small $x$ (i.e. $x\sim M_N R_{bag}/\gamma\hbar c$) the spread of the partons can 
meet or exceed the nucleon bag 
radius.  Furthermore, the actual distribution may be 
somewhat broader due to the propagation of the virtual partons between the
subsequent emissions along the ladder.  

So, in our picture, the sea quark and gluon large--$Q^2$ distributions have 
the same
transverse size as the parent nucleon, but the longitudinal size can be
significantly bigger than the parent.  Furthermore, the drop off in the parton
density in the longitudinal direction occurs at the characteristic radius of
$\sim \hbar c/xP_L$.  This picture of the nucleon is consistent with the
uncertainty principle based arguments of A.~H.~Mueller \cite{almueller89}, 
later user by Geiger to initialize the parton distributions in
his Parton Cascade Model \cite{geigerPCM}.

\subsection{Small--$x$ (BFKL) Partons}

In the small--$x$ regime, the parton density 
is high and $\alpha_s (Q^2) \ln (1/x) \gtrsim 1$.  The small--$x$ partons 
are mostly gluons.  In this regime, the leading
logs come from the $1/x$--type singularities, i.e. from the cut rungs. 
Typically the rungs are thought of as ``reggeized,'' meaning that each rung 
in the ladder is allowed to radiate gluons, leading to a large multiplication in
the parton density.  Since leading logs come from the $1/x$ type singularities, 
the largest 
contributions come about by strongly ordering the longitudinal momentum 
fraction as one moves down the ladder
\cite{genBFKL}:
\[
  1 \gg x_1 \gg \ldots \gg x_{n-1} \gg x_{n}.
\] 
BFKL--type evolution has only a weak dependence on the virtuality of the partons
as we move down the ladder, so we assume $q^2$ to be fixed:
$q_{n-1}^2\approx q_{n}^2\gg 1/R_{bag}^2$.
This does not significantly effect the results of the analysis \cite{bartels93}.

Now we must understand how the transverse momentum and energy of each parton leg
changes as we go down the ladder.  A well known effect of iterating the BFKL 
kernel (equivalent to moving down the ladder) is that the transverse momentum 
undergoes a random walk in $\ln(q^2_T)$ \cite{laenen94,genBFKL}.  In fact, 
after iterating through a 
sufficiently large number of rungs, the spread in the $q_T$ distribution is
given by:
\[
\left< \left( \ln (\frac{q_{nT}^2}{q_{1T}^2}) \right)^2 \right> = 
   C \ln (\frac{1}{x})
\]
where $C=\frac{N_c\alpha_s}{\pi} 28 \zeta (3) = 32.14\alpha_s$.  Thus, 
$q_{nT}^2$ can be orders of magnitude larger or smaller than $q_{1T}^2$.
We restate this as
\begin{equation}
	q_{nT}^2 \sim q_{1T}^2 e^{\pm 5.7 \sqrt{\alpha_s \ln (1/x)}}.
\label{eqn:BFKLqT}
\end{equation}
We will consider the extreme cases
of the transverse momentum and comment on the typical case,
$q^2_{nT}\sim q^2_{1T}$.

If the random walk results in a large transverse momentum, we will have
$q^2_{nT}\gg q^2_{1T}\sim -q^2\sim (x_n P_L)^2$.  Thus, the $n^{th}$ parton
will have 3--momentum in the transverse direction.  We know that the parton can
only propagate to a distance of roughly $R_{\|}\sim \hbar c/|q_0|=\hbar
c/\sqrt{q^2+q^2_T+(x P_L)^2}$ in the direction parallel to $\vec{q}$.  Since
$\hbar c/\sqrt{q^2+q^2_T+(x P_L)^2} \approx \hbar c/|q_T| \ll R_{bag}$, the
parton cannot travel far from the original source in the transverse direction.
On the other hand, the parton's longitudinal spread can be larger than the
longitudinal bag size.  The parton can propagate to a distance of 
$R_{\perp} \sim \hbar c/\sqrt{-q^2}$ in the
direction perpendicular to $\vec{q}$, so we can expect a longitudinal spread of
the parton distribution of $\Delta R_L \sim R_{bag}/\gamma + \hbar
c/\sqrt{-q^2}$.  Since $R_{bag}\gg\hbar c/\sqrt{-q^2}$, this additional spread
can not match the spread of the DGLAP partons.

If the random walk results in a small transverse momentum, we will have
$q^2_{nT}\ll q^2_{1T}\sim -q^2\sim (x_n P_L)^2$.  In this case, the $n^{th}$
parton will have 3--momentum in the longitudinal direction.  As in the case of
the DGLAP partons the additional transverse spread is 
$\Delta R_T\sim \hbar c/\sqrt{-q^2} \ll R_{bag}$ and so is negligible.  
The additional longitudinal spread is
$\Delta R_L\sim\hbar c/|q_0|=\hbar c/\sqrt{q^2+(x P_L)^2+q_T^2}$ and may be 
significantly larger than the bag radius because the parton is space--like.  

Summarizing both possibilities, the BFKL parton distributions have the same 
transverse spread, $\Delta R_T \sim R_{bag}$, but different longitudinal 
spreads.  The longitudinal spread may range from 
$\Delta R_L \sim R_{bag}/\gamma +\hbar
c/\sqrt{-q^2} \ll R_{bag}$ to $\Delta R_L \sim R_{bag}/\gamma +\hbar
c/\sqrt{q^2+(xP_L)^2+q_T^2}\gg R_{bag}$ for partons with space--like momentum.  
The fact that the spatial extent of the BFKL cloud is so large in the
longitudinal direction suggests that
the small--$x$ partons (which are mostly gluons) can see the color charge of any
other nucleon in the longitudinal tube centered on the parent nucleon.
This suggests that we should treat the nucleus as a whole as a source of color
charge in the spirit of McLerran--Venugopalan model \cite{McLerran}.  

The large longitudinal extant of the small--$x$ cloud has another consequence:
in a zero impact parameter nucleon--nucleon collision, we would find that
the soft (BFKL) partons interact much earlier than the harder (DGLAP) partons
because of their greater longitudinal spread.  This, coupled with the large
density of small--$x$ partons, leads to earlier entropy
production and stopping of the soft partons.  

\section{Conclusion}

We have made progress toward specifying the initial phase--space parton 
distributions of a relativistic nuclear collision.  
Regardless of the kinematical regime, the transverse spread of a parton 
distribution is dominated by the bag radius $\sim 1$ fm.  
The longitudinal spread of a parton distribution varies from 
roughly $\sim R_{bag}/\gamma+\hbar c/xP_L$ for moderate to large $x$ (i.e. for
DGLAP partons) and from  
$\Delta R_L\sim R_{bag}/\gamma + \hbar c/\sqrt{-q^2}$ to 
$\Delta R_T\sim R_{bag}/\gamma + \hbar c/\sqrt{q^2+(xP_L)^2+q_T^2}$
for small $x$ (i.e. BFKL partons).  Since the small $x$ partons have a large 
longitudinal spread
and a high density, we expect the small $x$ partons to interact much earlier
than the large $x$ partons in a typical nuclear collision.  This would cause
earlier entropy production and higher stopping than one expects in models 
that include only DGLAP parton distributions such as the PCM
\cite{geigerPCM}, HIJING \cite{HIJING}, and others.

Using time--ordered non--equilibrium methods, we derived phase--space 
evolution equations 
for QED, illustrating how to find them in QCD.  Unlike conventional transport 
approaches, our calculation does not rely on the gradient 
approximation.  Thus, it should work on all length and momentum scales.
These phase--space evolution equations describe the complete
evolution of a system from some time in the distant past to the present,
including all splittings, recombinations and scatterings of the particles.  
One can use these evolution equations perturbatively or to derive
semiclassical transport equations.
These evolution equations rely on the Generalized Fluctuation--Dissipation
Theorem.  This theorem states that a particle's density is the convolution
of the Wigner transform of its self--energy and a phase--space propagator.
The Generalized Fluctuation--Dissipation Theorem is quite general and can be
directly applied to QCD.

In conventional Feynman perturbation theory, we found the reaction rates (and 
hence the cross sections) can be written in a parton model form.  In other words,
they take the form of a reaction rate density convoluted with a phase--space
Parton Distribution Function.  This phase--space PDF is 
the parton number density and has the form of a phase--space source
folded with a phase--space propagator.  Our work with the Weizs\"acker--Williams 
Approximation demonstrates that the Parton Distribution Functions can be
defined in phase--space.

In order to illustrate how the propagators and sources work in
phase--space, we calculated the effective photon and electron distributions.
We found that both the retarded and Feynman propagators propagate particles to 
distances of $\sim R_\| =\hbar c/\mbox{min}(|q_0|,|\vec{q}|)$ parallel to the 
particle's momentum and to 
distances of $\sim R_\perp =\hbar c/\sqrt{|q^2|}$
perpendicular to the particle's momentum when $q^2\neq 0$.  When $q^2=0$, the
particles tend to follow their classical paths with deviations from this path
being of order $1/|q_0|$.
Furthermore, the retarded propagator can only send
particles forward in time and inside the light--cone 
while the Feynman propagator sends particles both 
forwards and backwards in time and both inside and outside of the light--cone. 
We also described a phase--space source that included a simple
``partonic'' splitting: the electron distribution of a point
charge.  These electrons are created when a virtual photon splits into an
electron--positron pair; the diagram for this process is the first segment of a
parton ladder.   We found that shape of the electron's source is controlled
by the parent photon's distribution.

We hope that we have provided insight into the behavior and calculation of
the phase--space densities.  Specifically, we hope the ``source--propagator''
picture of the Generalized Fluctuation--Dissipation Theorem and the resulting 
phase--space evolution equations can be coupled with appropriately defined 
phase--space parton densities.  The resulting theory could 
describe the various many--particle effects we expect in a nuclear collision at 
RHIC or the LHC and it could incorporate parton model phenomenology. 

\acknowledgements

The authors acknowledge conversations with the participants of the Institute 
for Nuclear Theory Fall 1996 program, specifically Miklos Gyulassy, 
Berndt M\"uller, Alejandro Ayala, Raju Venugopalan, and Alexander Makhlin.  
They also acknowledge 
Vladimir Zelevinsky, Scott Pratt, C.P.~Yuan, Ed Shuryak and George Bertsch 
for their valuable discussions.  This work was supported by the National 
Science Foundation under Grant PHY-9605207.

%
%
\appendix

\section{Electrodynamics with Fermions and Scalars}
\label{sec:lagrangian}

Throughout this paper we use QED and Scalar Electrodynamics 
to describe the interactions between the
electrons, photons and scalars.  In this appendix, we review the QED
lagrangian, equations of motion, equal time commutation relations and contour 
Feynman rules.  The lagrangians, etc., for QED and for Scalar Electrodynamics
are given in many places \cite{genFT,bog79}.  Even so, we restate them here both
to keep this work self--contained and to clarify our notation.  We do not
include the renormalization counterterms nor the gauge fixing terms for
the photons although they can be easily included.  
We work in the Lorentz gauge.

The lagrangian for scalar QED coupled with spinor QED is
\begin{equation}
\begin{array}{ccl}
{\cal L} &=&
   \frac{i}{2}\bar{\psi}(x)\slbothpartial\psi(x)-m_e\bar{\psi}(x)\psi(x)\\
   & &-\frac{1}{16\pi} F_{\mu\nu}(x)F^{\mu\nu}(x)\\
   & &+(\partial_{\mu}\phi^{*}(x))(\partial^{\mu}\phi(x))-M^2\phi^{*}(x)\phi(x)\\
   & &-e\bar{\psi}(x)\not\!\!{A}(x)\psi(x)
   -iZeA^{\mu}(x)(\phi^{*}(x)\bothpartial_{\mu}\phi(x)) +Z^2 \alpha_{em}
   A^2(x)\phi^{*}(x)\phi(x).
\end{array}
\label{eqn:lagrangian}
\end{equation}
Here $\psi_\alpha (x)$ is the fermion field, $\phi(x)$ is the complex
scalar field, $A_{\mu}(x)$ is the photon field, and
$F_{\mu\nu}(x)=\partial_{\mu}A_{\nu}(x)-\partial_{\nu}A_{\mu}(x)$.
The masses of the fermion and scalar fields are $m_e$ and $M$ respectively.  
The electrons couple to the photons with strength $e$ while the scalars couple
with $Ze$.

The second quantized field operators satisfy the standard equal time 
commutation relations:
\begin{mathletters}
\begin{eqnarray}
	\left[\hat{A}_{\mu}(t,\vec{x}),\dot{\hat{A}}_{\mu}(t,\vec{x}')\right]_{-}&=&
		4\pi g_{\mu\nu}\deltaftn{3}{\vec{x}-\vec{x}'}\\
	\left[\hat{A}_{\mu}(t,\vec{x}),\hat{A}_{\mu}(t,\vec{x}')\right]_{-}&=&
		\left[\dot{\hat{A}}_{\mu}(t,\vec{x}),
		\dot{\hat{A}}_{\mu}(t,\vec{x}')\right]_{-}=0\\
	\left[\hat{\psi}_{\alpha}(t,\vec{x}),\hat{\psi}_{\beta}^{\dagger}(t,\vec{x}')
		\right]_{+}&=&\delta_{\alpha\beta}\deltaftn{3}{\vec{x}-\vec{x}'}\\
	\left[\hat{\phi}(t,\vec{x}),\dot{\hat{\phi}}(t,\vec{x}')\right]_{-}&=&
		\deltaftn{3}{\vec{x}-\vec{x}'}\\
	\left[\hat{\phi}(t,\vec{x}),\hat{\phi}(t,\vec{x}')\right]_{-}&=&
		\left[\dot{\hat{\phi}}(t,\vec{x}),\dot{\hat{\phi}}(t,\vec{x}')
		\right]_{-}=0
\end{eqnarray}
\end{mathletters}

The lagrangian (\ref{eqn:lagrangian}) leads to the following equations of 
motion:
\begin{mathletters}\begin{eqnarray}
	4\pi j^\nu (x) & = & \partial_\mu F^{\mu\nu}(x)\\ 0 & = &
	(i\not\!\partial-e\not\!\!A(x))\psi(x)-m_e\psi(x)\\ 0 & = &
	(\partial_\mu+ieA_\mu(x))(\partial^\mu+ieA^\mu(x))\phi(x)+M^2\phi(x).
\end{eqnarray}\end{mathletters}
The electromagnetic current operator is
\[
	\hat{j}^\mu (x)= e\hat{\bar{\psi}}(x)\gamma^{\mu}\hat{\psi}(x)+
		iZe\hat{\phi}^{*}(x)\slbothpartial\hat{\phi}(x).
\] 
These equations are solved by the Green's
functions in Section \ref{sec:greens} in the limit as $e\rightarrow 0$.

We list the contour Feynman rules for spinor and scalar QED below: 
\begin{enumerate}
\item The vertex Feynman rules are summarized in Table \ref{table:vertex}.
\item The contour propagators are summarized in Table \ref{table:props}.
\item Every closed fermion loop yields a factor of $(-1)$.
\item Every single particle line that forms a closed loop or is linked by the
	same interaction line yields a factor of $iG^{<}$.
\end{enumerate}
Notice that the second scalar coupling has higher order than the
rest of the couplings.  So we neglect this coupling in
the derivation of the evolution equations of Section \ref{sec:evolve}. 

\section{The Cross Section in Terms of Phase--Space Densities}
\label{sec:crosssection}

In this appendix, we discuss the cross--section in terms of phase--space 
quantities.  Since the cross section is measured by scattering a beam of 
particles off a target, we define the cross section in terms of the
projectile/target reaction rate density and the projectile flux.  The beam is 
uniform in the beam direction and in time on the 
scale of the projectile/target interaction.  Thus, the beam can only directly 
probe the transverse structure of the interaction region.  Even this transverse
information is washed out in the typical experiment, since the beam is usually
uniform in the transverse direction on the length scale of the interaction.  
In the limit of a transversely uniform beam, we
recover the conventional definition of the cross section.  Since we
consider only  simple scattering problems, we work in Feynman perturbation
theory where we can specify both the initial and final states of the
reactions.

The beam is a collection of single particle wavepackets distributed
uniformly throughout the transverse area $A$ of the beam.  For the sake of
illustration, we take these particles to be scalars.  The 
Wigner function of these incident wavepackets is
\begin{equation}
	f(x,p)=\frac{1}{2Vp_0}\int\dnpi{4}{p'}e^{-x\cdot p'}
	f(p+p'/2)f^{*}(p-p'/2)
\end{equation}
where the wavefunction $f(p)$ is given by\footnote{The delta function that
puts the particle on--shell is absorbed into $f(p)$}
\begin{equation}
 	\ket{i}=\int \dnpi{4}{p} f(p) \ket{\vec{p}}.
	\label{eqn:ket}
\end{equation}
We will assume the beam to be uniform in the longitudinal direction with length
$L$ and to be turned on for macroscopic time $T$.  The quantities 
$A$, $T$, and $L$ are much
larger than the projectile/target interaction region.  

The projectile/target interaction region is characterized by a reaction rate
density ${\cal W}_{i\rightarrow f}(x)$.  We assume the reaction rate density to
be localized in both space and time.  This reflects the small spatial extent of
the target and the short interaction time compared to the beam lifetime.  The
reaction rate is trivially related to the reaction probability:
\begin{equation}
	|S_{i\rightarrow f}|^2=\int d^4 x \;{\cal W}_{i\rightarrow f}(x).
\end{equation}
Thus, the reaction rate is easily identifiable in the calculations in sections
\ref{sec:pdist},\ref{sec:edist}.  For example, in the process 
$\gamma B\rightarrow B'$ in Fig. \ref{fig:dis}b, the reaction rate density is 
${\cal W}_{\gamma B\rightarrow B'}(x,q)$.  For the process 
$AB\rightarrow A'B'$ in Fig. \ref{fig:dis}a., it is
\begin{equation}
	{\cal W}_{AB\rightarrow A'B'}(x)=\int d^4 r \frac{d^4 q}{(2\pi)^4}
	J^{\mu\nu}_{A}(x+r/2) D^{c}_{\mu\nu\mu'\nu'}(r,q) J^{\mu'\nu'}_{B}(x-r/2).
\end{equation}
Note that the reaction rate density is a function of the average space--time
location of all the vertices in the process.

The cross section is the effective area of the target, so we define the
cross section as the integral over the beam face of the fraction of incident
particles that interact with the target per unit area:
\begin{equation}
	\sigma=\int_{A} d^2x_T 
	\left(\frac{\# \;\mbox{scattered particles}}{\mbox{unit area}}\right)\left/
	\left(\frac{\# \;\mbox{incident particles}}{\mbox{unit
	area}}\right).\right. 
\end{equation}
The number of incident particles per unit area crossing the target plane is the
particle flux:
\begin{equation}
	\frac{\# \;\mbox{incident particles}}{\mbox{unit area}}={\cal N}_{inc}
	\int^{L/2}_{-L/2} dx_L \;\hat{n}\cdot\vec{j}(x) 
	\equiv
	{\cal F}(\vec{x}_{T}).
\end{equation} 
Here $\hat{n}$ is a unit normal to the target plane and ${\cal N}_{inc}$ is the
number of particles in the beam.  The single particle current is given in terms
of the incident particle Wigner function by \cite{carruth83}
\begin{equation}
	\vec{j}(\vec{x})=\int\dn{3}{p}\dn{}{p^2}\vec{v} f(x,p).
\end{equation}
We need not average over time because the beam is uniform on the time 
scale of the reaction.
The number of scattered particles per unit area is found by multiplying the
number of incident particles by the reaction probability per unit area:
\begin{equation}
	\frac{\# \;\mbox{scattered particles}}{\mbox{unit area}}={\cal N}_{inc}
	\int^{L/2}_{-L/2} dx_L \int^{T/2}_{-T/2} dx_0 \;{\cal W}_{i\rightarrow f}(x)
	\equiv{\cal N}_{inc} \bar{{\cal W}}_{i\rightarrow f}(\vec{x}_T).
\end{equation}
Thus, the cross section is 
\begin{equation}
	\sigma=\int_A d^2 x_{T} 
	\frac{{\cal N}_{inc} \;\bar{{\cal W}}_{i\rightarrow f}(\vec{x}_T)}
	{{\cal F}(\vec{x}_T)}.
\label{eqn:crosssection}
\end{equation}

In equation (\ref{eqn:crosssection}), all longitudinal and temporal
structure of the interaction is washed out by the beam.  Furthermore, in a
any practical  experiment, the wavepackets are delocalized in the transverse
direction on the length scale of the
interaction region.  Thus, the transverse 
structure of ${\cal F}(\vec{x}_{T})$ is gone
and the flux reduces to ${\cal F}={\cal N}_{inc}|\vec{v}|/A$, where $|\vec{v}|$
is the mean projectile velocity.  The flux can then be pulled out of the
transverse integral in (\ref{eqn:crosssection}).  The transverse integral of
the reaction probability per unit area is 
${\cal N}_{inc} |S_{i\rightarrow f}|^2$, so the
cross section becomes 
\begin{equation}
	\sigma=\frac{A |S_{i\rightarrow f}|^2}{|\vec{v}|}.
\end{equation}
This is the conventional momentum space cross section in our choice of
normalization.

%
%
\section{Wavepackets}
\label{append:wvpkts}

Throughout this paper, we use wavepackets in the initial and final states
of a reaction to provide spatial localization or delocalization.  In this 
appendix, we detail
the construction of an initial or final state wavepacket and discuss the limits
of either a completely localized or delocalized wavepacket.

\subsection{On--Shell Gaussian Wavepacket}

An initial (or final) state ket can be written with wavepackets:
\begin{equation}
 	\ket{i}=\int \dnpi{4}{p} f(p) \ket{\vec{p}}.
	\label{eqn:ket2}
\end{equation}
The corresponding Wigner function of the particles is
\begin{equation}
\begin{array}{rl}
	f(x,p)=&\displaystyle\int\dnpi{4}{p'}e^{-ix\cdot p'} 
		\ME{i}{\hat{\phi}^{*}(p-p'/2)\hat{\phi}(p+p'/2)}{i}\\
		=&\displaystyle\frac{1}{2Vp_0}\int\dnpi{4}{p'}e^{-x\cdot p'}
		f(p+p'/2)f^{*}(p-p'/2).
\end{array}
\label{eqn:wigpartdens}
\end{equation}
Particles in either the initial or final states are on--shell, so they can be
expanded in momentum eigenstates.  We choose our wavepacket to be a Gaussian
superposition of momentum eigenstates with a momentum spread $\sigma$:
\[
	\phi(p)={\cal N} \deltaftn{}{p^2-M^2} 
	\exp{[-(\vec{p}-\vec{p}_i)^2/2\sigma^2]}
\] 


The Wigner transform of this wavepacket can not be done analytically except 
in the limit when $|\vec{p}_i|\gg\sigma$.  In this limit, 
$\vec{p}_i\approx\vec{p}\gg\vec{p'}$
so our wavepacket is localized in momentum giving the following Wigner density
of particles:
\begin{equation}
	f(x,p)=\frac{|{\cal N}|^2}{8\pi p_0^2}\deltaftn{}{p^2-M^2}
	\exp{\left[-\frac{(\vec{p}-\vec{p}_i)^2}{2\sigma^2}\right]}
	(2\sigma\sqrt{2\pi})^3 \exp{\left[-2\sigma^2
	(\vec{v}x_0-\vec{x})^2\right]}.
\end{equation}
Here $\vec{v}=\vec{p}/p_0$ is the velocity of the wavepacket.  
Thus, the particle's Wigner function is a Gaussian in both momentum and space. 
The spread in momentum is the inverse spread in space.  The centroid of the 
Gaussian follows the particle's classical trajectory.
The energy of the packet is set by the delta function out front.  We have not
constrained the particle in energy so this
density contains both positive and negative energy contributions.  

\subsection{Delocalizing the Wavepacket in Space: Free Wavepacket} 

In accordance with the uncertainty principle,   the wavepacket becomes 
completely delocalized in space in the limit of complete localization in 
momentum (i.e. $\sigma\rightarrow 0$).
In this limit, the spatial Gaussian approaches unity and the momentum Gaussian
becomes a delta function.  After working out the normalization, we find 
\begin{equation}
	f^{\rm free} (x,p)=\frac{1}{2Vp_0}\twopideltaftn{4}{p-p_i}.
\label{eqn:imfree}
\end{equation} 
This is no surprise since we squeezed the state into a momentum eigenstate.

\subsection{Localizing the Wavepacket in Space: Classical Wavepacket}
\label{append:classdens}

A classical particle is localized in both space and momentum, a seeming
violation of the uncertainty principle.  In real life, this is not a problem
since the reason classical particles appear localized is that we probe
them on length (or momentum) scales too coarse to resolve the interesting
quantum features.  In the case of our Gaussian wavepacket, this amounts to 
probing the distribution on length scales much larger than $1/\sigma$.  
In this case, the space Gaussian is too localized to resolve and we can replace
it with a delta function.  Additionally, we assume that $\sigma$ is large, so 
we can replace the momentum Gaussian with a delta function as well.  

Making these approximations, we
find the Wigner density of a classical particle:
\begin{equation}
	f^{\rm classical} (x,p) = \frac{1}{2}(2\pi)^4 
	\deltaftn{3}{\vec{p}-\vec{p}_i}
	\deltaftn{}{p^2-M^2}\deltaftn{3}{\vec{v}x_0-\vec{x}}
\label{eqn:classdens}
\end{equation}
Here we have inserted the correct normalization for the wavepacket.
This density corresponds to an on--mass--shell particle that
follows its classical trajectory $\vec{v}x_0=\vec{x}$.  Again, we left in both
positive and negative energy contributions. 

%
%
\section{The Classical Current}
\label{append:current}
In this appendix, we derive the classical
current used in the effective photon distribution calculation.
For the sake of illustration, we take our point particle to be a
scalar particle.  The derivation goes in three steps: first we define the 
Wigner current of a scalar particle, then we
derive the photon/scalar interaction vertex in phase--space, and finally we 
localize the initial and final states of the scalar to give the classical
current.    

\subsection{Wigner Current}

We begin by restating equation (\ref{eqn:current}):
\begin{equation}
	\Jcurrent{\mu}{\nu}{x}{q}{A}=\int\dnpi{4}{\tilde{q}}
	e^{-i\tilde{q}\cdot x}\ME{A'}{j^{\mu}(q+\tilde{q}/2)}{A}
	\ME{A}{j^{\dagger \nu}(q-\tilde{q}/2)}{A'}.
	\label{eqn:currentdef}
\end{equation}
We write the initial and final state bra's and ket's according to
equation (\ref{eqn:ket}).  
Rewriting equation (\ref{eqn:currentdef}) in terms of initial and final Wigner
densities, 
\begin{equation}
	\Jcurrent{\mu}{\nu}{x}{q}{A}=\int\dnpi{4}{p_i}\dnpi{4}{p_f}
	\denslabel{f}{x}{p_i}{A}\denslabelstar{f}{x}{p_f}{A'}
	\twopideltaftn{4}{p_i-p_f-q}\Gamma_{\mu \nu}(q,p_i,p_f).
	\label{eqn:currwvpkt}
\end{equation}
We assume that the initial and final wavepackets are localized in momentum and
some-what delocalized in space.  Shortly, we assume that we probe the
current on length scales much larger than even this delocalized space
distribution.

\subsection{Scalar Vertex}

$\Gamma_{\mu \nu}(q,p_i,p_f)$ is not quite the Wigner transform of the
$\gamma AA'$ vertex, although it does arise from performing the Wigner
transform in equation (\ref{eqn:currentdef}).  It is defined by
\begin{equation}
	\twopideltaftn{4}{p_i-p_f-q}\Gamma_{\mu \nu}(q,p_i,p_f) =
	4V^2 p_f^0 p_i^0
	\int\dnpi{4}{\tilde{q}}\ME{\vec{p}_f}{j_{\mu}(q+\tilde{q}/2)}{\vec{p}_i}
	\ME{\vec{p}_i}{j^{\dagger}_{\nu}(q-\tilde{q}/2)}{\vec{p}_f}
\end{equation}
Using the matrix element
\[
	\ME{\vec{p}_f}{j_{\mu}(q)}{\vec{p}_i}=eZ\twopideltaftn{4}{p_i-p_f-q}
	\frac{(p_i+p_f)_{\mu}}{2V\sqrt{p_f^0 p_i^0}},
\] 
we get
\begin{equation}
	\Gamma_{\mu \nu}(q,p_i,p_f)  = 
	\alpha_{em} Z^2(p_i+p_f+\frac{1}{2}(\tilde{p_i}+\tilde{p_f}))_{\mu}
	(p_i+p_f-\frac{1}{2}(\tilde{p_i}+\tilde{p_f}))_{\nu}
	\label{eqn:currstep2}
\end{equation}
The relative momenta, $\tilde{p}_i$ and $\tilde{p}_f$, become
derivatives on $x$ in the current (\ref{eqn:currwvpkt}).  We
assume the wavepackets to be uniform the the reaction's length scales, so we 
ignore the derivatives and arrive at the phase--space scalar vertex
\begin{equation}
	\Gamma_{\mu \nu}(q,p_i,p_f)=\alpha_{em} Z^2 (p_i+p_f)_{\mu}(p_i+p_f)_{\nu}.
	\label{eqn:vertexthingee}
\end{equation}

\subsection{Classical Current}

We are now in a position to derive equation (\ref{eqn:classcurrent})
for the classical current density in phase--space.  
First, we take the final state to be a momentum eigenstate and sum over it. 
Since the final state is localized in momentum around $p_f$, this is not a bad
approximation.  Second, we take the initial state to be a classical wavepacket.
In other words, we assume that the initial state is localized in momentum and
delocalized in space but we probe it on such large length scales that we
still see a spatially localized wavepacket.  So, putting equations
(\ref{eqn:imfree}), (\ref{eqn:classdens}) and (\ref{eqn:vertexthingee}) into 
(\ref{eqn:currwvpkt}) and summing over final states, we get 
\[
	\Jcurrent{\mu}{\nu}{x}{q}{}=2\pi \alpha_{em} Z^2\: 
	v_{\mu} v_{\nu} \:\deltaftn{3}{\vec{x}-x_0\vec{v}}
	p_{i0}\delta ((p_f+q)^2-M^2).
\]
Using $p_f^2=M^2$ and $v_\mu\approx p_{f\mu}/p_{i0}$ and assuming 
$q^2/p_{i0}\ll q\cdot v$, we get the classical current: 
\begin{equation}
	\Jcurrent{\mu}{\nu}{x}{q}{classical}=2\pi \alpha_{em} Z^2\: 
	v_{\mu} v_{\nu} \:
	\delta (q\cdot v)\deltaftn{3}{\vec{x}-x_0\vec{v}}.
\label{eqn:yeaclasscurrent}
\end{equation}
Note that this current allows for emission of both positive and negative energy
photons.  To use the retarded propagators in Section \ref{sec:pdist}, we need a
$\theta(q_0)$ in equation (\ref{eqn:yeaclasscurrent}).  We can do this by
suitable choosing $\vec{p}_f$ and $\vec{p}_i$ and restricting the initial and
final states to have only positive energy.

%
%

\section{Phase--Space Effective Photon Distribution of a Stationary Point Charge}
\label{append:static}

In this appendix, we describe limit of $\vec{v}=0$ of the photon distribution of 
the point charge in Subsection~\ref{sec:classEPD}.  Since the spatial dependence of
the effective photon distribution is controlled by the Wigner transform of the
vector potential, $A_{\mu}(x)$, we only discuss $A_{\mu\nu}(x,q)$ here.  
When $\vec{v}=0$, the photon
vector potential becomes $A_\mu(x)=(\frac{e}{|\vec{x}|},\vec{0})$ so 
$A_{\mu\nu}(x,q)$ is the Wigner transform of the Coulomb potential. 

We take the point charge to be resting at the origin and emiting photons with
four--momentum $q_\mu=(q_0,\vec{q})$.  Putting $\vec{v}= 0$ in 
Eq.~(\ref{eqn:Amunu}), we find
\begin{equation}
\begin{array}{rl}
   A_{00}(x,q)&=\displaystyle 32\pi^2\alpha_{em}\delta(q_0)\frac{1}{|\vec{q}|}
      {\cal A}(2|\vec{x}||\vec{q}|\cos(\theta),2|\vec{x}||\vec{q}|\sin(\theta))\\
   A_{ij}&=0
\end{array}
\end{equation}
where $\theta$ is the angle between $\vec{x}$ and $\vec{q}$ and the
dimensionless function ${\cal A}$ is given in Eq.~(\ref{eqn:dimlessA}).  
Clearly the photon field is time independent and is composed of only zero 
energy photons.  Furthermore, by virtue of the $1/|\vec{q}|$ singularity, the 
photon field is mostly composed of low momentum photons. 

In Fig~\ref{fig:coulomb}, we plot the dimensionless function ${\cal A}$ as a
function of $\vec{x}$ in the plane defined by $\vec{x}$ and $\vec{q}$.  Note
that the central region of the distribution is circular, but becomes elliptical
as one moves away from the center.  In the transverse direction (i.e. the
direction perpendicular to the photon three--momentum), the
distribution approaches zero, but never goes negative.  The width in the
transverse direction is approximately $250$~fm.  In the longitudinal
direction, the distribution drops to zero at about $x_L\approx 250$~fm and
oscillates about zero for larger distances.  These oscillations are expected
for a Wigner transformed quantity and simply reflect the fact that $x_L$ and
$q_L$ are Fourier conjugate variables.

Because the photon source is a point source, the 
the shape of the Coulomb distribution comes directly from the shape of the 
the retarded propagator in Subsection~\ref{sec:retprop}.
Thus, we can estimate the width of the photon distribution using the 
estimates of the retarded propagator in Subsection~\ref{sec:retprop}.
In the both the longitudinal and transverse directions, the propagator width 
is $\sim \hbar c/|q_L|=250$~fm, which is exactly the
width we measure from the plots.

%
%
\section{Effective Electron Phase--space Distribution with $m_e\neq 0$}
\label{append:massive}

In this appendix, we calculate the effective electron distribution for
electrons with a mass much larger than their momentum.  This calculation is not
included in Section \ref{sec:edist} because it is not relevant for partons. 

When the electron momentum is much smaller than its mass, we can use
Remler's propagators for massive particle \cite{remler90}.  His propagator
takes one of two forms depending on whether the electron momentum is
space--like or time--like.  His propagator is discussed in
Appendix~\ref{append:prop}.  We show a sample electron density for both the
time--like momentum and space--like momentum cases.  
The momenta of the electrons
are chosen to satisfy the requirements that $p\cdot v >0$ and $k_{T {\rm max}}$ 
be real.  These requirements are equivalent to the requirement that 
$p\cdot v \geq m_e/\gamma$.  Since $p\ll m_e$, we must also have $\gamma \gg 1$.

In Section \ref{sec:esource} we show that 
the electron's source is controlled by the parent photon distribution so
we show the parent photon distributions next to the electron 
distributions in all subsequent plots. 
We do not restrain the photons to have $q_0>0$ as in Section \ref{sec:edist}, 
so our sources include negative energy contributions.  Because we use retarded 
photon propagators in our source, these calculations only serve to
illustrate how Remler's propagators function.  In fact, had we restricted
$q_0>0$, there would not be enough momentum--space to perform the $\vec{k}_T$
integrals and the electron distribution would be zero. 

\subsection{Feynman Propagator for Particles with Space--Like Momentum}

The propagator for electrons with space--like momentum is:
\[\begin{array}{rl}
\Gcaus{\Delta x}{p}=&\displaystyle\int^{\infty}_{-\infty} \dn{}{\tau} 
      \deltaftn{4}{\Delta x-\frac{p}{\sqrt{-p^2}}\tau}e^{-2m_e|\tau|}
      \frac{1}{2m_e\sqrt{-p^2}(m_e^2-p^2)}\\
      &\displaystyle\times\left\{\sqrt{-p^2}
      \cos{(2\tau\sqrt{-p^2})}
      +m_e\sin{(2|\tau|\sqrt{-p^2})}\right\}.
\end{array}\]  
Remler's propagator for space--like electrons has a very simple 
interpretation. First, the delta function forces the electron to follow 
its classical trajectory, but with the electron
velocity defined as $v_\mu=p_\mu/\sqrt{-p^2}$.   
The exponential in proper time strongly damps
propagation that extends farther in time than $1/2m_e$ along the classical
trajectory.  The fact that the proper time can extend forward or backwards in
time simply reflects the boundary conditions of the Feynman
propagator.  Next, the sine and cosine 
cause the expected Wigner oscillations.  The rest of the terms simply give 
normalization.  Finally, this propagator allows propagation outside of
the light--cone, but such propagation is strongly damped.  This may seem
strange, but should come as no surprise:  the coordinate space propagator 
for massive particles propagates particles  outside 
of the light--cone \cite{bog79}.

\subsection{Massive Electrons Distribution for Electrons with Space--Like 
Momentum}

We now perform the integrals over $\dn{2}{k_T}$ and $\dn{4}{x}$ in equation
(\ref{eqn:elecdist}).  
The $\dn{4}{x}$ is a trivial delta function integral and the integral 
over $\dn{2}{k_T}$ can be done numerically.  On the left in Fig. 
\ref{fig:edist1} we have a sample cut through the 
phase--space density for electrons with a typical space--like 
4-momentum ($p_{\mu}=(0.05,0.008,0.06,0.0)$ MeV/c).  
On the right is one of the underlying photon distributions.
We chose the source velocity so that $\gamma = 12.47$.  This velocity is a 
compromise between having enough
momentum space available for the electron and rendering the plot unreadable
because of the Lorentz contraction.  

Now we examine these plots.  First, we see the contributions 
from retarded emission and propagation (upper left electron pancake) and from 
advanced propagation (lower right pancake).  Let us concentrate on the 
retarded electrons.  At some time in the past, the photons split into the 
electrons and the positrons.  The electrons than propagate forward 
along their classical trajectory until they reach the location of the left 
pancake.  Notice that this pancake has nearly the same size 
as the photon distribution on the right.  The other photon distribution, 
corresponding to the other root of positron momentum, has a slightly different 
tilt and width, but the difference in the plots is not noticeable.
The electron pancakes are slightly larger than the 
photon pancake, presumably because of 
momentum broadening from the emitted positron.  The 
advanced electrons have exactly
the same shape and size as their retarded brethren, but they followed a 
time--reversed classical trajectory, coming from some time in the future.  

\subsection{Feynman Propagator for Particles with Time--Like Momentum}

Remler's propagator for massive particles with time--like momentum is 
\[\begin{array}{rl}
\Gcaus{\Delta x}{p}=& \displaystyle\int^{\infty}_0 \dn{}{\tau}
      \frac{1}{2m_e\sqrt{p^2}} \left\{
      \frac{\sin{(2\tau(\sqrt{p^2}-m_e))}}{(\sqrt{p^2}-m_e)}
      \deltaftn{4}{\Delta x-\frac{p}{\sqrt{p^2}}\tau}\right. \\
      &- \displaystyle\left.
      \frac{\sin{(2\tau(\sqrt{p^2}+m_e))}}{(\sqrt{p^2}+m_e)}
      \deltaftn{4}{\Delta x+\frac{p}{\sqrt{p^2}}\tau}\right\}.
\end{array}\]
Remler's time--like propagator does not have as simple an interpretation as his
space--like propagator.  Delta functions still keep the
particle on its classical trajectory, but the integrals in ``proper time'' are 
Fourier sine transformed along this classical trajectory.  Thus, a simple peak
in the underlying photon distribution will get Fourier transformed into a series
of peaks and valleys in the electron distribution.  Furthermore, the advanced 
and retarded branches enter with different signs, so we have large {\em
negative} contributions from the advanced branch.

\subsection{Massive Electrons Distribution for Electrons with Time--Like 
Momentum}

Despite the difficulty in interpreting the propagator, the $d^4 x$ and
$\dn{2}{k_T}$ integrals can be done.   A sample cut through the 
phase--space distribution is shown in Fig. \ref{fig:edist2}.  These electrons
have a typical time--like 4--momentum ($p_\mu=(0.05,0.005,0.04,0)$~MeV/c)
and the source has a $\gamma=12.47$ and is moving to the right.  
Again, the source velocity was 
picked as a compromise between readability of the plot and available momentum
space for the electron.

We see the difficulty in interpreting the electron distribution.  the
Fourier transform took a simple photon peak and produced a series of large
electron peaks.  The retarded branch corresponds to the envelope of large
positive peaks on the upper left.  The advanced branch corresponds to the 
envelope of large negative peaks in the lower right.    
Each peak in the pair of envelopes appears to be a Lorentz pancake, but the
envelope as a whole is significantly broader than the underlying photon peak. 
Presumably, averaging the distribution over unit areas in phase--space would
result in a much tighter distribution.

%
%
\section{Free Scalar Propagators in Phase--Space}
\label{append:prop}

In this section, we state all of our phase--space propagators,
discuss the symmetries of the massless propagators in some detail 
and outline the derivation of the retarded and Feynman 
scalar propagators.  The massive Feynman propagator is discussed by Remler
\cite{remler90} so our discussion is brief.  
The Dirac and vector propagators differ
from the scalar propagators by the inclusion of either spin projectors (in the case
of Dirac particles) or polarization projectors (in the case of vector
particles) so we do not need to discuss them.

We define the Wigner transform of any translationally invariant propagator as
\begin{equation}\begin{array}{ccl}
   \Gprop{x}{p} 
   & = & \displaystyle\int\dnpi{4}{p'} e^{-i x\cdot p'} G(p+p'/2)
      G^{\dagger}(p-p'/2) \\ 
   & = & \displaystyle\int\dn{4}{x'} e^{i x'\cdot p} G(x+x'/2)
      G^{\dagger}(x-x'/2)
\label{eqn:propdef}
\end{array}\end{equation}
The vacuum propagators that we use are \cite{bog79}:
\begin{mathletters}
\begin{eqnarray}
G^{\pm}(p)&=&-\left( p^2-m^2\pm i\epsilon p_0\right)^{-1}\label{eqn:Gpm}\\
G^{\stackrel{c}{a}}(p)&=&-\left( p^2-m^2\pm i\epsilon \right)^{-1}
\end{eqnarray}
\end{mathletters}


\subsection{Massless Scalar Propagators}

\subsubsection{Symmetries} 

A time reversal transform in coordinate space is equivalent to a 
simultaneous reflection in time and energy in phase--space.  Under time
reversal the $+$ and $-$  propagators 
change into one another while the Feynman and anti--Feynman propagators
remain unchanged:
\begin{mathletters}
\begin{eqnarray}
   \Gplus{x_{0},\vec{x}} {p_{0},\vec{p}}&=&\Gmin {-x_{0},\vec{x}}
      {-p_{0},\vec{p}}\\
   G^{\stackrel{a}{c}} (x_{0},\vec{x},p_{0},\vec{p}) &=&
      G^{\stackrel{a}{c}} (-x_{0},\vec{x},-p_{0},\vec{p}).
\end{eqnarray}

A parity transform in coordinate space is equivalent to a simultaneous
reflection in a space coordinate and the corresponding momentum coordinate.
Under a parity transformation, all of the propagators remain unchanged:
\begin{eqnarray}
   G^{\pm} (x_{0},\vec{x},p_{0},\vec{p}) &=&
      G^{\pm} (x_{0},-\vec{x},p_{0},-\vec{p})\\
   G^{\stackrel{a}{c}} (x_{0},\vec{x},p_{0},\vec{p}) &=&
      G^{\stackrel{a}{c}} (x_{0},-\vec{x},p_{0},-\vec{p}).
\end{eqnarray}

The Feynman propagators have another (rather amusing) {\em argument
switching} symmetry.  Here all the space--time components are switched with the
the corresponding momentum--energy components:
\begin{equation}
   G^{\stackrel{a}{c}}(x,p)=G^{\stackrel{a}{c}}(p,x)
\end{equation}

Finally, the Feynman and anti--Feynman propagator are related through
a  complete reflection of all of the space or momentum coordinates:
\begin{eqnarray}
   \Gcaus{x}{p}&=&\Gacaus{-x}{p}\\
   \Gcaus{x}{p}&=&\Gacaus{x}{-p}.
\end{eqnarray}
\end{mathletters}

\subsubsection{Propagators}

We now present the massless Feynman and retarded propagators.  The advanced
and anti--Feynman propagators can be recovered using the symmetry
relations above.
Since all of the massless scalar propagators are
dimensionless and Lorentz invariant, we expect that they will be
functions of $x\cdot p$ and $x^2 p^2$ and possibly theta functions in
energy or time.  In fact, the propagators are:
\begin{mathletters}
\begin{eqnarray}
   \Gcaus{x}{p}&=&
   	\frac{1}{4\pi} \left[ \mbox{ sgn} (x^2) + 
      	\mbox{ sgn} (p^2) + 2 \mbox{ sgn} (x\cdot p)\right]\nonumber\\
      	& &\times\displaystyle\left\{ \theta (\lambda^2)
      	\frac{\sin{(2\sqrt{\lambda^2})}}{\sqrt{\lambda^2}} -
      	\theta (-\lambda^2) \frac{\exp{(-2\sqrt{-\lambda^2})}}
      	{\sqrt{-\lambda^2}}\right\}\\
      \label{eqn:Gcausdef}
   \Gplus{x}{p}&=&\frac{1}{\pi}\theta (x_0)\theta (x^2) \theta (\lambda^2) 
      \frac{\sin{(2\sqrt{\lambda^2})}}{\sqrt{\lambda^2}} 
      \label{eqn:Gretdef}
\end{eqnarray}
\end{mathletters}
Here the Lorentz invariant $\lambda^2$ is given by 
$\lambda^2=(x\cdot p)^2-x^2 p^2$.
Since we discuss how the propagators work in the Sections \ref{sec:retprop}
and \ref{subsec:feynprop}, we do not do so here.

\subsubsection{Derivation of $G^{+}$}

The Wigner transform of $G^{+}$ is easiest to do in coordinate space.  In
coordinate space, $G^{+}(x)=\frac{1}{2\pi}\theta(x_0)\delta(x^2)$, so the 
Wigner transform integral is a series of delta function integrals.  Performing
the first delta function integral in (\ref{eqn:propdef}), and simplifying the 
theta functions, we find
\[
   \Gplus{x}{p}=\frac{\theta(x_0)}{(2\pi)^2}\int^{2x_0}_{-2x_0} d x'_0
      \sqrt{4x^2+{x'_0}^{2}}\int_{4\pi}d\Omega_{\vec{x}'} 
      e^{i x'\cdot p}\delta(x'\cdot x).
\]
Using $2\pi\delta(x)=\int^{\infty}_{-\infty}d \alpha e^{ix\alpha}$, we can do
the angular integral, giving us a Bessel function:
\[
   \Gplus{x}{p}=4\pi\theta(x_0)\theta(x^2)\int_{-1}^{1} d \alpha 
      e^{i\alpha \eta}J_0(\xi\sqrt{1-\alpha^2}).
\]
Here $\eta=2(p_0 |\vec{x}|-x_0 \hat{x}\cdot\vec{p})$ and 
$\xi=2\sqrt{x^2(\vec{p}^2-(\vec{p}\cdot\hat{x})^2)}$.  This integral is in any
standard integral table\cite{grad80}.  After a bit of simplification, 
one  gets the
result (\ref{eqn:Gretdef}).  This result can be checked by performing the
Wigner transforms in momentum space, but the contour integrals needed for this
calculation are quite tedious.

\subsubsection{Derivation of $G^{c}$}  

The simplest derivation of $\Gcaus{x}{p}$ is far more complicated than 
the derivation of $\Gplus{x}{p}$.  We start by finding the transport--like 
equation of motion for the Wigner propagator.\footnote{The
constraint--like equation could also be used, but $G^c$ is easier to derive
using the transport--like equation.}  
The derivation is simple and very similar to the derivation for the retarded 
equation of motion in Section \ref{sec:transport}.  So  we only state 
the result:
\[
   p\cdot\partial \:\Gcaus{x}{p}=\frac{1}{\pi^2}
      \left[\pi\delta(x^2) \sin{(2x\cdot p)}-
      {\cal P}\frac{1}{x^2} \cos{(2x\cdot p)}\right].
\]
Now we define a projector onto the space perpendicular to the particle's
momentum, $g_{\perp\mu\nu}=g_{\mu\nu}-p_\mu p_\nu/p^2$.  This allows us to
change variables to $x_{\perp\mu}=g_{\perp\mu\nu} x^{\nu}$ and $\tau=x\cdot
p/\sqrt{|p^2|}\mbox{ sgn}({p^2})$.  In terms of these variables, we find
$\lambda^2=-p^2x_{\perp}^2$ and the equation of motion becomes
\[
   \partial_\tau G^{c} (\tau,x_{\perp},p)=
      \frac{\mbox{ sgn}(p^2)\sqrt{|p^2|}}{\pi^2}
      \left[\pi\delta(|k^2|\tau^2-\lambda^2)\sin{(2\sqrt{|k^2|}\tau)}
      -{\cal P}\frac{\cos{(2\sqrt{|k^2|}\tau)}}{|k^2|\tau^2-\lambda^2}\right].
\]
So, instead of doing the Wigner transform directly, we only have to solve this
ordinary differential equation.

We find the solution by integrating this differential equation.  The delta
function integral is simple and the principle value integral can be done by
contour integration.  We find 
\[\begin{array}{rl}
   \displaystyle G^{c} (\tau,x_{\perp},p)= G^{c} (\infty,x_{\perp},p)-
   	\frac{1}{\pi} & \displaystyle\left\{ 
      \theta(\lambda^2)\frac{\sin{(2\sqrt{\lambda^2})}}{\sqrt{\lambda^2}}
      \left[\frac{1}{2}(\theta(p^2)-\theta(x^2))+
      \mbox{ sgn}(p^2)\theta(-\tau)\right]\right.\\
      & \displaystyle
      \left. -\mbox{ sgn}(p^2)\theta(-\tau)
      \frac{e^{-2\sqrt{-\lambda^2}}}{\sqrt{-\lambda^2}}\right\}
\end{array}\]
We must now divine the boundary condition at $\tau\rightarrow\infty$.

To find the boundary condition, we actually have to go back to the Wigner
transform of the propagator starting from momentum--space version of equation
(\ref{eqn:propdef}).  We again change variable from $x$ to $\tau$ and 
$x_\perp$.  We also change from $p'$ to $p'_\perp=g_{\perp\mu\nu}{p'}^\nu$ and
$p\cdot p'=\mbox{ sgn}(p^2)\sqrt{|p^2|} k$.  With this, we perform the $k$
contour integral.  The integral is straight forward, but tedious.  However, when
we take the limit as $\tau\rightarrow\infty$, the result simplifies
dramatically:
\[
   G^{c} (\infty,x_{\perp},p)=\frac{1}{\pi^2\sqrt{|p^2|}}\int\dn{3}{p_\perp}
      \cos{(2x_\perp\cdot p_\perp)}\delta(p^2+p_\perp^2).
\]
The delta function integral is trivial and the last pair of integrals requires
integral tables, but in the end we find:
\[
   G^{c}(\infty,x_{\perp},p)=\frac{1}{\pi}\left\{
      \theta(\lambda^2)\mbox{ sgn}(p^2)
      \frac{\sin{(2\sqrt{\lambda^2})}}{\sqrt{\lambda^2}}
      +\theta(-\lambda^2)\theta(-p^2)
      \frac{e^{-2\sqrt{-\lambda^2}}}{\sqrt{-\lambda^2}}\right\}.
\]
Plugging this into the solution of our differential equation, we find
equation (\ref{eqn:Gcausdef}).  This result can be checked by performing a
series of contour integrals in momentum or coordinate space.

\subsection{Massive Scalar Feynman Propagator}

Remler \cite{remler90} has found the Wigner transform of the massive 
Feynman propagator.  This 
transform is difficult but, when one uses the approximation 
${p'}^2\approx(p\cdot p')^2/p^2$, where $p$ is the average momentum and 
$p'$ is the relative momentum, the integrals become simple contour 
integrals.  We state Remler's result here:
\begin{equation}
\label{eqn:remlerprop}
\Gcaus{x}{p}=\left\{
\begin{array}{ll}
   \begin{array}{ll}
      {\displaystyle\int^{\infty}_{-\infty}} & \dn{}{\tau} 
      \deltaftn{4}{x-\frac{p}{\sqrt{-p^2}}\tau}e^{-2m|\tau|}
      \frac{1}{2m\sqrt{-p^2}(m^2-p^2)} \\
      & \times\left\{\sqrt{-p^2}\cos{(2\tau\sqrt{-p^2})}
      +m\sin{(2|\tau|\sqrt{-p^2})}\right\}
   \end{array}
   &{\rm for } \; p^2<0\\
   &\vspace*{.10cm}\\
   \begin{array}{ll}
      {\displaystyle\int^{\infty}_0} \dn{}{\tau}
      \frac{1}{2m\sqrt{p^2}} & \left\{
      \frac{\sin{(2\tau(\sqrt{p^2}-m))}}{(\sqrt{p^2}-m)}
      \deltaftn{4}{x-\frac{p}{\sqrt{p^2}}\tau}\right. \\
      & \left.-\frac{\sin{(2\tau(\sqrt{p^2}+m))}}{(\sqrt{p^2}+m)}
      \deltaftn{4}{x+\frac{p}{\sqrt{p^2}}\tau}
      \right\}
   \end{array}
   &{\rm for } \; p^2>0\\
\end{array}
\right.
\end{equation}
Note that, because of the approximation made, this propagator is
oversmoothed in the direction transverse to the particle's momentum and we
expect these propagators to be accurate only for length scales much larger than
the size of the smoothing.  Since the resulting propagators vary on length
scales of order $1/m$, we should only use these propagators for momenta with 
$p\ll m$. Note also that the sine and exponential functions in the two terms in 
(\ref{eqn:remlerprop}) have the property that they become proportional
to $\deltaftn{}{p^2-m^2}$ as $\tau\rightarrow\infty$.  Thus, this
propagator reduces to the classical
propagator\cite{remler90}.  Finally, we note
that the $\delta$--functions constrain the particle to more along
its classical trajectory, even though its four--momentum (and hence
its four--velocity) is being modulated by the sine and exponential functions.

%
%

\newpage

\begin{center} TABLES \end{center}

\begin{minipage}[t]{4.5in}
\begin{table}
\caption{The vertex Feynman rules for scalar and spinor QED.}
\vspace*{4.5mm} 
\begin{tabular}{|c|c|c|}
      \parbox[b]{1.5in}{\hspace*{.1in}3~point \\ \hspace*{.1in} photon--scalar 
      	\\ \hspace*{.1in} vertex \vspace*{.3in}} &
      \includegraphics{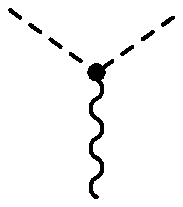} &
      \parbox[b]{1.7in}{\hspace*{.1in}
      	$eZ\bothpartial_{\mu}=eZ(\leftpartial_{\mu}-
      	 \rightpartial_{\mu})$\vspace*{.3in}}\\
   \hline
      \parbox[b]{1.5in}{\hspace*{.1in}4~point \\ \hspace*{.1in}photon--scalar 
      	\\ \hspace*{.1in} vertex \vspace*{.3in}} &
      \includegraphics{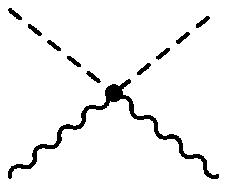} &
      \parbox[b]{1.7in}{\hspace*{.1in}$2ie^2Z^2g_{\mu\nu}$ \vspace*{.3in}}\\
   \hline
      \parbox[b]{1.5in}{\vspace*{.25in}\hspace*{.1in}fermion--photon \\ 
      	\hspace*{.1in}vertex \vspace*{.3in}} &
      \includegraphics{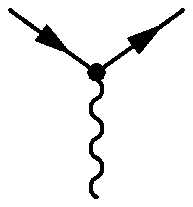} & 
      \parbox[b]{1.7in}{\hspace*{.1in}$-ie\gamma_{\mu}$ \vspace*{.3in}}\\
\end{tabular}
\label{table:vertex}
\end{table}
\end{minipage}


\begin{minipage}[t]{5in}
\begin{table}
\caption{The contour scalar, photon, and electron propagators.}
\vspace*{4.5mm} 
\begin{tabular}{|c|c|c|}
      \parbox[b]{1.1in}{\hspace*{.1in}scalar line \vspace*{.05in}} &
      \includegraphics{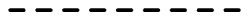} & 
      \parbox[b]{2.5in}{$G(x_1,x_2)$ 
      \vspace*{.025in}}\\
   \hline
      \parbox[b]{1.1in}{\hspace*{.1in}photon line \vspace*{.05in}} &
      \includegraphics{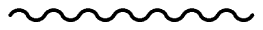} & 
      \parbox[b]{2.5in}{$D_{\mu\nu}(x_1,x_2)=4\pi g_{\mu\nu}G(x_1,x_2)$ 
        \vspace*{.025in}}\\ 
   \hline
      \parbox[b]{1.1in}{\hspace*{.1in}fermion line \vspace*{.05in}} &
      \includegraphics{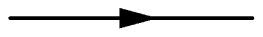} & 
      \parbox[b]{2.5in}{$S_{\alpha\beta}(x_1,x_2)=
        -(\not\!\!q + m)_{\alpha\beta}G(x_1,x_2)$ 
        \vspace*{.025in} }\\ 
\end{tabular}
\label{table:props}
\end{table}
\end{minipage}

\newpage

\begin{center} FIGURE CAPTIONS \end{center}


\begin{figure}
   \caption[fig:dis]{\footnotesize (a) Cut diagram for current A to exchange 
      a photon with current B.  (b) Cut diagram for current B to absorb a free
      photon.  In both figures, the photon/current B interaction is unspecified
      and is represented with a blob.}
   \label{fig:dis}
\end{figure}


\begin{figure}
   \caption[fig:photondist]{\footnotesize Both figures are plots of the
   	dimensionless function ${\cal A}$ corresponding to the effective photon 
   	distribution of a point charge with 3-velocity $\vec{v}=(v_L,\vec{0}_T)$ 
   	with $v_L=0.9c$. 
   	The photons in these slices of the phase--space distribution have
   	$q_{\mu}=(m_{e},m_e/v_L,\vec{0}_T)$ (left) and 
   	$q_{\mu}=(m_{e},m_e/v_L,0.56\mbox{ MeV/c},0) $ (right).
   	In both plots, only the negative and zero contours are labeled.  The 
   	positive contours increase in increments of $0.25$.}
   \label{fig:photondist}
\end{figure}


\begin{figure}
   \caption[fig:splitting]{\footnotesize (a) Cut diagram for creating an 
      electron--positron pair by photon splitting.  The electron interacts 
      with the  
      probe particle, $B$.  The square vertex represents the photon source.  
      (b) Cut diagram for a free electron interacting with the probe particle.}
   \label{fig:splitting}
\end{figure}


\begin{figure}
	\caption[fig:source]{\footnotesize On the left:
		the electron source for electrons with momentum 
		$p_{\mu}=(2.0, 2.05, \vec{0}_T)$~MeV/c electrons. 
		In this figure, only the zero contours are labeled.  
		The positive contours are (in arbitrary units) 1.0, 2.5, 5.0, 7.5 and 10.0.
		On the right: the virtual photon distributions corresponding to one of 
		dominant contributions to electron source.  
		These photons have  $\left<q_{+\mu}\right>=
		(0.956, 1.063, 0.045, 0.045)$~MeV/c.  
		The other root has similar momentum and a similar distribution.
		In this figure, only the negative and zero contours are labeled.
		The positive contours increase in increments of 0.25 (in arbitrary units).}
	\label{fig:source}
\end{figure}


\begin{figure}
	\caption[fig:edist4]{\footnotesize Coordinate space 
		distribution of  $p_{\mu}=(2.0, 2.05, \vec{0})$~MeV/c 
		electrons.  Only 
		the negative and zero contours are labeled.  The positive contours are 
		in increments of 1.0 (in arbitrary units).  The sign of the contours in 
		each region are denoted by $\pm$ signs.}
	\label{fig:edist4}
\end{figure}


\begin{figure}
   \caption[fig:2phot2]{\footnotesize Cut diagram for lepton pair production 
      from a two photon 
      interaction.  $R$ is the space--time point of the center of the collision 
      region.}
   \label{fig:2phot1}
\end{figure}


\begin{figure}
   \caption[fig:2phot2]{\footnotesize The diagrams that contribute, at 
      lowest order, to the $\gamma\gamma\rightarrow e\bar{e}$ effective vertex.}
   \label{fig:2phot2}
\end{figure}
 

\begin{figure}
	\caption[fig:overlap]{\footnotesize  The ellipses represent the edge
		of the photon distributions, each  with four--momentum
		$q=(m_e,m_e/v_L,\vec{0}_T)$.  The shaded region is the geometrical
		overlap of the photon distributions and  sets the size of the
		$e\bar{e}$ production region.  The arrows point in the direction of the 
		photons' source's 3--momentum.}
	\label{fig:overlap}
\end{figure}


\begin{figure}
	\caption[fig:contour]{\footnotesize The contour in the complex time plane 
		used in the evaluation of operator expectation values.  The upper
		branch corresponds to causal ordering and the lower branch to
		anticausal ordering. The arrows denote the contour ordering enforced by 
		the $\tilde{T}$ operator.}
	\label{fig:contour}
\end{figure}


\begin{figure}
	\caption[fig:DSE]{\footnotesize The Dyson--Schwinger equations 
		of the propagators.  Double lines represent the dressed Green's functions
		and single lines represent the non--interacting Green's functions.  The 
		particle self--energies are the large square vertices.}
	\label{fig:DSE}
\end{figure}


\begin{figure}
	\caption[fig:selfenergy]{\footnotesize The scalar and electron self 
		energies and the photon polarization tensor.  Bare vertices are 
		represented by dots and dressed vertices by blobs.  The self--energies 
		and the polarization tensor are all represented by large square vertices.}
	\label{fig:selfenergy}
\end{figure}


\begin{figure}
	\caption[fig:contour]{\footnotesize Cut diagram for probing the particle
		densities in the Generalized
		Fluctuation--Dissipation Theorem.  Time flows downward and, since the 
		the probe interaction is in the future, we leave the nature of the probe 
		unspecified.}
	\label{fig:GFDThm}
\end{figure}


\begin{figure}
	\caption[fig:contour]{\footnotesize Cut diagram for the time--ordered 
		(nonequilibrium) photon density.  Time flows downward and, since the 
		probe interaction is in the future, we leave the nature of the probe 
		unspecified.}
	\label{fig:noneqEPD}
\end{figure}


\begin{figure}
	\caption[fig:contour]{\footnotesize Cut diagram for the time--ordered 
		(nonequilibrium) electron density.  Time flows downward and, since the 
		the probe interaction is in the future, we leave the nature of the 
		probe unspecified.}
	\label{fig:noneqEED}
\end{figure}


\begin{figure}
	\caption[fig:contour]{\footnotesize Cut diagram for probing the $n^{th}$
		generation of partons in a typical cascade.  Time flows downward in this
		diagram and the probe, being somewhere in the future, is left unspecified.}
	\label{fig:noneqNthgen}
\end{figure}


\begin{figure}
	\caption[fig:rung]{\footnotesize Typical rung of the LLA ladder.}
	\label{fig:rung}
\end{figure}


\begin{figure}
	\caption[fig:coulomb]{\footnotesize Plot of the dimensionless function 
		${\cal A}$ corresponding to the Wigner transform of the Coulomb field
		of a static point charge.  The photons in this plot have
		$q_\mu=(0, 0.788, \vec{0}_T)$~MeV/c.  The longitudinal axis is 
		defined by the photon three--momentum.}
	\label{fig:coulomb}
\end{figure}


\begin{figure}
	\caption[fig:edist1]{\footnotesize On the left:  coordinate space 
		distribution of space--like 
		($p_{\mu}=(0.05, 0.008, 0.06, 0)$~MeV/c)  
		electrons.  The $\pm's$ indicate the sign of the function in a 
		particular region.  The
		contours go in steps of 25 (in arbitrary units).
		On the right: the photon distribution for photons with 
		$q_{+\mu}=(-6.63, -6.65 ,0.0648, 0.00478)$~MeV/c.
		The other root has $q_{-\mu}=(-6.33, -6.36, 0.0648, 0.00478)$~MeV/c 
		and its distribution is 
		similar.  Again, the $\pm's$ indicate the sign of the function in a 
		particular region. 
		Here, the contours go in steps of 0.25 (in arbitrary units).}
	\label{fig:edist1}
\end{figure}
 

\begin{figure}
	\caption[fig:edist2]{\footnotesize Coordinate space 
		distribution of time--like 
		($p_{\mu}=(0.05, 0.005, 0.04, 0)$~MeV/c) 
		electrons.  The $\pm's$ indicate the sign of the function in a particular 
		region.  The contours are (in arbitrary units) 
		150, 10, 5, 1, 0.2, 0, -1, -5, -10, and -150.
		One of the roots of the underlying photon distribution is shown at 
		the right.  These photons have momentum 
		$q_\mu=(-8.39, -8.42, 0.182, 0.142)$~MeV/c.
		Here, the $\pm's$ indicate the sign of the function in a particular 
		region and the contours go in steps of 0.25 (in arbitrary units).}
	\label{fig:edist2}
\end{figure}

\newpage

\begin{center} FIGURES \end{center}


\begin{center}
   \includegraphics{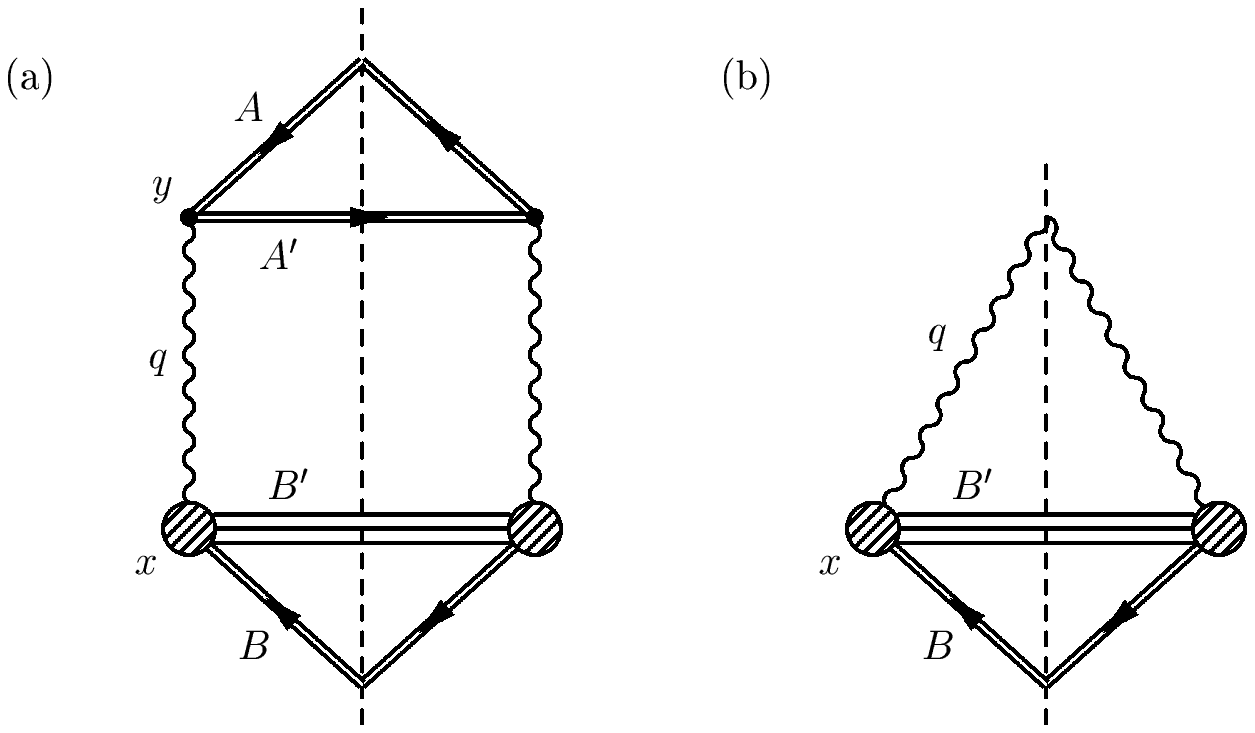}\\
	Fig.~1   
\end{center}

\newpage

\begin{center}
	\includegraphics[width=6in]{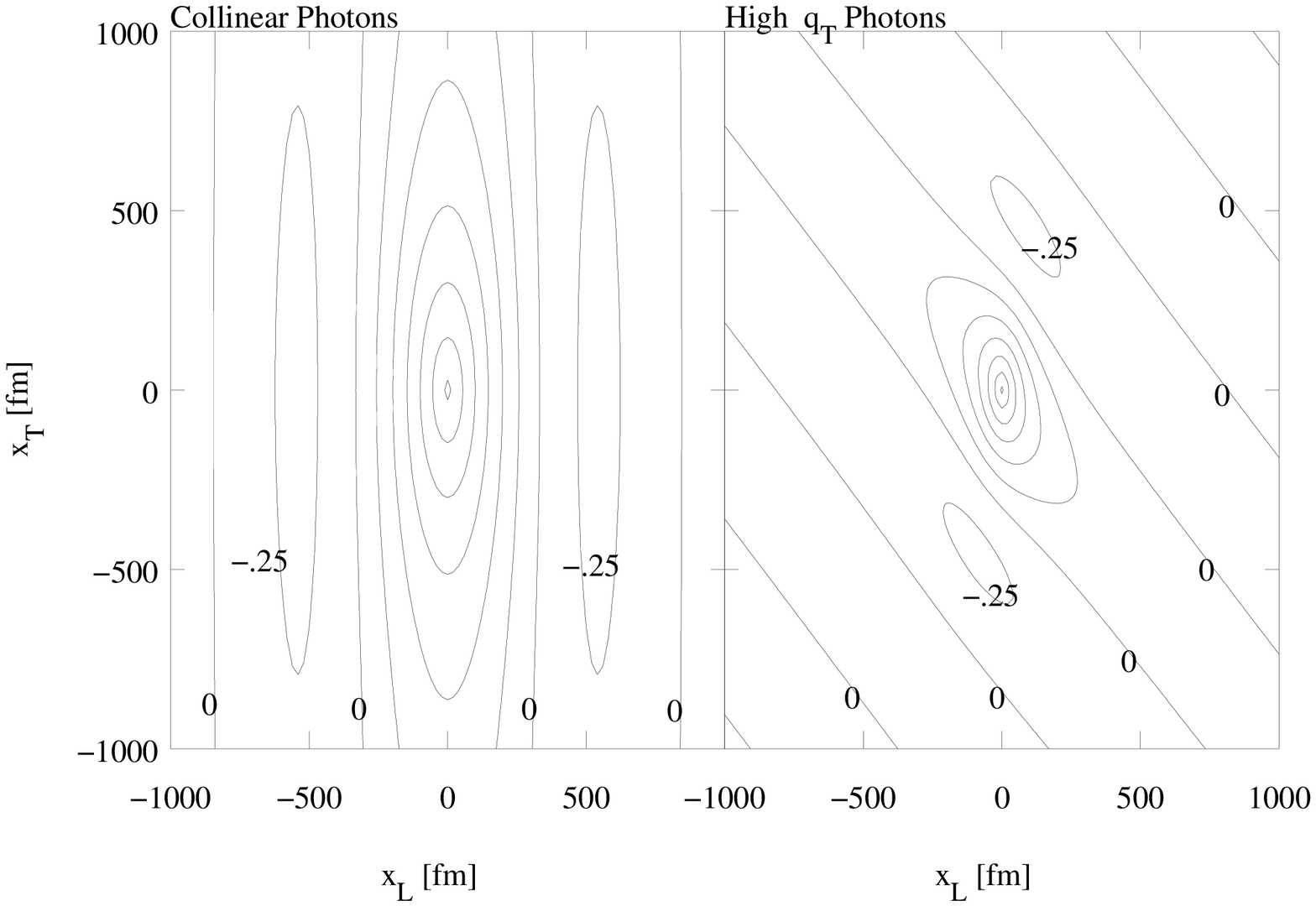}\\
	Fig.~2
\end{center}

\newpage

\begin{center}
   \includegraphics{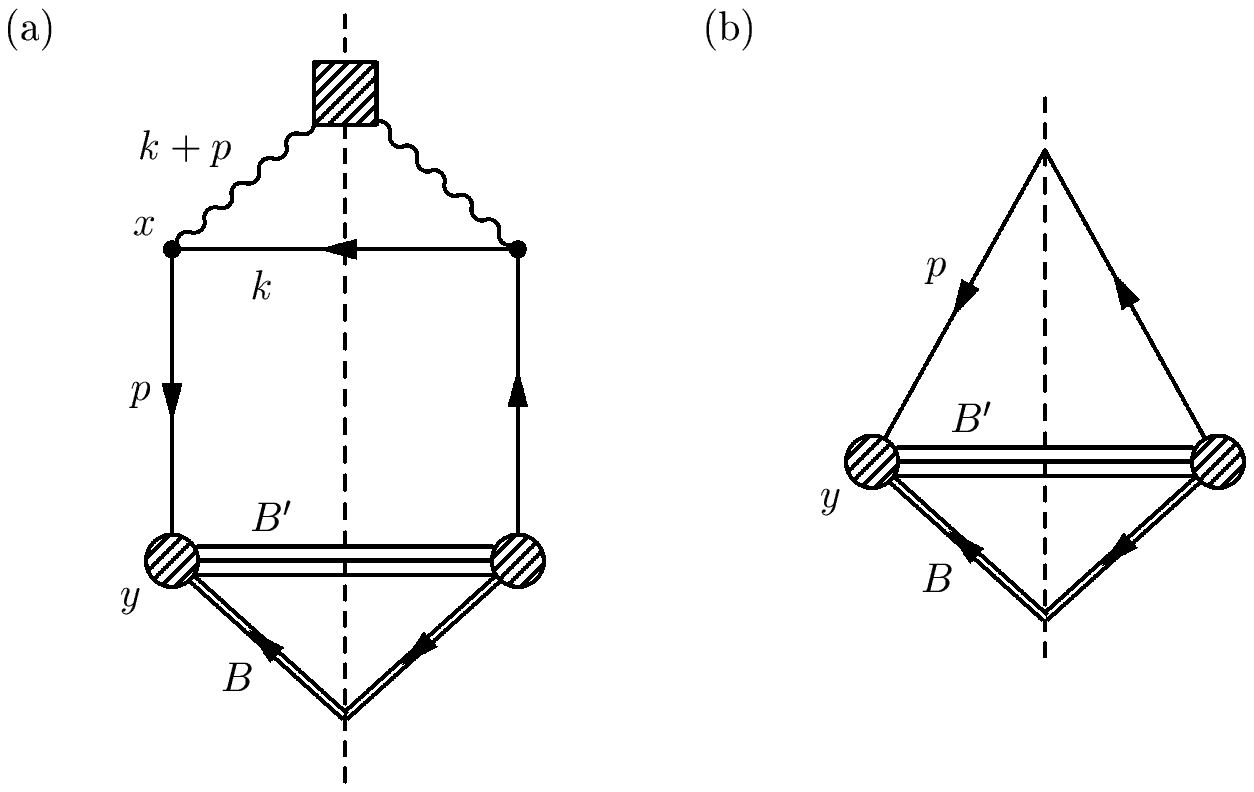}\\
   Fig.~3
\end{center}

\newpage

\begin{center}
	\includegraphics[width=6in]{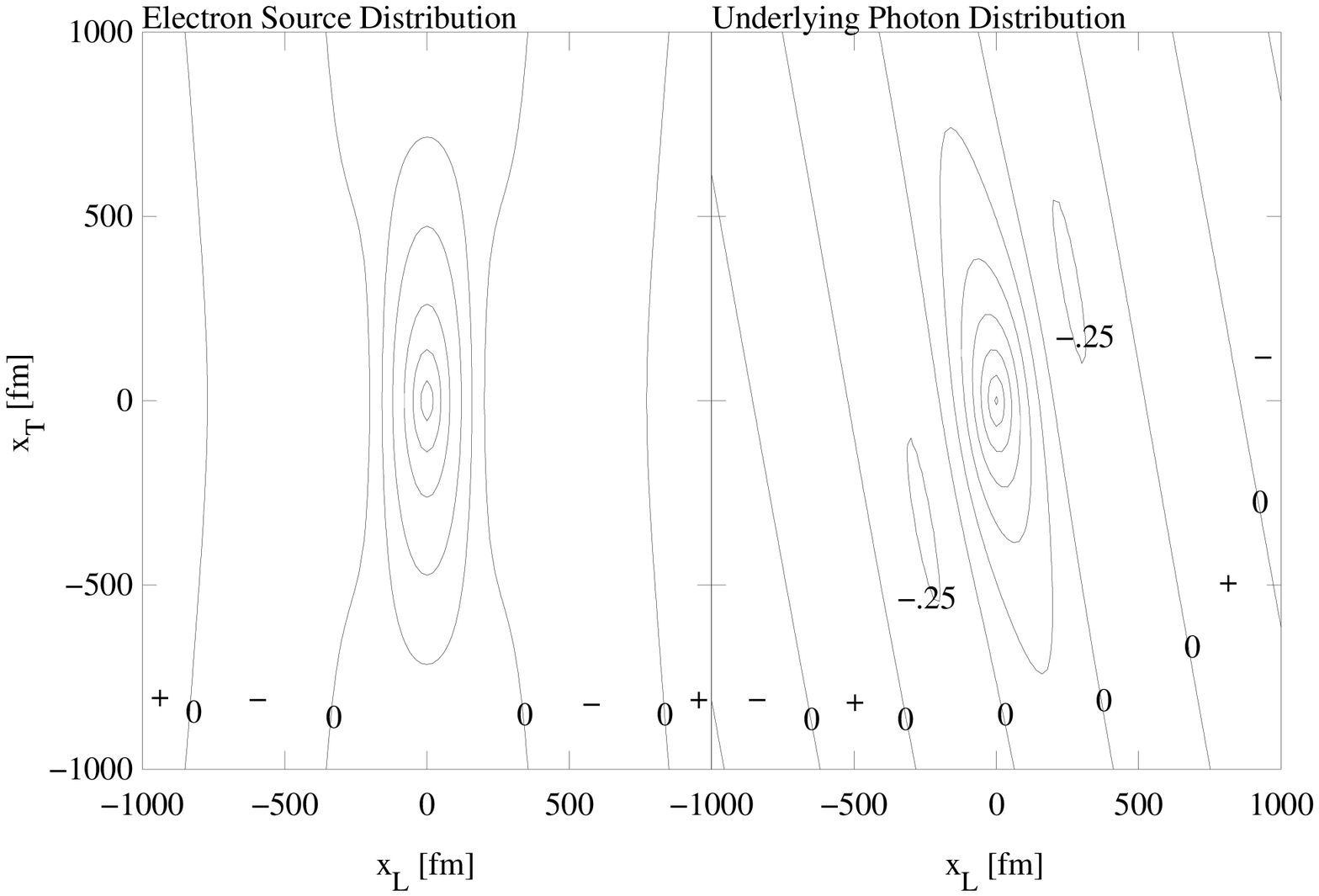}\\
	Fig.~4
\end{center}

\newpage

\begin{center}
	\includegraphics[width=3.5in]{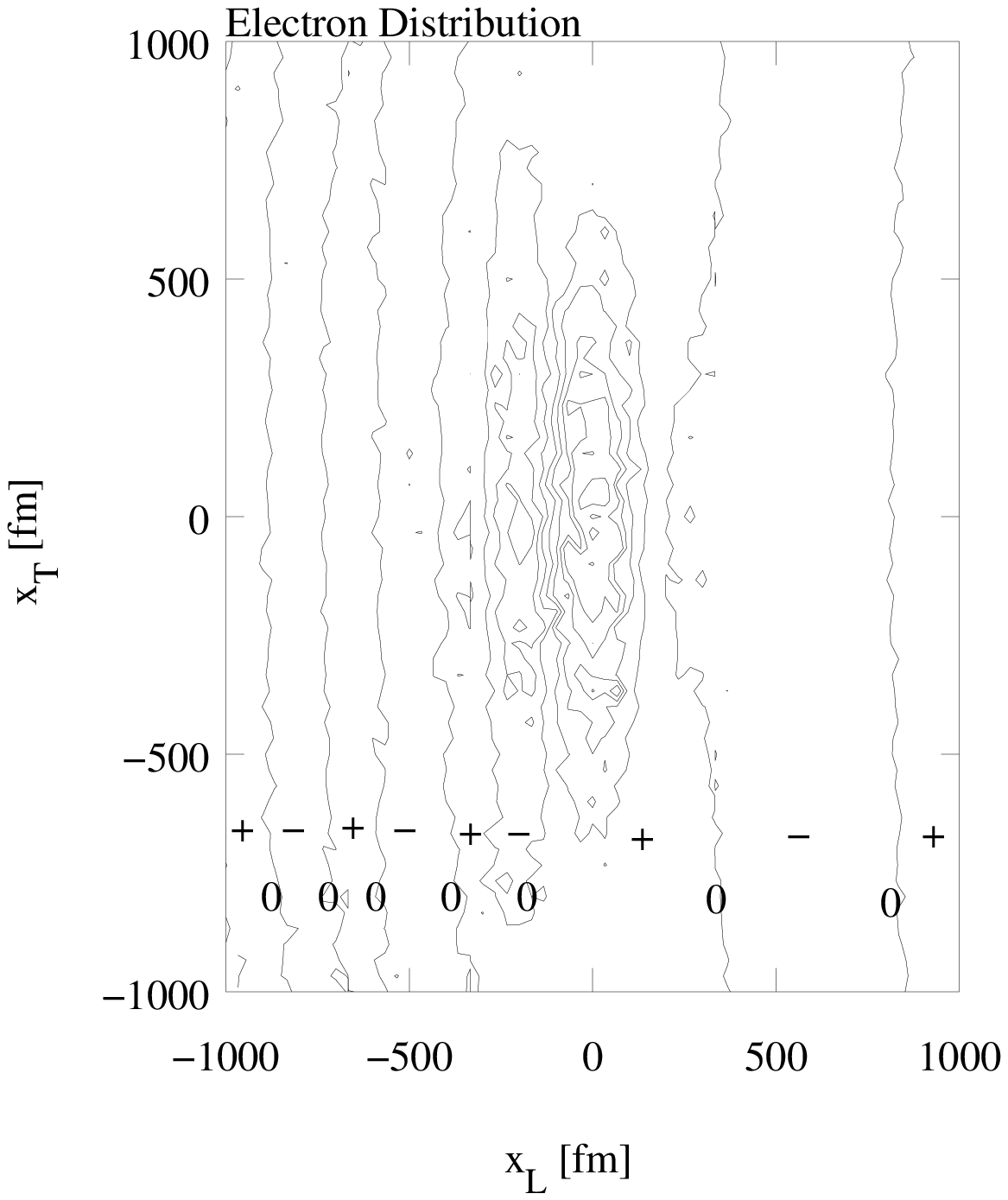}\\
	Fig.~5
\end{center}

\newpage

\begin{center}
   \includegraphics{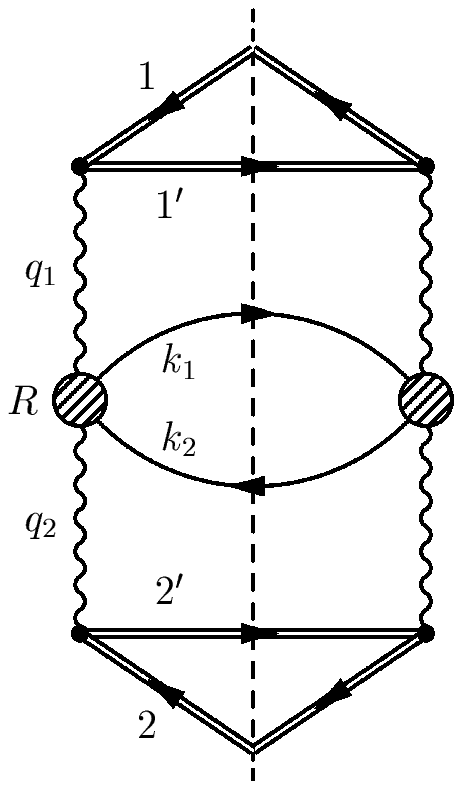}\\
   Fig.~6
\end{center}

\newpage

\begin{center}
   \includegraphics{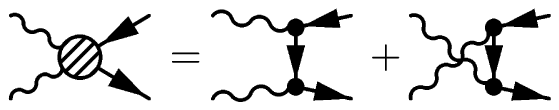}\\
   Fig.~7
\end{center}
 
\newpage

\begin{center}
	\includegraphics[angle=-90,totalheight=2.5in]{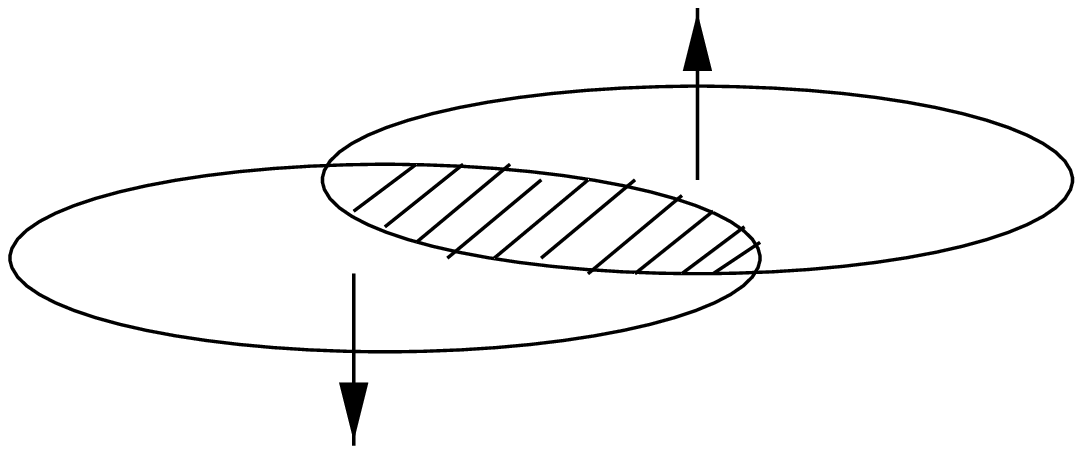}\\
	\vspace{.5cm} Fig.~8
\end{center}

\newpage

\begin{center}
	\includegraphics[angle=-90,width=7.5cm]{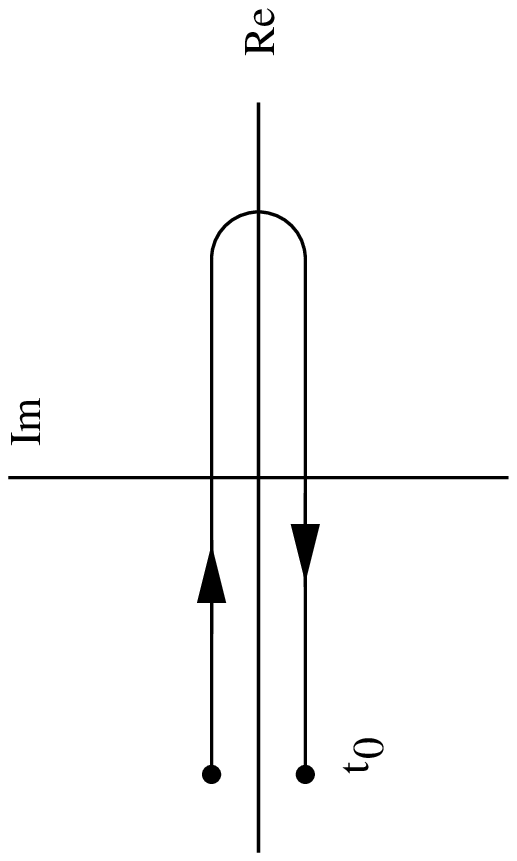}\\
	\vspace{.5cm} Fig.~9
\end{center}

\newpage

\begin{center}
	\includegraphics{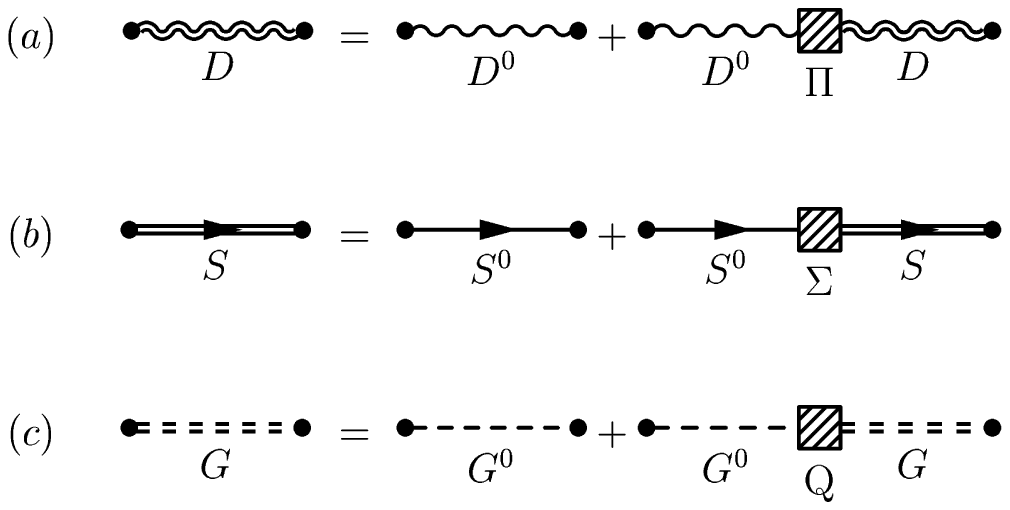}\\
	Fig.~10
\end{center}

\newpage

\begin{center}
	\includegraphics{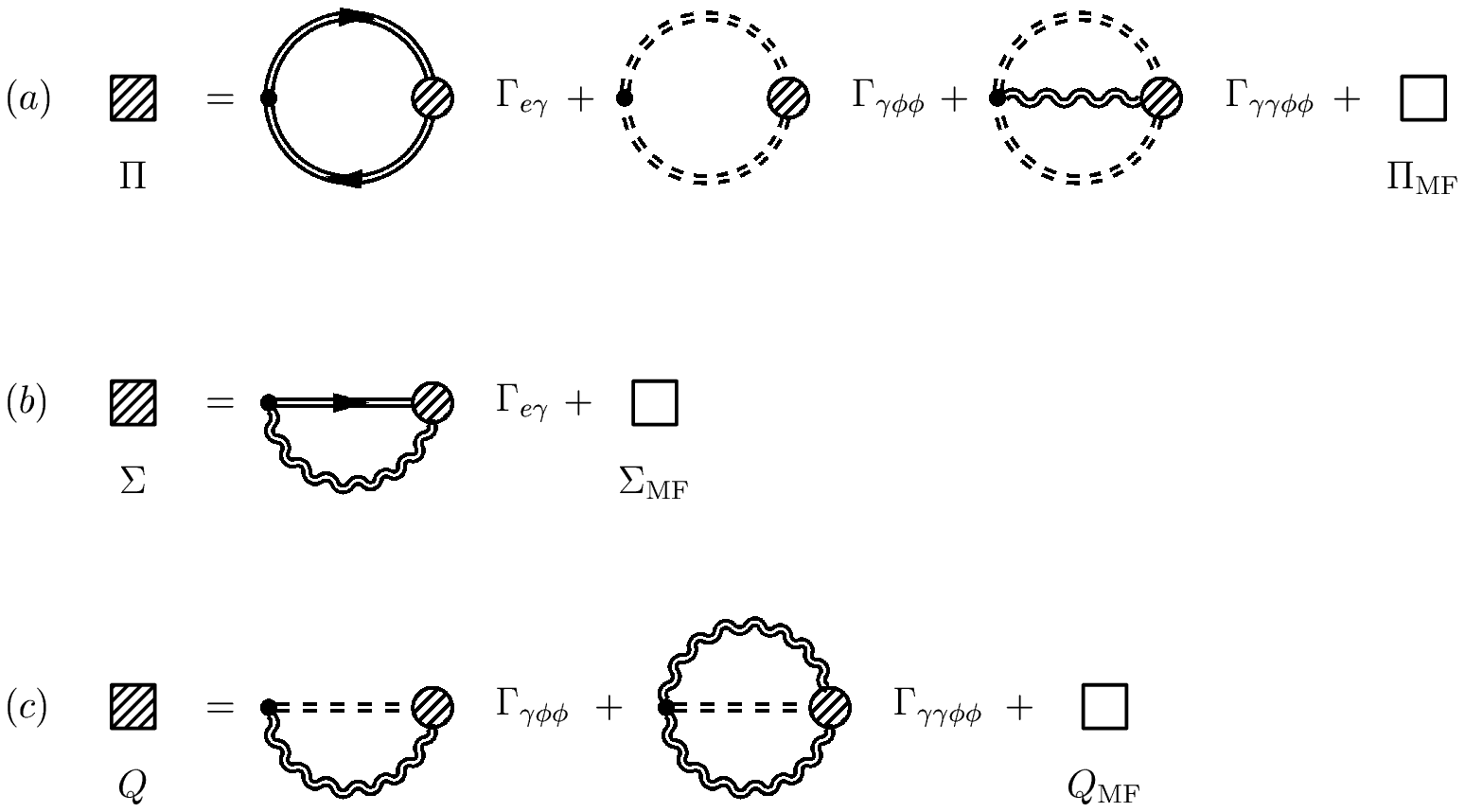}\\
	\vspace{.5cm} Fig.~11
\end{center}

\newpage

\begin{center}
	\includegraphics{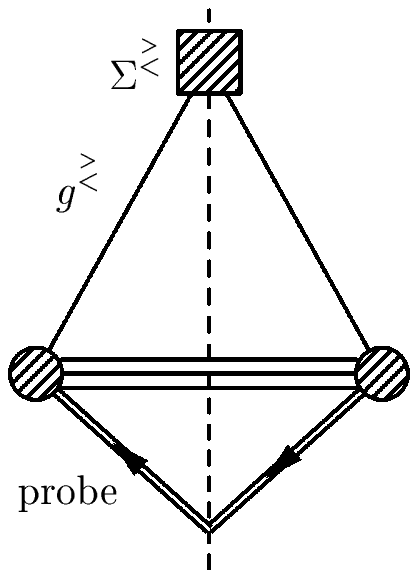}\\
	Fig.~12
\end{center}

\newpage

\begin{center}
	\includegraphics{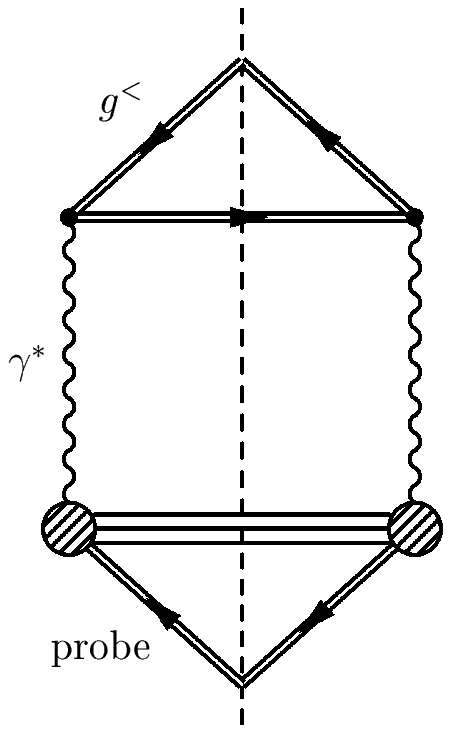}\\
	Fig.~13
\end{center}

\newpage

\begin{center}
	\includegraphics{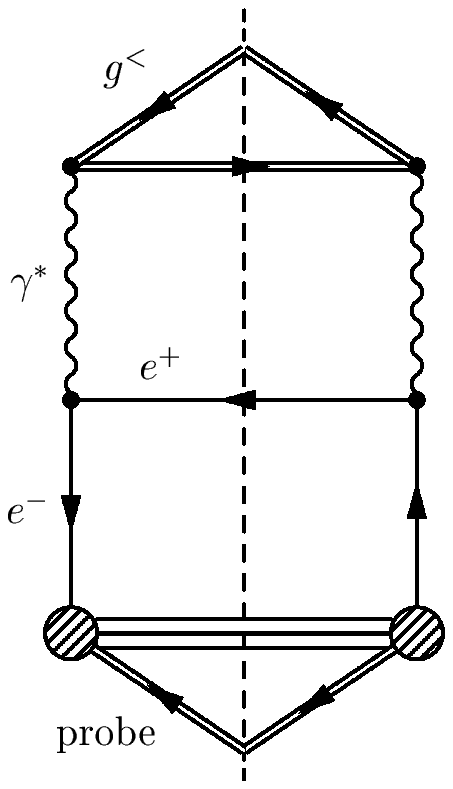}\\
	Fig.~14
\end{center}

\newpage

\begin{center}
	\includegraphics[width=5.5cm]{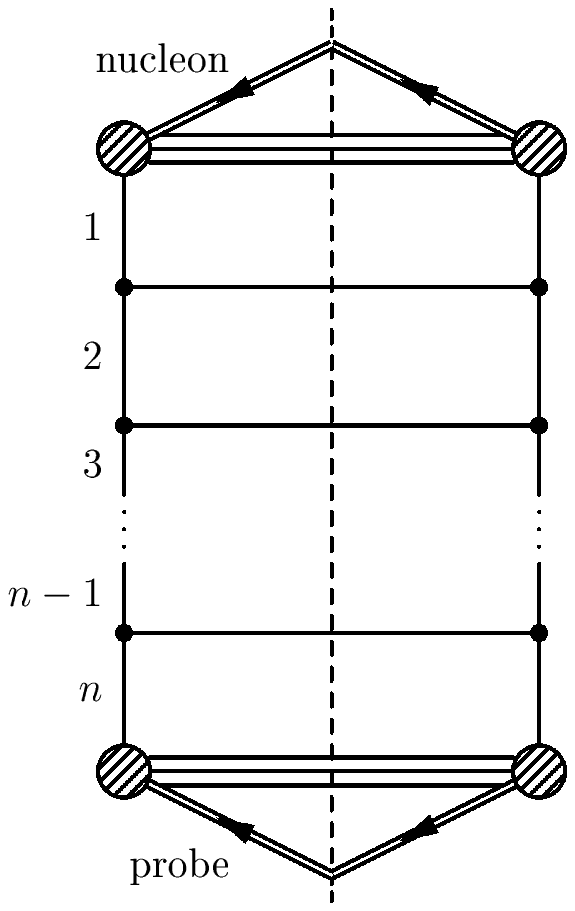}\\
	Fig.~15
\end{center}

\newpage

\begin{center}
	\includegraphics{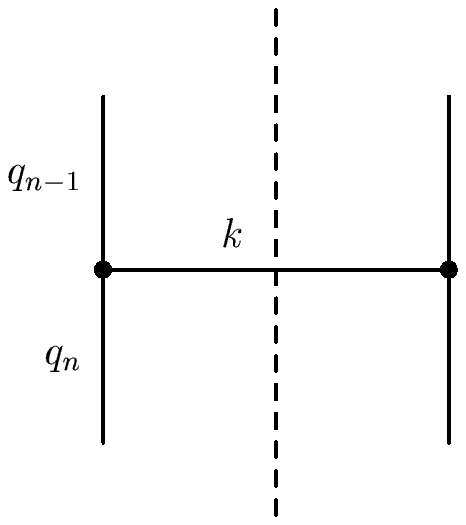}\\
	Fig.~16
\end{center}

\newpage

\begin{center}
	\includegraphics[width=3.5in]{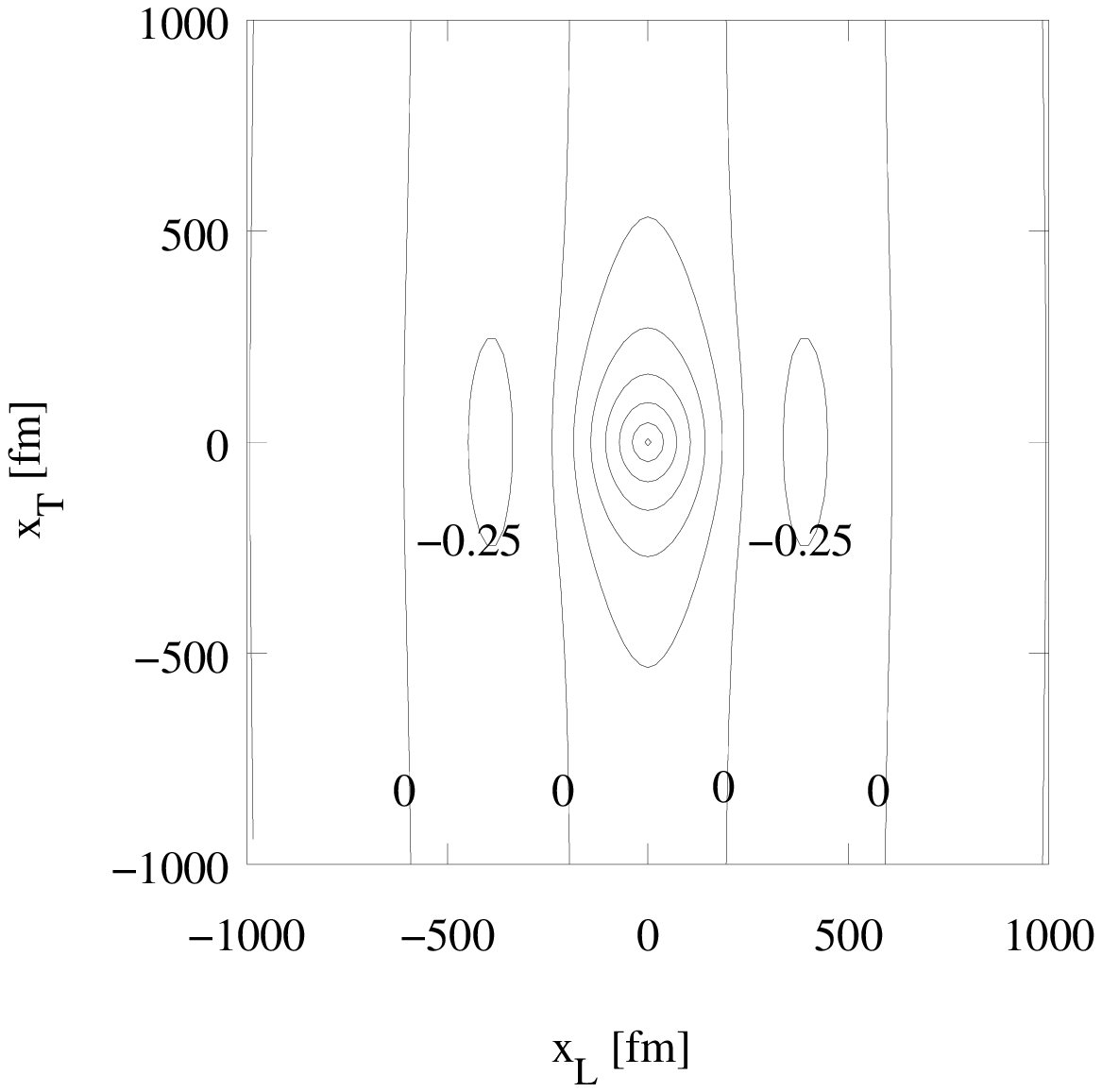}\\
	Fig.~17
\end{center}

\newpage

\begin{center}
	\includegraphics[width=6in]{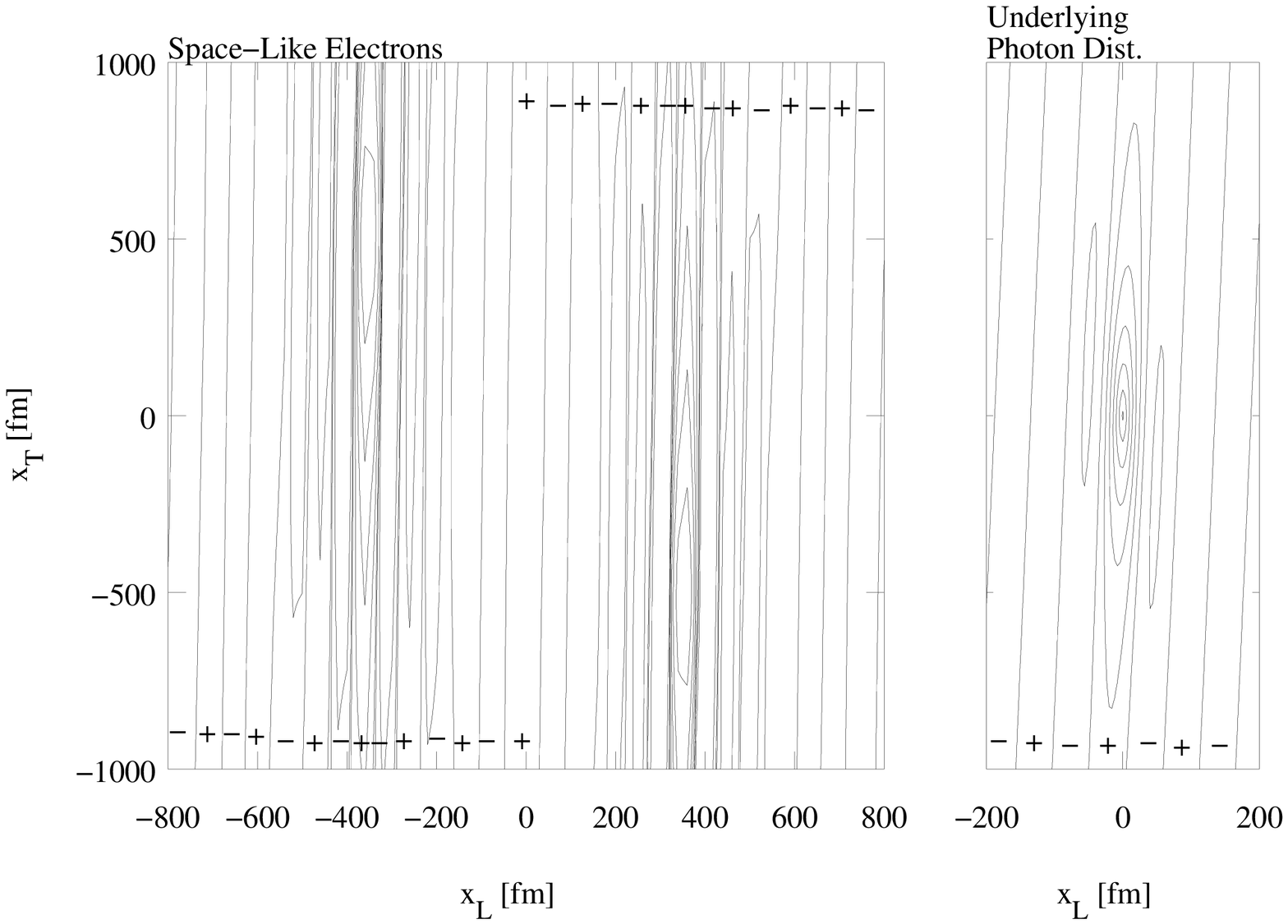}\\
	Fig.~18
\end{center}
 
\newpage

\begin{center}
	\includegraphics[width=6in]{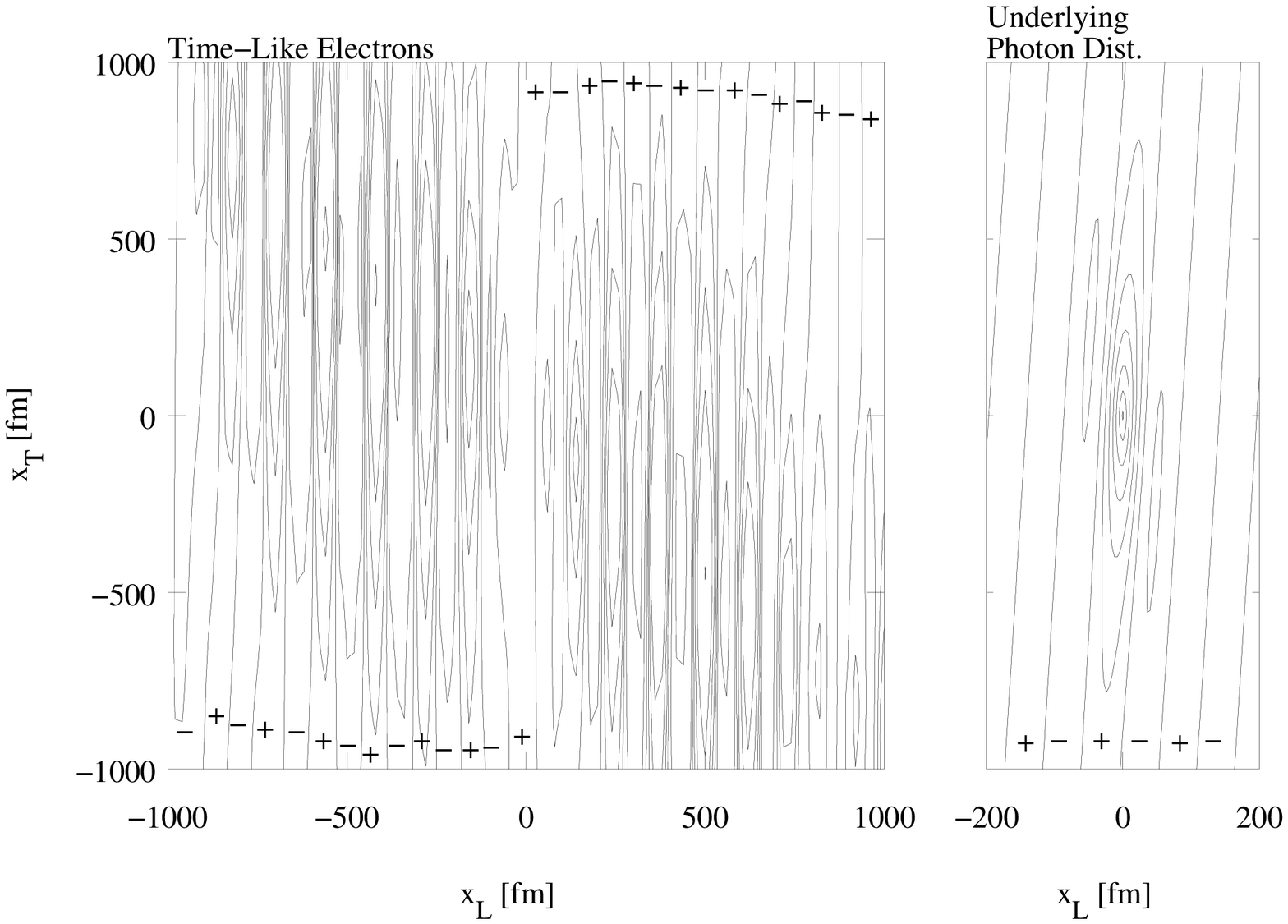}\\
	Fig.~19
\end{center}

\end{document}